\documentclass[conference]{IEEEtran}
\usepackage{cite}
\usepackage[pdftex]{graphicx}
\usepackage[utf8]{inputenc}
\usepackage{tikz}
\usepackage{xspace}
\usepackage{xcolor}
\usepackage{xstring}
\usepackage{color, colortbl}
\usepackage{multicol,multirow}

\usepackage{enumitem}
\usepackage{booktabs}
\usepackage[caption=false,font=footnotesize]{subfig}

\usepackage{fontawesome}
\usepackage{amsmath}
\usepackage{bm}
\usepackage{nicematrix}
\usepackage[pangram]{blindtext}
\usepackage{balance}
\usepackage{totpages}
\usepackage{listings}
\usepackage{multirow}
\usepackage{booktabs}
\usepackage{makecell}
\usepackage{enumitem}
\usepackage{array} 
\usepackage{seqsplit}
\usepackage{float}
\usepackage{diagbox}
\usepackage{threeparttable}
\usepackage[hidelinks,breaklinks=true]{hyperref}
\usepackage[numbers, sort&compress]{natbib}

\usepackage{cleveref}
\usepackage{microtype}

\usepackage[framemethod=tikz]{mdframed}

\usepackage{tcolorbox}

\usepackage{pifont}
\usepackage{soul}

\usepackage{tabularx}

\usepackage{hyperref}

\newcommand{\ie}{\textit{i}.\textit{e}.}
\newcommand{\eg}{\textit{e}.\textit{g}.}

\def\Snospace~{\S{}}

\newif\ifdraft\drafttrue
\newif\ifnotes\notestrue
\ifdraft\else\notesfalse\fi

\input{glyphtounicode}
\newcommand{\squishlist}{
\begin{itemize}[noitemsep,nolistsep]
  \setlength{\itemsep}{-0pt}
}
\newcommand{\squishend}{
  \end{itemize}
}

\newcommand{\PP}[1]{
\noindent{\bf \IfEndWith{#1}{.}{#1}{#1.}}
}

\newcommand{\PPNS}[1]{
\noindent{\bf \IfEndWith{#1}{.}{#1}{#1.}}
}

\newcommand{\aref}[1]{\hyperref[#1]{Appendix~\ref*{#1}}}

\newcommand*\justify{%
  \fontdimen2\font=0.4em
  \fontdimen3\font=0.2em
  \fontdimen4\font=0.1em
  \fontdimen7\font=0.1em
  \hyphenchar\font=`\-
}

\renewcommand{\texttt}[1]{%
  \begingroup
  \ttfamily
  \begingroup\lccode`~=`/\lowercase{\endgroup\def~}{/\discretionary{}{}{}}%
  \begingroup\lccode`~=`[\lowercase{\endgroup\def~}{[\discretionary{}{}{}}%
  \begingroup\lccode`~=`.\lowercase{\endgroup\def~}{.\discretionary{}{}{}}%
  \catcode`/=\active\catcode`[=\active\catcode`.=\active
  \justify\scantokens{#1\noexpand}%
  \endgroup
}

\newcolumntype{P}[1]{>{\centering\arraybackslash}p{#1}}

\newtcolorbox{boxI}{
    colback = lightgray!10, 
    colframe = black, 
    boxrule = 0.5pt, 
    toprule = 0.5pt, 
    arc = 2pt,
    left = 1pt,
    right = 1pt,
    bottom = 0pt,
    top = 0pt
}

\newcounter{observcntr}
\newcommand*{\observ}[1]{%
    \stepcounter{observcntr}%
    \begin{center}
        \begin{boxI}
        \textbf{Observation~\arabic{observcntr}: }{#1}. 
        \end{boxI}
    \end{center}
}

\begin{document}


\title{\textsc{PhishGen}: Understanding LLM Capabilities in Automated Phishing Website Generation}

\author{
    \IEEEauthorblockN{Fujiao Ji}
    \IEEEauthorblockA{University of Tennessee, Knoxville\\
        fji1@utk.edu}
    \\[1.2em]
    \IEEEauthorblockN{Seonhye Park}
    \IEEEauthorblockA{Sungkyunkwan University\\
        qkrtjsgp08@skku.edu}
    \and
    \IEEEauthorblockN{Yerim Kim}
    \IEEEauthorblockA{Sungkyunkwan University\\
        ykeem@skku.edu}
    \\[1.2em]
    \IEEEauthorblockN{Hyoungshick Kim}
    \IEEEauthorblockA{Sungkyunkwan University\\
        hyoung@skku.edu}
    \and
    \IEEEauthorblockN{Yena Cho}
    \IEEEauthorblockA{Sungkyunkwan University\\
        yena0511@skku.edu}
    \\[1.2em]
    \IEEEauthorblockN{Doowon Kim}
    \IEEEauthorblockA{University of Tennessee, Knoxville\\
        doowon@utk.edu}
}



\maketitle

\begin{abstract}
Phishing relies on visual deception, and building convincing phishing pages has historically required web-development skills or pre-built kits. LLMs erode this barrier by generating functional websites from natural-language prompts, yet prior work establishes only that LLMs can produce phishing-like content and does not answer how far LLM-driven generation actually reaches, owing to four limitations: (1) functionality is assessed by inspection rather than execution; (2) visual fidelity is scored by automated metrics or a few internal coders rather than the humans such pages aim to deceive; (3) evaluation is confined to commercial models; and (4) no study benchmarks generated pages against phishing kits deployed in the wild.

We present \textsc{PhishGen}, a framework that evaluates the appearance fidelity and end-to-end functionality of LLM-generated phishing pages. \textsc{PhishGen} constructs prompts across three input modalities (text, screenshots, HTML) in seven combinations with twelve safeguard-bypassing rewriting strategies, generates pages using 14 LLMs spanning commercial, open-source, and uncensored categories, and assesses each via automated similarity metrics, a reference-based detector, a crowdsourced user study with real end-users, and execution-based functionality testing, benchmarking all results against 64 real-world phishing kits.

Our findings complicate common assumptions. Although 96.2\% of safeguard-bypassing outputs render, their perceived fidelity is modest (2.02/5), and only 38.7\% complete the full credential-harvesting pipeline, with silent backend persistence failures (the `Unsaved' error) as the largest failure mode. Contrary to the belief that uncensored models pose the greatest threat, GPT~4.1 reaches 86.3 to 94.1\% functional success while uncensored models rarely exceed 45.4\%. Phishing kits remain more complete per sample, yet three strong LLMs surpass the kit-level rate (70.9 to 89.7\%), showing that LLMs trade average completeness for scalable, low-barrier generation of working phishing pages. We further identify generator-level fingerprints that current similarity-based defenses miss, and distill recommendations for hybrid detectors.

\end{abstract}

\section{Introduction} \label{sec:intro}

Phishing succeeds through visual deception: attackers craft websites that closely mimic legitimate services to harvest user credentials at scale~\cite{Zhang2022crawlphish,Lim2025Phisher,Oest2018Inside}. Producing convincing phishing pages requires either manual web-development effort or pre-built phishing kits, both of which impose non-trivial barriers for adversaries~\cite{Hugo2021Catching}. However, large language models (LLMs) are lowering these barriers for adversaries. Models such as GPT~\cite{GPT:online} can now generate functional websites from natural-language prompts, enabling adversaries with minimal technical expertise to generate phishing pages~\cite{Kim2025LLMOnline, Weinz2025Email, roy2024chatbots, deng2023mindweb}.

Recent work~\cite{roy2024chatbots, lu2025webgen, Yun2024webcode} has shown that LLMs \emph{can} produce content resembling phishing pages. 
Four gaps, however, limit what these findings reveal about real-world risk and about how far LLM-driven generation actually reaches.
\textit{First}, functionality has been assessed by inspection rather than execution. Prior evaluations establish that requested components (\eg, login forms) appear in the output, not that they behave as a credential harvester when exercised.
A functioning credential harvester requires frontend rendering, backend execution, and credential persistence to operate in concert, and functionality should be measured against that standard.
\textit{Second}, visual fidelity, the degree to which a generated page reproduces the appearance of the targeted site, has been scored by automated similarity metrics or by a handful of internal coders, never by the real humans such pages are ultimately built to deceive. 
\textit{Third}, assessment has centered on a narrow set of commercial models, leaving open-source and uncensored LLMs underexplored even as tightening commercial safeguards make them an increasingly attractive option for adversaries.
\textit{Finally}, no prior study compares LLM-generated artifacts against phishing kits deployed in the wild, making practical threat severity difficult to calibrate.

To address these gaps, we present \textsc{PhishGen}, a systematic framework that (1)~constructs prompts across three input modalities (text, screenshots, HTML) in 7 combinations with 12 rewriting strategies to bypass safeguards and boost diversity, (2)~generates phishing pages using 14 LLMs across commercial, open-source, and uncensored categories, and (3)~evaluates each page on appearance fidelity--through automated metrics, a reference-based detector, and a crowdsourced user study with real end-users--and end-to-end functionality, benchmarking both against real-world phishing kits deployed by attackers. Our work is guided by four research questions:



\begin{itemize}[leftmargin=*, itemsep=2pt, parsep=0pt,
    topsep=2pt, partopsep=0pt]
    \item \textbf{RQ\,1:} How closely do LLM-generated websites resemble their legitimate counterparts, and how effectively can LLMs replicate complex web designs?
    \item \textbf{RQ\,2:} What structural and stylistic patterns consistently emerge across LLM-generated phishing websites, regardless of the underlying model or input modality?


    \item \textbf{RQ\,3:} Are LLM-generated websites functionally effective, \ie, can they harvest and store victim credentials?
    \item \textbf{RQ\,4:} How do LLM-generated phishing websites compare to real-world phishing kits in visual fidelity and functionality?
\end{itemize}



\PP{Key Findings.}For \textbf{RQ\,1 (Screenshot-based Visual Fidelity)}, although 96.2\% of safeguard-bypassing outputs render successfully, their perceptual fidelity remains modest (mean value of User-perceived Website Similarity: 2.02/5); LLMs reproduce brand text more than logos (84.4\% vs. 18.2\%), and fidelity degrades on visually complex brands. Commercial models show the strongest visual fidelity, while uncensored LLMs maximize generation success at the cost of visual quality, with negligible difference between regular and explicit ``phishing'' prompts.

For \textbf{RQ\,2 (HTML-based Visual Fidelity)}, generated pages reproduce most of the DOM structure, yet CSS styling exhibits a systematic asymmetry: visual styling of buttons is reproduced faithfully, while typographic properties of forms (\eg, \texttt{font-family}, \texttt{line-height}) are routinely omitted.

For \textbf{RQ\,3 (Functionality)}, only 38.7\% (10,889) of rendered pages complete the full credential-harvesting pipeline. Failures are dominated by backend persistence rather than frontend rendering. Contrary to the common assumption that uncensored models pose the greatest threat, GPT~4.1 attains functional success rates of 86.3\%-94.1\%, while uncensored models rarely exceed 45.4\% even with explicit ``phishing'' prompts. 

For \textbf{RQ\,4 (Comparison to Phishing Kits)}, real-world phishing kits remain more complete per sample: they reproduce brand logos more faithfully (42.2\% vs.\ 18.2\%), contain substantially heavier structures, and achieve higher functionality (57.8\% vs.\ 38.7\%), while textual brand mimicry is comparable (90.6\% vs.\ 84.4\%). However, three strong LLMs exceed the kit-level functional rate, reaching 70.9\%-89.7\%. Thus, LLMs trade lower average completeness for scalable, low-barrier generation of functional phishing pages.

In summary, this paper makes the following contributions:
\begin{itemize}[leftmargin=*, itemsep=2pt, parsep=0pt,
    topsep=0pt, partopsep=0pt]
    \item We design \textsc{PhishGen}, a measurement pipeline that \emph{jointly} measures appearance fidelity and end-to-end functionality of LLM-generated websites, spanning three categories of LLMs, four brands across major phishing-targeted categories, and seven input-modality configurations with diverse prompts.

    \item We characterize generated pages through visual fidelity, using screenshot-based automated metrics, a user study, and HTML-centric structural analysis, as well as end-to-end functionality, including failures in frontend rendering, backend execution, and credential persistence across model categories.

    \item We benchmark LLM-generated phishing pages against 64 real-world phishing kits, grounding our assessment of their practical threat and revealing where automated generation falls short of or exceeds deployed attacks.

    \item We identify candidate generator-level fingerprints--including the DOM--CSS structural-stylistic asymmetry and the textual-over-visual brand mimicry bias--that existing DOM- and visual-similarity-based defenses miss, and distill actionable recommendations for hybrid detectors.


\end{itemize}

\section{Background and Problem Statement} \label{sec:background_related_work}


\subsection{Background}


\PP{LLMs}We consider three classes: \textit{commercial} LLMs (\eg, GPT~\cite{GPT:online}), which offer APIs but face cost and access constraints, \textit{open-source} LLMs (\eg, Llama~3.2~\cite{metallam86:online}), which developers can use, modify, and distribute, and \textit{uncensored} LLMs~\cite{Lin2024Malla}, which are derived from open-source LLMs but operate without vendors' restrictions to generate harmful content.



\PP{Credential-Harvesting Functionality}
A phishing website is \emph{functional} only if it completes the credential-harvesting pipeline from input capture to server-side processing and attacker-side collection. We use this end-to-end pipeline as the basis for evaluating functional correctness in \autoref{sec:method}.

\subsection{Problem Statement}
While prior work~\cite{roy2024chatbots, lu2025webgen, Yun2024webcode} shows that LLMs \textit{can} generate phishing content, what remains open is \textit{how far} LLM-driven generation reaches and \textit{where} it breaks down as follows: 
\looseness=-1

\begin{itemize}[leftmargin=*, itemsep=2pt, parsep=0pt,
    topsep=0pt, partopsep=0pt]
    \item \emph{End-to-end Functionality.} Prior work verifies that functional components \emph{appear} in the output~\cite{roy2024chatbots}, rather than whether the credential-harvesting pipeline actually \emph{executes} end-to-end.

    \item \emph{Model Coverage.} A controlled apples-to-apples assessment of commercial, open-source, and uncensored models under identical targets, prompts, and evaluation remains limited.

    \item \emph{Website Characterization.} The structural and perceptual characteristics of LLM-generated sites remain underexplored, including their DOM/CSS organization, branding strategies, textual-vs.-visual biases, and visual deceptiveness as perceived by independent external users.

    \item \emph{Kit-Level Real-World Grounding.} LLM-generated sites lack systematic benchmarking against real-world phishing kits under common visual, functional, and structural measures.

\end{itemize}

\subsection{Our Threat Model and Scope} 
\label{subsec:problem}

\PP{Threat Model.}
We adopt a controlled, measurement-oriented model that isolates the intrinsic capability of LLMs to generate operational phishing sites. The adversary seeks to impersonate a widely used legitimate service (\eg, Facebook), relying on LLMs and publicly accessible resources to produce pages that deceive victims into submitting their credentials, which are then delivered to an attacker-controlled endpoint. At the prompt level, the adversary may apply diverse strategies (\eg, role-play) and legitimate brand resources to elicit both frontend and backend credential-harvesting logic. Adversarial effort is confined to prompting. 
An unsatisfactory output may be discarded and regenerated under a revised prompt, but the returned code is not hand-edited, reflecting an attacker with limited programming background.
This scoping attributes observed successes and failures to the model together with the prompting effort that drives it, rather than to manual web development skill.

\PP{Out of Scope.}We deliberately exclude the downstream stages of a phishing campaign: hosting, domain registration, takedown evasion, and any human post-editing of the generated code. This scoping isolates the generation step, allowing us to attribute observed successes and failures to the LLM's intrinsic capability rather than to external attacker effort or infrastructure.

\section{Experimental Setup and Analysis Plan} 
\label{sec:method}
\begin{figure}[t]
\centering
\includegraphics[width=\columnwidth]{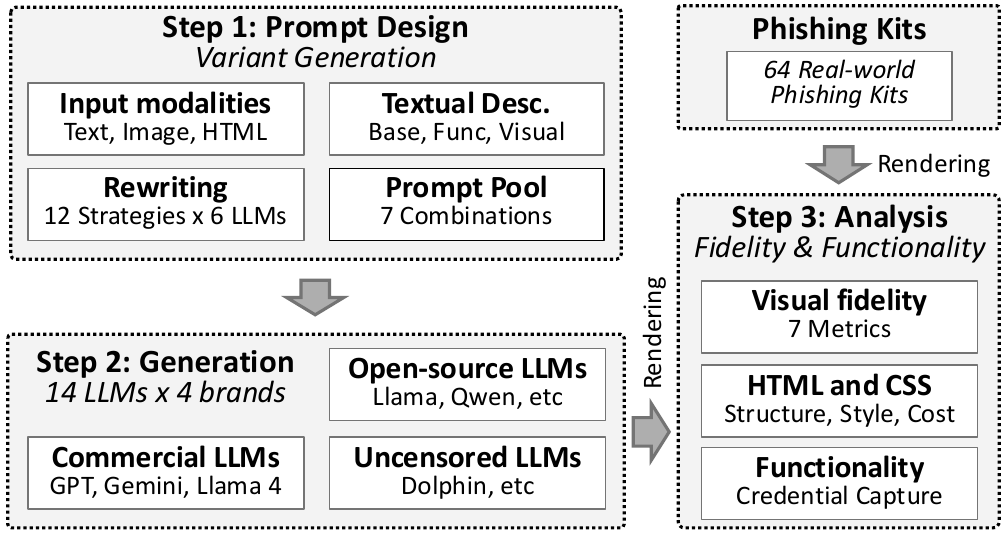}
\caption{Analysis Overview of LLM-Generated Phishing.}
~\label{fig:prompt_design}
\end{figure}

\autoref{fig:prompt_design} illustrates our systematic experimental framework (\textsc{PhishGen}). The framework comprises three main phases: (1) systematic prompt design and variation to elicit phishing website generation across realistic scenarios (\autoref{subsec:prompt}), (2) controlled generation of phishing sites using LLMs under identical experimental conditions (\autoref{subsec:brand} and \autoref{subsec:web_gen}), and (3) comprehensive analysis of LLM performance, the characteristics of generated phishing websites, and their differences from real-world phishing kits (\autoref{subsec:analysis} and \autoref{subsec:kits}). 


\subsection{Prompt Design for Real-world Scenarios}
\label{subsec:prompt}


Guided by the threat model, we construct a controlled prompt suite to study how prompt design affects LLM-generated phishing websites and to emulate attacker interactions with LLMs. We therefore vary the input modality, rewriting strategy, and rewriting model. Together, these variations capture diverse attacker resources and prompting behaviors, and reduce potential bias introduced by any single rewriting model.

\PP{Three Input Modalities}Attackers can use different input modalities when prompting LLMs to replicate target websites: (1) textual descriptions (base prompts, functional specifications, and visual characteristics), (2) screenshots or other visual references, and (3) HTML code. Effective phishing sites should visually resemble legitimate services to gain user trust while replicating critical workflows to capture victims' credentials.


\begin{itemize}[leftmargin=*, itemsep=-1pt, topsep=2pt]
    \item \textbf{Textual Descriptions:} Attackers can employ three strategies: basic, visual, and functional descriptions. Basic descriptions request a brand mimic without explicit specifications (\eg, ``generate a website that looks like Facebook.com''). Visual and functional descriptions are either manually composed or extracted from target websites by LLMs, covering visual characteristics (\eg, layout, color scheme) and/or functional behaviors (\eg, authentication flows, form validation), and are then incorporated into prompts. For each, we build two variants:
    one consists of screenshot-derived descriptions generated by LLMs (captions), the other is manually constructed from captions and prior work~\cite{roy2024chatbots, lu2025webgen}. An example of Text+Function is shown in~\aref{appendix_subsec:example_prompt}.

    \item \textbf{Screenshots:} Attackers may augment text with screenshots of legitimate sites as auxiliary input, guiding LLMs to replicate the target's layout, styling, and interactive elements (\eg, ``generate a website that looks like the attached screenshot'').

    \item \textbf{HTML Code:} Attackers may provide the HTML of target websites as auxiliary input, letting LLMs reuse or adapt existing page structures, styles, or functional components. To manage prompt length, we clean HTML using \textit{HTMLRAG} \cite{tan2025htmlrag} before passing the HTML content as input.

\end{itemize}

By varying the depth and combination of these input modalities, we construct \texttt{7} prompt combinations (denoted as $N_{\text{comb}}$), as shown in \aref{appendix_subsec:prompt}. 


\PP{Rewriting Techniques}To bypass LLM safeguards, increase volume and diversity, we employ 12 prompt rewriting strategies derived from prior work~\cite{wei2019eda, prabhumoye2017linguistic, kriz2018simplification, li2024empowering, min2020syntactic, poliak2018collecting}, detailed in~\aref{appendix_subsec:prompt}-\autoref{tab:prompt_12}. To mitigate single-model bias, we rewrite with six commercial and open-source LLMs (see~\aref{appendix_subsec:llms}-\autoref{tab:llm_webgen}), with minimal human intervention--manually reviewing outputs and regenerating unsatisfactory ones. Rewriting is applied only to manually crafted textual descriptions, as screenshot-derived descriptions (captions) are detailed and overlap with auxiliary input modalities (\ie, screenshots). Moreover, rewriting all prompt examples would be prohibitively expensive given time and resources.

%

%



\PP{Prompts for Uncensored LLMs}To characterize the behavior of uncensored LLMs, we construct a parallel set of prompts by inserting the keyword phishing (\eg, ``generate a phishing website that looks like Facebook.com''). For clarity, we refer to prompts without this keyword as \textit{regular prompts} and those containing it as \textit{phishing prompts} throughout the paper.

\PP{Final Prompt Dataset.}The rewriting process yields $N_{\text{rewrite}}^{\text{G}} = 72$ variations of the general prompt (12 rewriting strategies $\times$ 6 LLMs). 
To construct the final dataset, we combine these with two baseline prompts: the original general prompt and the caption-based prompt. 
The total number of prompts is defined as $N_{\text{prompt}} = (N_{\text{rewrite}}^{\text{G}} + 2) \times N_{\text{comb}}$, where $N_{\text{comb}}$ denotes the number of prompt combinations applied. 
For image-supported LLMs ($N_{\text{comb}} = 7$), this yields 518 prompts. For text-only LLMs ($N_{\text{comb}} = 4$), it results in 296 prompts.

\subsection{Selection of Target Brands and LLMs} \label{subsec:brand}
\PP{Target Brand Selection}
According to APWG Q2 trends~\cite{APWG73:online}, e-commerce, payment, social media, and financial institutions are frequently targeted. To explore the underlying capabilities of LLMs in these areas, we select one widely impersonated brand from each of the four most-targeted APWG categories: Amazon (e-commerce), PayPal (payment), Facebook (social media), and Wells Fargo (financial). Lower-ranked or region-specific brands are valuable but may confound evaluation because poorer outputs could reflect limited pretraining exposure rather than weaker generation capability. We therefore restrict our study to these well-known brands.
For each brand, we collect login page resources to serve as generation references. Notably, the Wells Fargo login page exhibits significantly greater design complexity than the other targets, providing a more challenging test case (see \autoref{fig:brand_loginpage}). 



%

\PP{LLM Selection for Phishing Generation}
Prior work~\cite{roy2024chatbots, lu2025webgen, Devising2023Heiding} examines a narrow subset of commercial models (\eg, GPT~\cite{GPT:online}), limiting insight into the evolving threat landscape. As commercial vendors strengthen safety guardrails, adversaries may shift to open-source and uncensored models to evade content restrictions and cut costs. 
We thus organize our study around three model categories: commercial, open-source, and uncensored LLMs, and select representative LLMs for website generation. Model details are provided in \aref{appendix_subsec:llms}.

\subsection{Phishing Website Generation} 
\label{subsec:web_gen}

Based on the three model categories above, we use crafted prompts to instruct each model to generate HTML and PHP code for functional phishing websites, as PHP is commonly used in phishing kits~\cite{Lee2024Beneath,Hugo2021Catching,liu2023knowledge}. 
Since visual appearance is critical, we render the generated HTML and capture screenshots with Playwright \cite{Playwright:online} (Chromium on Linux).
By comparing generated screenshots with the targeted brands' legitimate websites, we can obtain a more comprehensive understanding of the visual quality of the generated content.




During generation, we first check whether each output can be successfully extracted according to the expected format; failed outputs are directly regenerated. Successfully extracted websites are then rendered and screened using FastDup~\cite{visualla13:online}, CLIP~\cite{CLIP66:online}, and manual inspection. Based on the similarity distribution and cluster analysis (threshold: 0.6, with clear separation), we identify 9,383 samples for regeneration. Two researchers jointly inspect these cases for issues such as missing brand elements or incorrect layouts and refine only the prompts by adding targeted constraints (\eg, requiring inline CSS) before regenerating the corresponding websites consistently across all models. All subsequent analyses use the revised dataset.

\subsection{Phishing Kits} \label{subsec:kits}

We use 4,409 phishing kits shared by \cite{Lim2025Phisher}, which were collected from APWG eCX between July 2021 and March 2025. Because these kits target a wide range of brands, we first filter them based on file content, retaining only kits containing keywords associated with the four target brands. We then render the remaining kits as screenshots in a controlled environment. Because keyword-based filtering may still retain kits targeting other brands, we cluster the kits based on screenshot similarity (FastDup) and discard clusters associated with non-target brands. Finally, we manually inspect the remaining screenshots, resulting in a dataset of 64 phishing kits targeting the four brands. Detailed statistics are reported in \autoref{tab:kit_info}.

\subsection{Analysis Plan for Characteristics} \label{subsec:analysis}

Visual appearance shapes user trust, whereas functional mechanisms determine whether sensitive information can actually be captured. We therefore analyze generated phishing websites, as well as real-world phishing kits, along \textit{appearance fidelity}, characterizing visual mimicry of the target brand from both the rendered screenshot and the HTML source, and \textit{functionality}, capturing whether the credential-harvesting pipeline operates correctly end-to-end.

\PP{Screenshot-based Characterization.}Given the diversity of LLM-generated pages, no single metric captures their resemblance to legitimate counterparts. We render each page with Playwright and evaluate it along seven complementary axes: 

The (1) \textit{safeguard bypass rate} quantifies how often a model produces any output, as some LLMs refuse outright (\eg, ``\textit{I'm sorry, but I can't assist with that request.}''). Among returned responses, some are visually empty or broken (\eg, \autoref{fig:tet_output}); we use FastDup~\cite{visualla13:online} clustering (threshold: 0.9), and report the fraction of non-failed outputs as the (2) \textit{valid website rate}.

Since visual quality ultimately rests on human perception, we measure (3) \textit{UWS} (User-perceived Website Similarity) via a crowdsourced user study refining the Website Appearance Scale (WAS) of Roy~\textit{et~al.}~\cite{roy2024chatbots}. 
Participants rate each evaluated page against its legitimate counterpart shown side-by-side, following \cite{Hao2024Usenix}, without knowing whether it is legitimate, phishing, or LLM-generated. Because we assess perceived visual similarity rather than phishing detection, we use paired comparisons to directly measure visual fidelity.
We recruit 61 Prolific participants under an IRB-approved protocol. A pilot of the original WAS rubric reveals that lay participants disregard its complex descriptions and fall back on intuitive judgments, so we replace those anchors with concise descriptors (1: ``Not similar at all'' through 5: ``Nearly identical'') while preserving the 5-point Likert structure. We use {GPT~5} to pre-categorize a large sample pool into similarity tiers and then uniformly draw 2-5 images per (model, brand) combination for balanced coverage across quality tiers. Beyond the Likert rating, we collect per-trial confidence, rationale tags, and an embedded attention check; filter out responses with confidence~$\leq 2$; and report inter-rater reliability (IRR) via Krippendorff's~$\alpha$. Each sample receives at least two independent ratings, yielding 10-11 per (model, brand) cell, and UWS is the mean (full protocol and findings in~\aref{appendix_subsec:user_study_analysis} and~\aref{appendix_subsec:user_study}).

We measure (4)~\textit{homogeneity} as the fraction of screenshots in the largest FastDup cluster (threshold: 0.96); higher values indicate lower diversity. For brand mimicry, (5)~\textit{brand text matching} extracts text via PaddleOCR~\cite{PaddlePa11:online} and fuzzy-matches it against the target brand string ($\geq 0.9$), (6)~\textit{brand logo matching} uses the Google Cloud Vision API~\cite{VisionAI50:online} (confidence $\geq 0.7$); both thresholds are empirically selected based on clear performance separation, and (7)~\textit{Phishpedia detection rate} measures the fraction of generated pages detected by Phishpedia~\cite{Lin2021Phishpedia} from screenshots (threshold: 0.83). 
%
We choose Phishpedia because it exposes the pure brand-matching signal from screenshots, capturing the brand-impersonation metric.



\PP{HTML-based Characterization.}Screenshots capture what users see, but HTML governs what is actually rendered and how the page behaves. We therefore complement the screenshot-based analysis with an HTML-centric one along four axes. \textit{Generation cost} (time and monetary costs) reflects the computational scalability of each model for an attacker operating at scale. \textit{Structural analysis} quantifies the organization and complexity of the generated markup. \textit{HTML similarity} measures structural and semantic imitation of legitimate pages, indicating the effectiveness of mimicry at the source-code level. \textit{Form and button CSS styling} capture interactive realism, with particular attention to the design of credential-submission elements. Implementation details for structure, similarity, and styling are provided in~\aref{appendix_subsec:html_sim_plan} and~\aref{appendix:html_style_analysis}.

\PP{Functionality Characterization.}A phishing website is operationally useful only if it can capture, transmit, and deliver victim credentials to an attacker-controlled destination. We therefore define \textit{functionality} as the capability of the generated full-stack implementation to execute this credential-harvesting lifecycle end-to-end (\autoref{sec:background_related_work}). Because generated websites and phishing kits expose different points
of this lifecycle, we assess them accordingly. For generated websites, we (1) render the generated HTML, (2) inject synthetic credentials into input fields, (3) trigger submission, and (4) verify that scripts produce a local log file containing those credentials. A website is classified as \textit{functional} only when the entire client-server workflow succeeds. The full procedure--\textit{Browser Automation}, \textit{Submission Mechanisms}, and \textit{Evaluation Criteria}--is detailed in~\aref{appendix_subsec:function}. For phishing kits, we instead verify the exfiltration destination--the attacker-controlled address to which harvested credentials are delivered.

\section{Results of Screenshot-based Evaluation} \label{sec:result_screenshot}

\begin{table*}[!htbp]
\centering
\caption{Screenshot-based Visual Characteristics of LLM-Generated Phishing Websites. Safeguard bypass rates are computed over the total number of prompts, valid website rates over safeguard bypass responses, and all other metrics over valid websites.}
\label{tab:scr_cluster}
\footnotesize
\resizebox{1\linewidth}{!}{
    \begin{tabular}{rrrr rrrr rr r}
    \toprule
    & \multicolumn{1}{c}{\textbf{Type}} & \multicolumn{1}{c}{\textbf{LLMs}} & \multicolumn{1}{c}{\textbf{Total}} & \multicolumn{1}{c}{\textbf{\makecell{Safeguard \\Bypass}}} & \multicolumn{1}{c}{\textbf{\makecell{Valid \\Websites}}} & \multicolumn{1}{c}{\textbf{\makecell{User \\Eval.}}} & \multicolumn{1}{c}{\textbf{\textbf{Homo.}}} & \multicolumn{1}{c}{\textbf{ \makecell{Brand Text\\Matching}}} & \multicolumn{1}{c}{\textbf{\makecell{Brand Logo\\Matching}}} & \multicolumn{1}{c}{\textbf{\makecell{Phishpedia\\Detect}}} \\\midrule
    \multirow{14}{*}{\rotatebox{90}{\textbf{Amazon} (6,364)}} & \multirow{4}{*}{\rotatebox{90}{\textbf{Com.}}} 
        & \textbf{GPT~5}             & 518 & 433 \phantom{0}(83.6\%) & 433 (100.0\%)           & 3.00 & 367 (84.8\%) &  34 \phantom{00}(7.9\%) & 18 \phantom{0}(4.2\%)  & 14 (3.2\%)\\
      & & \textbf{GPT~4.1}           & 518 & 498 \phantom{0}(96.1\%) & 498 (100.0\%)           & 3.50 & 487 (97.8\%) & 498 (100.0\%)           & 455 (91.4\%)           & 431 (86.5\%)\\ 
      & & \textbf{Gemini}            & 518 & 511 \phantom{0}(98.6\%) & 509 \phantom{0}(99.6\%) & 2.91 & 488 (95.9\%) & 497 \phantom{0}(97.6\%) & 211 (41.5\%)           & 132 (25.9\%)\\
      & & \textbf{Llama~4}           & 518 & 518 (100.0\%)           & 505 \phantom{0}(97.5\%) & 2.80 & 457 (90.5\%) & 457 \phantom{0}(90.5\%) & 176 (34.9\%)           & 170 (33.7\%)\\
      \cmidrule{2-11}
    & \multirow{6}{*}{\rotatebox{90}{\textbf{Open}}}  
         & \textbf{Llama~3.2}         & 518 & 495 \phantom{0}(95.6\%)  & 491 \phantom{0}(99.2\%)  & 1.50 &  356 (72.5\%) & 457 \phantom{0}(93.1\%) & 5 \phantom{0}(1.0\%)    & 5 (1.0\%)\\ 
       & & \textbf{Qwen~3}            & 518 & 518 (100.0\%)            & 518 (100.0\%)            & 2.80 &  478 (92.3\%) & 492 \phantom{0}(95.0\%) & 102 (19.7\%)            & 100 (19.3\%)\\
       & & \textbf{Qwen}              & 518 & 518 (100.0\%)            & 518 (100.0\%)            & 2.30 &  472 (91.1\%) & 487 \phantom{0}(94.0\%) & 0 \phantom{0}(0.0\%)    & 0 (0.0\%)\\ 
       & & \textbf{Mistralai}         & 518 & 518 (100.0\%)            & 517 \phantom{0}(99.8\%)  & 1.73 &  401 (77.6\%) & 480 \phantom{0}(92.8\%) & 108 (20.9\%)            & 75 (14.5\%)\\ 
       & & \textbf{DeepSeekVL2}       & 518 & 517 \phantom{0}(99.8\%)  & 477 \phantom{0}(92.3\%)  & 2.00 &  386 (80.9\%) & 434 \phantom{0}(91.0\%) & 0 \phantom{0}(0.0\%)    & 0 (0.0\%)\\ 
       & & \textbf{Phi}               & 518 & 479 \phantom{0}(92.5\%)  & 459 \phantom{0}(95.8\%)  & 1.27 &  239 (52.1\%) & 332 \phantom{0}(72.3\%) & 0 \phantom{0}(0.0\%)    & 0 (0.0\%)\\ 
      \cmidrule{2-11}
    & \multirow{4}{*}{\rotatebox{90}{\textbf{Uncensor.}}}  
          & \textbf{Daredevil}         & 296 & 296 (100.0\%) & 278 \phantom{0}(93.9\%) & 1.90 & 206 (74.1\%) & 265 \phantom{0}(95.3\%) & 1 \phantom{0}(0.4\%)  & 1 (0.4\%)\\
       &  & \textbf{WhiteRabbitNeo}    & 296 & 296 (100.0\%) & 293 \phantom{0}(99.0\%) & 1.90 & 169 (57.7\%) & 234 \phantom{0}(79.9\%) & 7 \phantom{0}(2.4\%)  & 6 (2.0\%)\\ 
       &  & \textbf{OpenVoid}          & 296 & 296 (100.0\%) & 285 \phantom{0}(96.3\%) & 1.50 & 184 (64.6\%) & 262 \phantom{0}(91.9\%) & 24 \phantom{0}(8.4\%) & 25 (8.8\%)\\ 
       &  & \textbf{Dolphin}           & 296 & 296 (100.0\%) & 277 \phantom{0}(93.6\%) & 1.80 & 207 (74.7\%) & 201 \phantom{0}(72.6\%) & 0 \phantom{0}(0.0\%)  & 0 (0.0\%)\\ \midrule
     
      \multirow{14}{*}{\rotatebox{90}{\textbf{Facebook} (6,364)}} & \multirow{4}{*}{\rotatebox{90}{\textbf{Com.}}} 
      & \textbf{GPT~5}               & 518 & 454 \phantom{0}(87.6\%)  & 454 (100.0\%)            & 2.10 & 396 (87.2\%) & 104 \phantom{0}(22.9\%) & 62 (13.7\%)  &  81 (17.8\%)\\
      &  & \textbf{GPT~4.1}            & 518 & 518 (100.0\%)            & 518 (100.0\%)          & 2.90 & 481 (92.9\%) & 513 \phantom{0}(99.0\%) & 438 (84.6\%) &  453 (87.5\%)\\       
      & & \textbf{Gemini}            & 518 & 516 \phantom{0}(99.6\%)  & 515 \phantom{0}(99.8\%)  & 3.20 & 487 (94.6\%) & 496 \phantom{0}(96.3\%) & 386 (75.0\%) &  401 (77.9\%)\\
      & & \textbf{Llama~4}           & 518 & 518 (100.0\%)            & 501 \phantom{0}(96.7\%)  & 2.90 & 381 (76.0\%) & 465 \phantom{0}(92.8\%) & 237 (47.3\%) &  183 (36.5\%)\\
      \cmidrule{2-11}
    & \multirow{6}{*}{\rotatebox{90}{\textbf{Open}}}  
         & \textbf{Llama~3.2}        & 518 & 479 \phantom{0}(92.5\%)  & 457 \phantom{0}(95.4\%) & 1.30 & 251 (54.9\%) & 430 \phantom{0}(94.1\%) & 97 (21.2\%)  &  58 (12.7\%)\\
       & & \textbf{Qwen~3}           & 518 & 517 \phantom{0}(99.8\%)  & 507 \phantom{0}(98.1\%) & 2.60 & 421 (83.0\%) & 487 \phantom{0}(96.1\%) & 240 (47.3\%) &  299 (59.0\%) \\
       & & \textbf{Qwen}             & 518 & 518 (100.0\%)            & 493 \phantom{0}(95.2\%) & 1.80 & 305 (61.9\%) & 452 \phantom{0}(91.7\%) & 75 (15.2\%)  &  73 (14.8\%)\\ 
       & & \textbf{Mistralai}        & 518 & 518 (100.0\%)            & 513 \phantom{0}(99.0\%) & 1.90 & 310 (60.4\%) & 493 \phantom{0}(96.1\%) & 257 (50.1\%) &  171 (33.3\%) \\ 
       & & \textbf{DeepSeekVL2}      & 518 & 518 (100.0\%)            & 492 \phantom{0}(95.0\%) & 1.50 & 268 (54.5\%) & 429 \phantom{0}(87.2\%) & 141 (28.7\%) &  114 (23.2\%) \\ 
       & & \textbf{Phi}              & 518 & 516 \phantom{0}(99.6\%)  & 486 \phantom{0}(94.2\%) & 1.10 & 265 (54.5\%) & 412 \phantom{0}(84.8\%) & 130 (26.7\%) &  38 (7.8\%) \\
       \cmidrule{2-11}
    & \multirow{4}{*}{\rotatebox{90}{\textbf{Uncensor.}}}  
          & \textbf{Daredevil}         & 296 & 296 (100.0\%)           & 248 \phantom{0}(83.8\%) & 1.70 & 141 (56.9\%) & 230 \phantom{0}(92.7\%) & 61 (24.6\%)           &  31 (12.5\%) \\
       &  & \textbf{WhiteRabbitNeo}    & 296 & 291 \phantom{0}(98.3\%) & 286 \phantom{0}(98.3\%) & 2.00 & 136 (47.6\%) & 239 \phantom{0}(83.6\%) & 47 (16.4\%)           &  36 (12.6\%)\\ 
       &  & \textbf{OpenVoid}          & 296 & 296 (100.0\%)           & 232 \phantom{0}(78.4\%) & 1.50 & 114 (49.1\%) & 193 \phantom{0}(83.2\%) & 63 (27.2\%)           &  28 (12.1\%)\\ 
       &  & \textbf{Dolphin}           & 296 & 296 (100.0\%)           & 264 \phantom{0}(89.2\%) & 1.50 & 199 (75.4\%) & 199 \phantom{0}(75.4\%) & 16 \phantom{0}(6.1\%) &  9 (3.4\%)\\ \midrule
      \multirow{14}{*}{\rotatebox{90}{\textbf{PayPal} (6,364)}} & \multirow{4}{*}{\rotatebox{90}{\textbf{Com.}}} 
        & \textbf{GPT~5}             & 518 & 402 \phantom{0}(77.6\%) & 402 (100.0\%)             & 1.80 & 353 (87.8\%) & 11 \phantom{00}(2.7\%)  & 5 \phantom{0}(1.2\%) &  1 (0.2\%)  \\
      &  & \textbf{GPT~4.1}            & 518 & 481 \phantom{0}(92.9\%) & 481 (100.0\%)           & 3.20 & 424 (88.1\%) & 380 \phantom{0}(79.0\%) & 374 (77.8\%)         &  267 (55.5\%)  \\ 
      & & \textbf{Gemini}            & 518 & 512 \phantom{0}(98.8\%) & 511 \phantom{0}(99.8\%)   & 3.09 & 487 (95.3\%) & 498 \phantom{0}(97.5\%) & 285 (55.8\%)         &  263 (51.5\%)  \\
      & & \textbf{Llama~4}           & 518 & 517 \phantom{0}(99.8\%) & 517 (100.0\%)             & 2.45 & 431 (83.4\%) & 358 \phantom{0}(69.2\%) & 229 (44.3\%)         &  10 (1.9\%)  \\
      \cmidrule{2-11}
    & \multirow{6}{*}{\rotatebox{90}{\textbf{Open}}}  
         & \textbf{Llama~3.2}        & 518 & 452 \phantom{0}(87.3\%)  & 435 \phantom{0}(96.2\%)  & 2.27 & 255 (58.6\%) & 394 \phantom{0}(90.6\%) & 31 \phantom{0}(7.1\%)  &  3 (0.7\%)\\ 
       & & \textbf{Qwen~3}           & 518 & 517 \phantom{0}(99.8\%)  & 504 \phantom{0}(97.5\%)  & 2.36 & 457 (90.7\%) & 458 \phantom{0}(90.9\%) & 146 (29.0\%)           &  73 (14.5\%)\\
       & & \textbf{Qwen}             & 518 & 518 (100.0\%)            & 509 \phantom{0}(98.3\%)  & 1.60 & 415 (81.5\%) & 472 \phantom{0}(92.7\%) & 54 (10.6\%)            &  0 (0.0\%)\\ 
       & & \textbf{Mistralai}        & 518 & 518 (100.0\%)            & 511 \phantom{0}(98.6\%)  & 2.10 & 371 (72.6\%) & 492 \phantom{0}(96.3\%) & 155 (30.3\%)           &  26 (5.1\%)\\ 
       & & \textbf{DeepSeekVL2}      & 518 & 515 \phantom{0}(99.4\%)  & 514 \phantom{0}(99.8\%)  & 1.90 & 322 (62.6\%) & 374 \phantom{0}(72.8\%) & 42 \phantom{0}(8.2\%)  &  1 (0.2\%)\\ 
       & & \textbf{Phi}              & 518 & 467 \phantom{0}(90.2\%)  & 434 \phantom{0}(92.9\%)  & 1.00 & 196 (45.2\%) & 309 \phantom{0}(71.2\%) & 21 \phantom{0}(4.8\%)  &  2 (0.5\%)\\ 
      \cmidrule{2-11}
    & \multirow{4}{*}{\rotatebox{90}{\textbf{Uncensor.}}}  
          & \textbf{Daredevil}         & 296 & 296 (100.0\%)           & 264 \phantom{0}(89.2\%) & 2.80 & 187 (70.8\%) & 244 \phantom{0}(92.4\%) & 33 (12.5\%)           &  1 (0.4\%)\\
       &  & \textbf{WhiteRabbitNeo}    & 296 & 296 (100.0\%)           & 291 \phantom{0}(98.3\%) & 1.80 & 185 (63.6\%) & 231 \phantom{0}(79.4\%) & 24 \phantom{0}(8.2\%) &  1 (0.3\%)\\ 
       &  & \textbf{OpenVoid}          & 296 & 296 (100.0\%)           & 243 \phantom{0}(82.1\%) & 2.10 & 154 (63.4\%) & 213 \phantom{0}(87.7\%) & 20 \phantom{0}(8.2\%) &  0 (0.0\%)\\ 
       &  & \textbf{Dolphin}           & 296 & 296 (100.0\%)           & 267 \phantom{0}(90.2\%) & 1.40 & 194 (72.7\%) & 197 \phantom{0}(73.8\%) & 2 \phantom{0}(0.7\%)  &  0 (0.0\%)\\ \midrule

      \multirow{14}{*}{\rotatebox{90}{\textbf{WellsFargo} (6,364)}} & \multirow{4}{*}{\rotatebox{90}{\textbf{Com.}}} 
        & \textbf{GPT~5}            & 518 & 393 \phantom{0}(75.9\%) & 393 (100.0\%)            & 1.90 & 324 (82.4\%) &  12 \phantom{00}(3.1\%) & 0 \phantom{0}(0.0\%) &  2 (0.5\%) \\
      &  & \textbf{GPT~4.1}           & 518 & 478 \phantom{0}(92.3\%) & 478 (100.0\%)          & 2.60 & 377 (78.9\%) & 467 \phantom{0}(97.7\%) & 0 \phantom{0}(0.0\%) &  169 (35.4\%)\\ 
      & & \textbf{Gemini}           & 518 & 516 \phantom{0}(99.6\%) & 516 (100.0\%)            & 2.55 & 423 (82.0\%) & 508 \phantom{0}(98.4\%) & 0 \phantom{0}(0.0\%) &  182 (35.3\%) \\
      & & \textbf{Llama~4}          & 518 & 517 \phantom{0}(99.8\%) & 514 \phantom{0}(99.4\%)  & 1.30 & 414 (80.5\%) & 375 \phantom{0}(73.0\%) & 0 \phantom{0}(0.0\%) &  39 (7.6\%) \\
      \cmidrule{2-11}
    & \multirow{6}{*}{\rotatebox{90}{\textbf{Open}}}  
         & \textbf{Llama~3.2}       & 518 & 416 \phantom{0}(80.3\%)  & 407 \phantom{0}(97.8\%)   & 1.40 & 230 (56.5\%) & 388 \phantom{0}(95.3\%) & 0 \phantom{0}(0.0\%)  & 2 (0.5\%)\\ 
       & & \textbf{Qwen~3}          & 518 & 508 \phantom{0}(98.1\%)  & 503 \phantom{0}(99.0\%)   & 2.90 & 405 (80.5\%) & 473 \phantom{0}(94.0\%) & 0 \phantom{0}(0.0\%)  & 119 (23.7\%)\\
       & & \textbf{Qwen}            & 518 & 518 (100.0\%)            & 505 \phantom{0}(97.5\%)   & 2.30 & 390 (77.2\%) & 454 \phantom{0}(89.9\%) & 0 \phantom{0}(0.0\%)  & 52 (10.3\%) \\ 
       & & \textbf{Mistralai}       & 518 & 518 (100.0\%)            & 496 \phantom{0}(95.8\%)   & 1.40 & 301 (60.7\%) & 470 \phantom{0}(94.8\%) & 0 \phantom{0}(0.0\%)  & 42 (8.5\%)\\ 
       & & \textbf{DeepSeekVL2}     & 518 & 500 \phantom{0}(96.5\%)  & 500 (100.0\%)             & 1.20 & 324 (64.8\%) & 475 \phantom{0}(95.0\%) & 0 \phantom{0}(0.0\%)  & 4 (0.8\%)  \\ 
       & & \textbf{Phi}             & 518 & 414 \phantom{0}(79.9\%)  & 392 \phantom{0}(94.7\%)   & 1.80 & 187 (47.7\%) & 268 \phantom{0}(68.4\%) & 0 \phantom{0}(0.0\%)  & 3 (0.8\%) \\ 
      \cmidrule{2-11}
    & \multirow{4}{*}{\rotatebox{90}{\textbf{Uncensor.}}}  
          & \textbf{Daredevil}        & 296     & 296 (100.0\%) & 280 \phantom{0}(94.6\%) &   1.10 & 189 (67.5\%) & 271 \phantom{0}(96.8\%) & 0 \phantom{0}(0.0\%)       &  0 (0.0\%)\\
       &  & \textbf{WhiteRabbitNeo}   & 296 & 296 (100.0\%)           & 291 \phantom{0}(98.3\%) &   1.09 & 149 (51.2\%) & 249 \phantom{0}(85.6\%) & 0 \phantom{0}(0.0\%) &  1 (0.3\%)\\ 
       &  & \textbf{OpenVoid}         & 296 & 296 (100.0\%)           & 292 \phantom{0}(98.6\%) &   1.18 & 190 (65.1\%) & 248 \phantom{0}(84.9\%) & 0 \phantom{0}(0.0\%) &  2 (0.7\%)\\ 
       &  & \textbf{Dolphin}          & 296 & 296 (100.0\%)           & 255 \phantom{0}(86.1\%) &   1.40 & 208 (81.6\%) & 233 \phantom{0}(91.4\%) & 0 \phantom{0}(0.0\%) &  3 (1.2\%)\\ \midrule
       
    \multicolumn{3}{c}{\textbf{All} (including \autoref{tab:scr_eval_uncersor1})}  & 30,192 & 29,217 (96.8\%) & 28,121 (96.2\%)  & 2.02 & 20,533 (73.0\%) & 23,724 (84.4\%) & 5,128 (18.2\%) & 4,348 (15.5\%)\\ 
      \bottomrule
    
    \end{tabular}
}
\end{table*}

We assess the visual fidelity of LLM-generated phishing websites from rendered screenshots, which capture what users see and drive vision-based detectors. We address \textbf{RQ\,1} (\textit{Screenshot-based Visual Fidelity}) using seven metrics (\autoref{subsec:analysis}) and analyze how outputs vary with target brand, LLM, and prompt configuration. For the visual component of \textbf{RQ\,4} (\textit{Comparison to Phishing Kits}), we compare LLM-generated pages with 64 real-world phishing kits. \autoref{tab:scr_cluster} reports results for the three model categories under \textit{regular prompts}; uncensored LLMs under explicit \textit{phishing prompts} appear in \autoref{tab:scr_eval_uncersor1}; and per-model UWS distributions are shown in \autoref{fig:user_uws}.

\subsection{Multidimensional Visual Traits across LLMs and Brands} \label{subsec:visual}

\PP{Validity \& Homogeneity}Safeguards are rarely the decisive barrier: bypass succeeds for 96.8\% of prompts on average and, once bypassed, 96.2\% of responses are valid websites (\autoref{tab:scr_cluster} and \autoref{tab:scr_eval_uncersor1}). Refusals are highly non-uniform across the model$\times$brand: commercial LLMs refuse most, peaking on GPT~5 for payment/banking targets. Open-source models are more stable (typically $>92.0\%$), with Llama~3.2 (PayPal: 87.3\%, Wells Fargo: 80.3\%) and Phi (PayPal: 90.2\%, Wells Fargo: 79.9\%) as exceptions. Uncensored models, regardless of the ``phishing'' keyword, respond at 98.3\%-100.0\%. 
This shows that willingness and capability are \emph{decoupled}: uncensored models respond most readily yet produce the lowest and most variable validity. Furthermore, brand-specific risk sensitivity leads to uneven protection across targets, potentially incentivizing attackers to shift toward less-monitored brands. 

Beyond being valid, Top-1 dominant cluster results reveal strong visual homogeneity in LLM-generated phishing pages, with 73.0\% of screenshots falling into models' dominant visual clusters. Commercial LLMs show the strongest convergence (76.0-97.8\%), indicating reliance on reusable templates, while uncensored models cluster moderately (47.6\%-81.6\%). Open-source models vary substantially: Qwen~3 clusters tightly (80.5\%-92.3\%), while Phi is far more diverse (45.2\%-54.5\%).

\observ{Refusals expose alignment gaps along both model and brand dimensions, yet remain narrow and easily bypassed, and model generation willingness and capability are decoupled. Commercial LLMs cluster tightly, uncensored models cluster moderately, and open-source models span highly templated (Qwen~3) to highly diverse (Phi)}

\begin{figure}[t]
\centering
\includegraphics[trim=0.2cm 0.2cm 0.2cm 0.2cm, clip,width=0.98\linewidth]{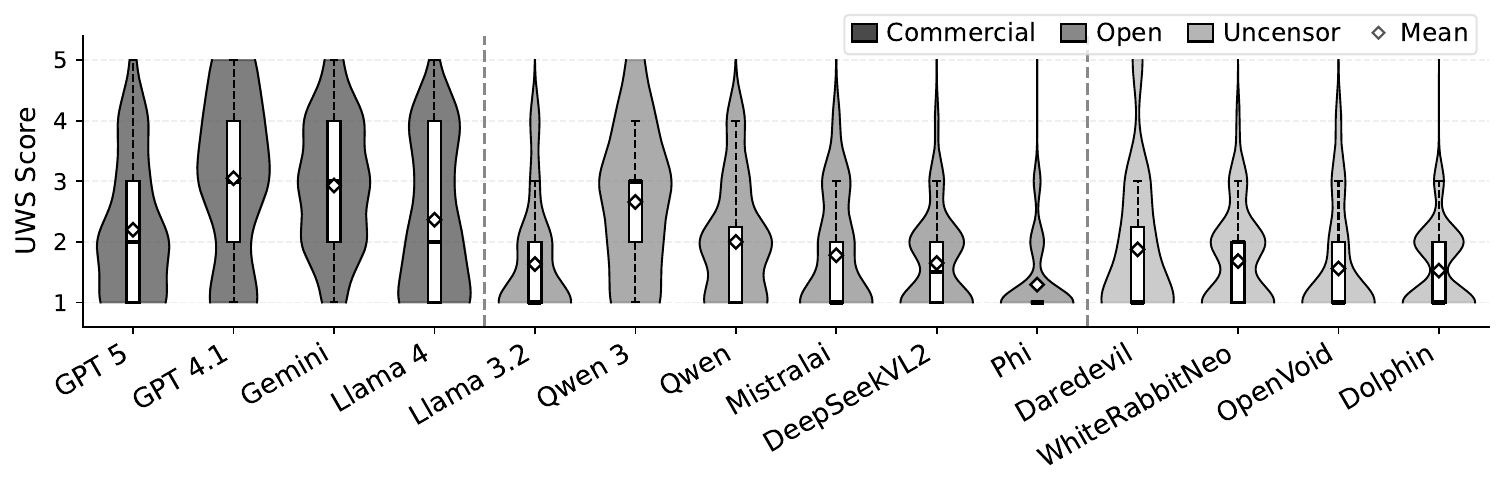}
\caption{User-perceived Website Similarity (UWS) among LLMs.}
~\label{fig:user_uws}
\end{figure}

\PP{User Evaluation}\autoref{tab:scr_cluster} reports per-(model, brand) UWS values, while \autoref{fig:user_uws} shows the per-model UWS distributions; full statistical tests are in~\aref{appendix_subsec:user_study_analysis}. On average, participants perceive LLM-generated phishing websites as only moderately similar to their legitimate counterparts (mean UWS 2.02). Commercial models significantly outperform open-source and uncensored ones (Kruskal-Wallis $H = 64.06$, $p < .001$; pairwise Dunn's with Bonferroni correction, $p < .001$), driven primarily by GPT~4.1 (mean 3.05, peaking 3.50 on Amazon) and Gemini (2.94), whose distributions concentrate in the upper range (\autoref{fig:user_uws}). GPT~5 is an exception as it bypasses safeguards least often and scores below both (mean 2.20). Open-source models exhibit lower and more concentrated distributions, while uncensored models cluster tightly at the lower end, all averaging below 1.90, with Daredevil on PayPal (2.80) as the highest entry, while Phi is consistently weakest. Qwen~3 outperforms most open-source and uncensored counterparts (Dunn's test, $p < .05$), with a median UWS comparable to leading commercial models, though its distribution is relatively concentrated around this median, reflecting stable but ceiling-limited performance compared to GPT~4.1 and Gemini.

Target complexity further shapes perceived similarity. Wells Fargo, whose login page combines a dense promotional layout with multiple secondary elements (\autoref{fig:brand_loginpage} in \aref{appendix_subsec:prompt}), receives the lowest mean UWS across all models, suggesting that visually richer targets expose the rendering limitations of lower-capability models more sharply (Kruskal-Wallis $H = 18.37$, $p < .001$; Wells Fargo vs.\ Amazon and PayPal $p_{\text{bonf}} < .05$, but not vs.\ Facebook).

To identify which visual cues drive these judgments, we analyze participants' qualitative rationales across 570 rated instances. ``Style (464: color, font, etc.)'' and ``Layout'' (459) dominate, indicating that stylistic and structural coherence anchor perceived similarity. ``Missing or Broken Logo'' is the third most cited factor, consistent with users relying on branding for visual verification. The same asymmetry surfaces in our automated metrics: models reproduce brand text in 84.4\% of cases but logos in only 18.2\%. Qwen~3 draws the fewest logo complaints among all models (13 instances), whereas uncensored LLMs receive the most. A sharper disparity emerges for ``Broken Image or Text'': Uncensored models account for 75 such reports against 22 for commercial models, despite producing fewer valid outputs overall, indicating that uncensored models are disproportionately prone to rendering failures that undermine their effectiveness as phishing lures.


\observ{User-perceived visual similarity tracks model \emph{capability}: the threat concentrates in a few high-end models, while uncensored models remain the least convincing. The primary bottleneck is visual fidelity and it becomes more pronounced for visually complex targets. In particular, stylistic coherence, layout fidelity, and accurate logo reproduction are the strongest drivers of user judgment}

\PP{Brand Mimicry} 
Across all valid websites, the large gap between brand text matching (84.4\%) and brand logo matching (18.2\%) reveals a strong text-over-logo bias in LLM-generated phishing pages. Specifically, GPT~4.1 and Gemini consistently achieve high brand text match rates (79.0\%-100.0\%), while Llama~4 follows closely (69.2\%-92.8\%). The clear outlier is GPT~5, whose brand text matching collapses to 2.7\%-22.9\%. Manual inspection indicates that it predominantly produces generic login pages rather than explicitly referencing brand information. Beyond text, GPT~4.1 is also the strongest logo imitator, achieving high brand logo matching (77.8\%-91.4\%) on Amazon, Facebook, and PayPal and thereby reproducing both textual and visual brand cues.

Most open-source and uncensored models also exhibit a clear text-first imitation pattern: textual brand cues are reproduced frequently, whereas logo matching remains substantially lower and varies strongly by brand. This disparity is particularly pronounced for Wells~Fargo, for which logo matching is 0\% across all evaluated models despite generally high text-matching rates, which also highlights pronounced brand-specific differences in reproducing brand identity.

\observ{LLMs adopt distinct brand imitation strategies: GPT~4.1 reproduces both textual and visual brand cues, GPT~5 largely suppresses explicit branding, and most other models primarily rely on textual cues, with Wells Fargo being consistently impersonated without any logo imagery}


\PP{Detection Rate}Phishpedia detects only $15.5\%$ of valid pages, leaving the large majority undetected. Across brands, detection varies by orders of magnitude: Facebook (${\sim}87.5\%$) and Amazon (${\sim}86.5\%$) are caught most often, whereas PayPal drops to single digits for most models and Wells~Fargo is missed almost entirely (typically $<1\%$, peaking at ${\sim}35\%$). Across models, commercial LLMs are the most detectable (GPT 4.1 exceeds 85\% on both Facebook and Amazon, while Gemini reaches 77.9\% on Facebook), open-source models fall in between with high variance, and uncensored models are essentially never flagged (near-$0\%$ across all four brands).

A natural concern is whether low detection simply reflects poor generation quality rather than detector limitations. Since Phishpedia relies on logo matching, its coverage is inherently bounded by whether generated pages reproduce a recognizable logo, even when otherwise convincing. For Facebook and Amazon, several models achieve $>95\%$ validity yet omit a reproducible logo and are therefore rarely flagged, whereas pages that do reconstruct the logo on the same brand (e.g., GPT~4.1 on Facebook, $84.6\%$ logo, $87.5\%$ detection) are caught far more often. This gap therefore reflects a mismatch between generated content and detector assumptions, not simply low page quality. The effect is strongest for Wells~Fargo, where commercial and open-source models produce valid websites at over $95\%$ with non-trivial user-evaluation scores yet almost never reconstruct the logo, leaving detection near-zero. Such failures are block-structured rather than sporadic, concentrating in entire brands where highly convincing pages systematically fall outside the detector's logo-based scope.


\observ{Phishpedia flags 15.5\% of valid phishing pages: commercial models often reproduce brand logos and are readily flagged on Facebook and Amazon; pages lacking these logo cues evade detection almost entirely}



\subsection{Multidimensional Visual Traits Across Prompt Combinations}

\begin{table*}[!htbp]
\centering
\caption{Screenshot-based Visual Characteristics of LLM-Generated Phishing Websites Across Prompt Combinations.}
\label{tab:llm_input_results}
\resizebox{0.96\linewidth}{!}{
    \begin{tabular}{llrrrr rr r}
    \toprule

    \multicolumn{1}{c}{\textbf{\makecell{Prompt \\Combinations}}} & \multicolumn{1}{c}{\textbf{Total}} & \multicolumn{1}{c}{\textbf{\makecell{Safeguard \\Bypass}}} & \multicolumn{1}{c}{\textbf{\makecell{Valid \\Websites}}} & \multicolumn{1}{c}{\textbf{\makecell{User \\ Eval.}}} & \multicolumn{1}{c}{\textbf{\textbf{Homo.}}} & \multicolumn{1}{c}{\textbf{ \makecell{Brand Text\\Matching}}} & \multicolumn{1}{c}{\textbf{\makecell{Brand Logo\\Matching}}} & \multicolumn{1}{c}{\textbf{\makecell{Phishpedia\\Detect}}} \\\midrule
    
    {Text}                & 5,328 (17.6\%)            & 5,246 (98.5\%)  & 5,197 (99.1\%)   & 1.86 & 2,886 (55.5\%) & 4,183 (80.5\%) & 707 (13.6\%)            & 511 (9.8\%)\\
    {Text+Function}       & 5,328 (17.6\%)            & 5,266 (98.8\%)  & 5,207 (98.9\%)   & 2.07 & 2,683 (51.5\%) & 4,290 (82.4\%) & 797 (15.3\%)            & 500 (9.6\%)\\
    {Text+Visual}         & 5,328 (17.6\%)            & 5,283 (99.2\%)  & 5,207 (98.6\%)   & 1.93 & 2,368 (45.5\%) & 4,629 (88.9\%) & 849 (16.3\%)            & 769 (14.8\%)\\
    {Text+HTML}           & 5,328 (17.6\%)            & 4,976 (93.4\%)  & 4,095 (82.3\%)   & 1.35 & 2,406 (58.8\%) & 3,627 (88.6\%) & 237 \phantom{0}(5.8\%)  & 199 (4.9\%)\\
    {Text+Scr}            & 2,960 \phantom{0}(9.8\%)  & 2,757 (93.1\%)  & 2,744 (99.5\%)   & 2.54 & 1,783 (65.0\%) & 2,254 (82.1\%) & 839 (30.6\%)            & 745 (27.2\%)\\
    {Text+Function+Scr}   & 2,960 \phantom{0}(9.8\%)  & 2,849 (96.2\%)  & 2,840 (99.7\%)   & 2.48 & 1,768 (62.3\%) & 2,318 (81.6\%) & 857 (30.2\%)            & 743 (26.2\%)\\
    {Text+Visual+Scr}     & 2,960 \phantom{0}(9.8\%)  & 2,840 (95.9\%)  & 2,831 (99.7\%)   & 2.35 & 1,718 (60.7\%) & 2,423 (85.6\%) & 842 (29.7\%)            & 881 (31.1\%)\\
    \midrule  
    {All Brands}          & 30,192                    & 29,217 (96.8\%) & 28,121 (96.2\%)  & 2.02 & 15,612 (55.5\%)& 23,724 (84.4\%) & 5,128 (18.2\%) & 4,348 (15.5\%)\\ 
      \bottomrule
    
    \end{tabular}
}
\end{table*}
\autoref{tab:llm_input_results} shows that prompt configurations substantially influence generation. Specifically, textual inputs (Text, Text+Visual, Text+Function) yield higher successful generation rates (98.5\%-99.2\%), whereas adding auxiliary inputs leads to a reduction (93.1\%-96.2\%). This suggests that these auxiliary inputs are more likely to trigger LLM safeguards. In particular, incorporating HTML results in one of the lowest success rates of 93.4\% and also more blank screenshots (valid website rate 82.3\% vs. $\geq98.6\%$ for other prompt combinations). 

Although screenshots reduce success rates, they greatly improve perceived visual quality. Specifically, screenshot-based prompts achieve higher UWS (2.35-2.54) than text-based prompts (1.35-2.07). However, text and HTML prompts produce more diverse outputs (45.5\%-58.8\% in the top-1 cluster), whereas screenshot-based prompts yield less diversity (60.7\%-65.0\%), indicating a shift toward more uniform and predictable outputs despite improved realism.

For brand mimicry, brand text matching is consistently high (80.5\%-88.9\%) across all prompt types, while brand logo matching is lower and highly prompt-dependent (5.8\%-30.6\%). Screenshot prompts roughly double logo reproduction: 5.8\%-16.3\% (without) vs. 29.7\%-30.6\% (with). This asymmetry indicates that LLMs prefer text cues for brand mimicry, making textual impersonation a dominant and resilient attack vector.


\observ{Textual prompts maximize generation success and diversity, while screenshot-based prompts enhance visual realism, especially logo fidelity, but reduce success rates and output diversity, with textual brand cues still strong}

\subsection{LLM-Generated vs. Real-World Phishing Kits}

\begin{table}[!t]
\centering
\caption{Visual Characteristics of Real-World Phishing Kits.}
\label{tab:kit_info}
\footnotesize
\resizebox{\columnwidth}{!}{
    \begin{tabular}{l rrrrr}
    \toprule
   
    \multicolumn{1}{c}{\textbf{\textbf{Brand}}} & \multicolumn{1}{c}{\textbf{\textbf{Total}}} & \multicolumn{1}{c}{\textbf{\textbf{Homo.}}} & \multicolumn{1}{c}{\textbf{ \makecell{Brand Text\\Matching}}} & \multicolumn{1}{c}{\textbf{\makecell{Brand Logo\\Matching}}} & \multicolumn{1}{c}{\textbf{\makecell{Phishpedia\\Detect}}} \\\midrule
    Amazon      & 13 & 5 (38.5\%)           &  13 (100.0\%)           & 6 (46.2\%)           & 6 (46.2\%)   \\
    Facebook    & 26 & 2 \phantom{0}(7.7\%) &  24 \phantom{0}(92.3\%) & 16 (61.5\%)          & 14 (53.8\%) \\
    PayPal      & 6  & 1 (16.7\%)           &  4 \phantom{0}(66.7\%)  & 5 (83.3\%)           & 5 (83.3\%) \\
    WellsFargo  & 19 & 5 (26.3\%)           &  17 \phantom{0}(89.5\%) & 0 \phantom{0}(0.0\%) & 9 (47.4\%) \\ 

    \midrule
    All Brands  & 64 & 13 (20.3\%) & 58 (90.6\%) & 27 (42.2\%) & 34 (53.1\%) \\
    \bottomrule           
    \end{tabular}
}
\end{table}

\autoref{tab:scr_cluster} and~\autoref{tab:kit_info} directly compare the visual fidelity of LLM-generated pages against in-the-wild phishing kits. Compared to the generated templates, real-world kits reproduce brand logos more faithfully (42.2\% vs.\ 18.2\%), since they bundle authentic logo image assets rather than reconstructing the mark from code, and this logo advantage carries directly into higher reference-based detection (53.1\% vs.\ 15.5\%). Both converge on brand-text matching (90.6\% vs.\ 84.4\%): each tends to embed the brand's textual cues across the screenshot, confirming that reproducing brand names is trivial for both LLMs and real-world phishing kits.

Across brands, the LLM-vs-kit gap is not uniform. Facebook kits are the most diverse of the four brands, with homogeneity far below both the other kit brands and the LLM-generated pages (7.7\% vs.\ up to 38.5\% for Amazon kits and 76.0\%-94.6\% for commercial LLMs), making them the least visually consistent. PayPal departs from the general pattern on brand text: its kits reproduce textual cues less often than the corresponding LLM pages (66.7\% vs.\ 69.2\%-97.5\% for commercial models), the only brand where LLM text impersonation clearly exceeds that of real kits. PayPal also shows the highest brand-logo matching among kits (83.3\%, against 46.2\% Amazon, 61.5\% Facebook, and 0.0\% Wells Fargo), reproducing logos more faithfully than any other brand's kits.


\observ{Both sources readily mimic brand text, but kits reproduce logos more reliably and are thus easier to detect overall. This varies by brand: kits dominate for PayPal, while LLMs do so for Facebook.}

\section{HTML-based Evaluation}  \label{sec:html_analysis}

To understand \textbf{RQ\,2} and \textbf{RQ\,4} from an HTML-centric perspective, we analyze the HTML source code of LLM-generated websites and real-world phishing kits, as HTML governs actionable behaviors, visual appearance, and structural composition beyond what rendered screenshots reveal. Specifically, we study HTML from four aspects: (i) the cost of generating HTML source code; (ii) the overall HTML structure to understand their structural organization; (iii) HTML similarity between generated websites or real-world kits and corresponding legitimate websites to quantify structural resemblance; and (iv) the CSS styling of login forms and buttons, which are critical components for user interactions.

\subsection{Cost Analysis of Generation} \label{subsec:cost}
\begin{figure}[t]
\centering
\includegraphics[width=0.9\linewidth]{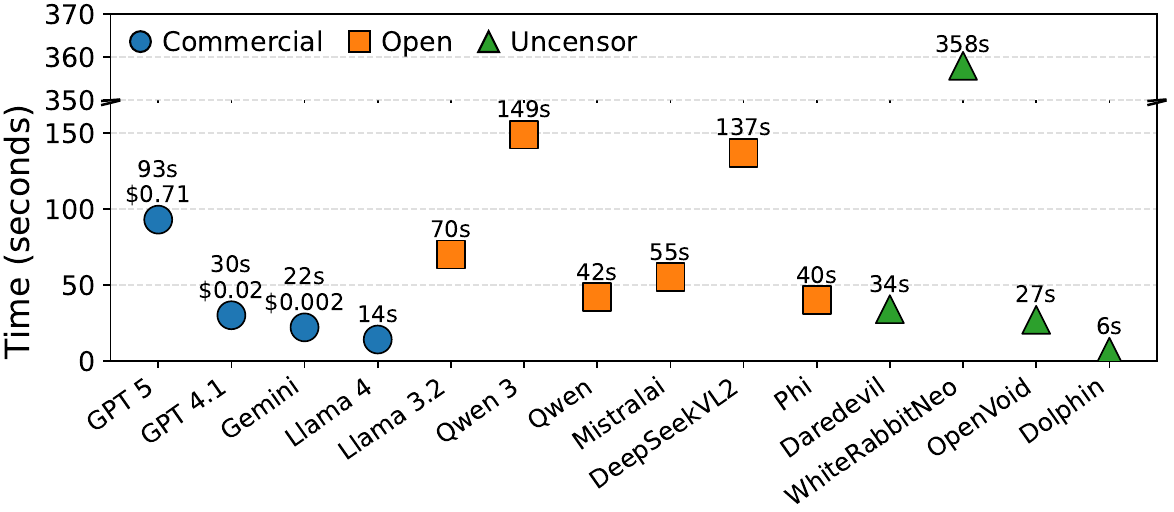}
\caption{Average Time and Monetary Cost Per Sample.}
~\label{fig:cost}
\end{figure}

\autoref{fig:cost} shows the average time and monetary cost incurred by different LLMs. We do not distinguish between uncensored models with and without phishing-related prompts, as their generation times are similar. 
Among commercial LLMs, GPT~5 exhibits the longest generation time (93 s) and highest monetary cost (\$0.71), while Llama~4 is the fastest. Gemini achieves a balance between a comparatively short generation time (22 s) and a minimal monetary cost (\$0.002). 
Among open-source LLMs, Qwen (42 s) and Phi (40 s) are time-efficient, whereas Qwen~3 (149 s) is the most time-consuming. 
Uncensored models show extreme variance in generation time. WhiteRabbitNeo is the slowest, requiring 358 seconds per sample, while Dolphin is the fastest at only 6 seconds. 


\observ{High cost and latency limit GPT~5 and WhiteRabbitNeo at scale, whereas free, low-latency models enable rapid, scalable phishing website generation}

\subsection{HTML Structure}

\begin{table}[!t]
\centering
\caption{HTML Structure of Generated Phishing Websites and Real-World Phishing Kits.}
\label{tab:html_statistics}

\resizebox{1\linewidth}{!}{%
\setlength{\tabcolsep}{1.6pt}
\begin{tabular}{l l w{r}{0.7cm}
                w{r}{0.9cm} w{r}{0.5cm} w{r}{0.8cm}
                w{r}{0.4cm} w{r}{0.4cm} w{r}{0.3cm}
                w{r}{1cm} w{r}{0.9cm} w{r}{0.9cm}}
\toprule
& \multirow{2}{*}{\textbf{Name}}&\multirow{2}{*}{\textbf{Num.}}  &
  \multicolumn{3}{c}{\textbf{Token}} &
  \multicolumn{3}{c}{\textbf{DOM Depth}} &
  \multicolumn{3}{c}{\textbf{Tags (Unique)}} \\
\cmidrule(lr){4-6}\cmidrule(lr){7-9}\cmidrule(lr){10-12}
&  &
  &\multicolumn{1}{c}{max} & \multicolumn{1}{c}{min} & \multicolumn{1}{c}{avg}
   &max & min&avg
   &\multicolumn{1}{c}{max} & \multicolumn{1}{c}{min} & \multicolumn{1}{c}{avg} \\
\midrule
\multirow{4}{*}{\rotatebox{90}{\textbf{Com.}}}
  & GPT~5          & 1,682 &  2,026 & 158 & 655 & 14 & 5 & 8 & 145 (33) & 13 (10) & 46 (19) \\
  & GPT~4.1        & 1,975 &  2,983 & 162 & 584 & 11 & 5 & 7 & 128 (28) & 15 (10) & 42 (16) \\
  & Gemini         & 2,055 & 37,342 &  82 & 811 & 13 & 3 & 7 & 261 (31) &   3 (3) & 49 (17) \\
  & Llama~4        & 2,070 &  1,885 &  69 & 375 & 13 & 2 & 6 & 105 (25) &   7 (4) & 31 (15) \\
\midrule
\multirow{6}{*}{\rotatebox{90}{\textbf{Open}}}
  & Llama~3.2      & 1,842 &  4,893 &  44 & 508 & 18 & 3 & 7 & 203 (25) &   6 (5) & 37 (17) \\
  & Qwen~3         & 2,060 &  4,551 &  28 & 586 & 12 & 3 & 7 & 387 (31) &   4 (4) & 44 (16) \\
  & Qwen           & 2,072 &  3,635 &  92 & 400 & 12 & 4 & 6 & 161 (26) & 11 (10) & 36 (16) \\
  & Mistralai      & 2,072 &  3,642 &  71 & 400 & 13 & 2 & 6 & 160 (25) &   6 (3) & 31 (15) \\
  & DeepSeekVL2    & 2,050 &  2,027 &  86 & 261 & 10 & 3 & 6 & 132 (23) &  10 (8) & 23 (14) \\
  & Phi            & 1,876 &  2,649 &  31 & 259 & 19 & 2 & 6 & 137 (27) &   3 (2) & 22 (13) \\
\midrule
\multirow{4}{*}{\rotatebox{90}{\textbf{Uncensor.}}}
  & Daredevil      & 1,184 &  3,610 &  34 & 588 & 13 & 3 & 6 & 159 (25) &   4 (4) & 37 (14) \\
  & WhiteRN. & 1,179 &  3,787 &  44 & 294 & 14 & 2 & 6 & 165 (25) &   5 (4) & 25 (15) \\
  & OpenVoid       & 1,184 &  3,932 &  24 & 534 & 12 & 2 & 6 & 158 (22) &   2 (2) & 33 (13) \\
  & Dolphin        & 1,184 &  1,156 &  16 & 199 &  9 & 2 & 5 &  67 (22) &   2 (2) & 17 (11) \\
\midrule
\multirow{4}{*}{\rotatebox{90}{\textbf{Uncensor.$^{*}$}}}
  & Daredevil      & 1,184 &  3,610 &  37 & 551 & 13 & 3 & 6 & 159 (26) &   5 (5) & 35 (14) \\
  & WhiteRN. & 1,183 &  3,606 &  24 & 283 & 12 & 2 & 6 & 157 (26) &   3 (2) & 24 (14) \\
  & OpenVoid       & 1,181 &  3,715 &  22 & 537 & 13 & 2 & 6 & 158 (23) &   3 (3) & 34 (13) \\
  & Dolphin        & 1,184 &    935 &  14 & 206 & 10 & 2 & 5 &  75 (22) &   2 (2) & 18 (12) \\ \midrule
\multirow{2}{*}{\rotatebox{90}{\textbf{Kits}}} & \multirow{2}{*}{----} & \multirow{2}{*}{64} & \multirow{2}{*}{962,950} & \multirow{2}{*}{23} & \multirow{2}{*}{28,551} & \multirow{2}{*}{33} & \multirow{2}{*}{0}  & \multirow{2}{*}{16} & \multirow{2}{*}{943 (32)} & \multirow{2}{*}{0 (0)} & \multirow{2}{*}{150 (21)} \\
&&&&\\
\bottomrule
\multicolumn{12}{l}{%
  $^{*}$~Phishing prompts with ``phishing'' keywords directly included in the prompt.} \\
\multicolumn{12}{l}{%
  max: maximum,\; min: minimum,\; avg: average.\;}\\
\multicolumn{12}{l}{%
  Tags (Unique): total tags with unique tag types in parentheses.} \\
\end{tabular}%
}
\end{table}

We focus on the 29,217 meaningful websites generated by LLMs and 64 real-world phishing kits. \aref{appendix_subsec:top10tag}-\autoref{fig:tag_wordcloud} shows the top-10 most frequently occurring HTML tags, including \texttt{<html>}, \texttt{<head>}, \texttt{<title>}, and \texttt{<form>}, etc., indicating that both the generated pages and real-world phishing kits are primarily composed of standard structural elements. \autoref{tab:html_statistics} presents the maximum, minimum, and average token length, DOM depth, total tag count, and number of unique tags. Note that tag counts include CSS and JavaScript elements, whereas token length and DOM depth exclude these components, as these components are often lengthy and may disproportionately bias these measurements.

\autoref{tab:html_statistics} shows that DeepSeekVL2, Phi, WhiteRabbitNeo, and Dolphin produce comparatively shorter HTML outputs with fewer total tags, indicating more compact and less diverse structures. Despite these differences in length and tags, DOM depth remains relatively stable across models (typically 5--8), implying similar hierarchical nesting strategies when generating webpage layouts. Furthermore, real phishing kits are far more complex than any LLM's output. Specifically: average 28,551 tokens vs. models' 199-811, DOM depth 16 vs. 5-8, and 150 total tags vs. 17-49. Models produce lightweight skeleton pages; real kits pack in heavy, obfuscated, fully-resourced code. However, structural statistics alone are insufficient to characterize websites generated from different LLMs. We therefore analyze HTML similarity in the following section.

\observ{Both the prevalence of basic HTML structures and the stability of DOM depth across LLMs indicate convergence toward a limited set of structural templates, whereas real phishing kits are far more complex}



\subsection{HTML Similarity} \label{subsec:html_sim}
\begin{figure}[!t]
\centering
\includegraphics[width=0.98\linewidth]{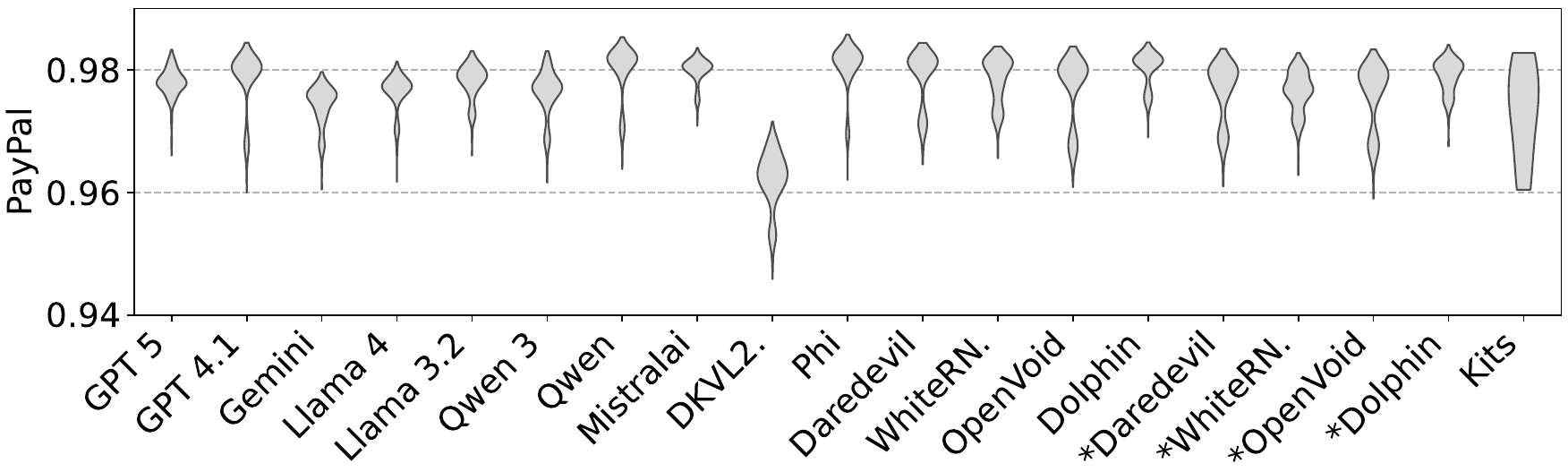}
\caption{MarkupLM-based HTML Cosine Similarity for PayPal.}
\label{fig:html_sim_paypal}
\end{figure}

We measure MarkupLM~\cite{Li2021MarkupLMPO}-based HTML similarity between LLM-generated HTML and corresponding legitimate HTML. \aref{apx:html_similar}-\autoref{fig:html_sim} shows the cosine similarity among four brands, while \autoref{fig:html_sim_paypal} reports the PayPal results. Results show that both LLM-generated HTML and real-world phishing kits are highly similar to legitimate websites, with cosine similarity scores ranging from 0.92-0.99 across four brands, indicating a strong replication of both structural and semantic patterns. PayPal is particularly easy to imitate, as most LLMs achieve a high median similarity above 0.98 except for DeepSeekVL2. DeepSeekVL2 consistently appears as an outlier, producing comparatively less accurate structural imitations. Moreover, explicitly adding `phishing' into prompts does not reduce similarity, as the distribution remains nearly unchanged. The consistently low variance across models further indicates template-level consistency in LLM-generated HTML.

\observ{LLM-generated HTML matches legitimate pages as closely as real phishing kits (cosine 0.92–0.99), making DOM feature-based detection difficult}







\subsection{CSS Styling} \label{subsec:html_style}

\begin{figure}[!t]
    \centering
    \includegraphics[width=0.9\linewidth]{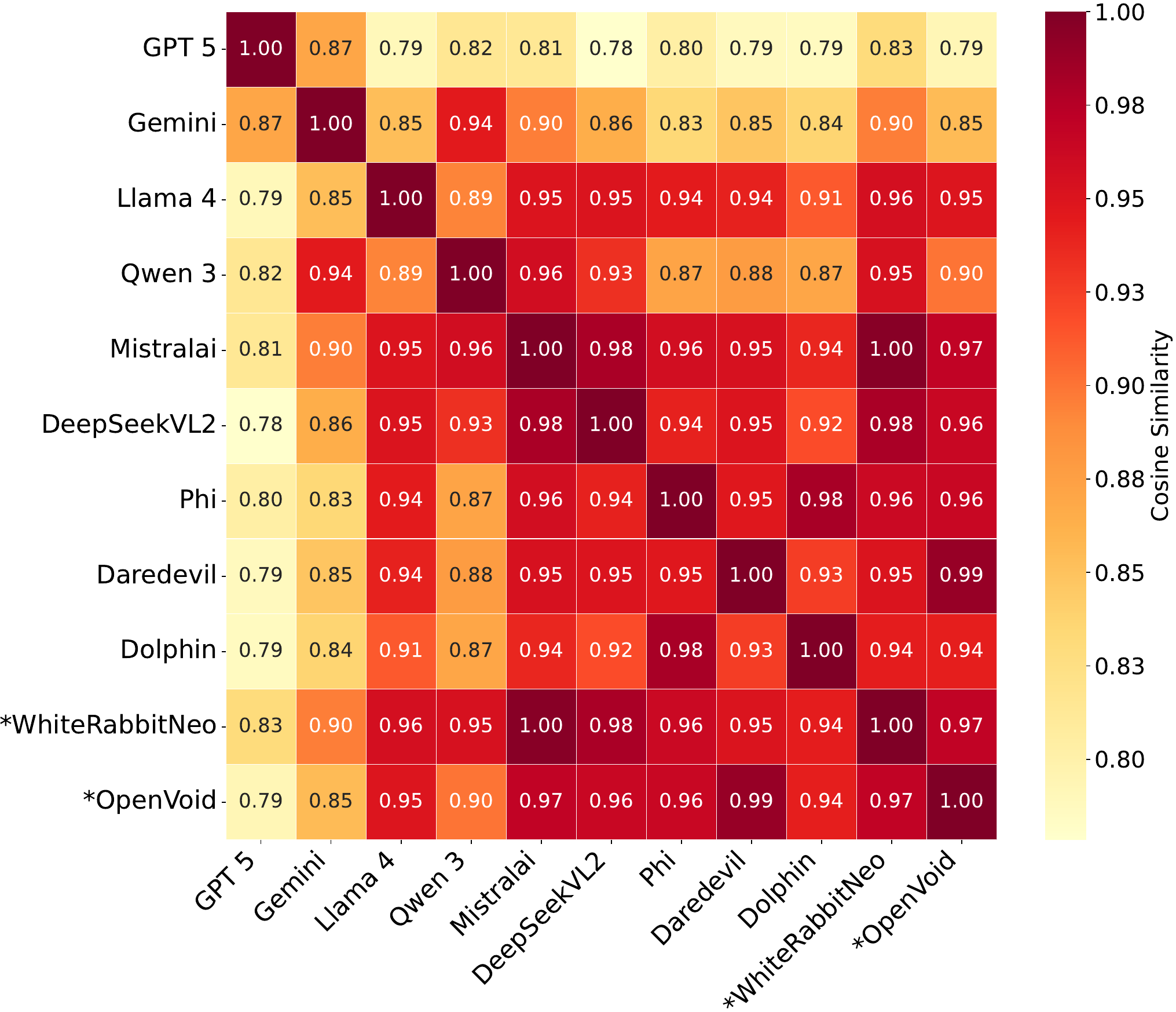}
  
    \caption{Cosine Similarity of CSS Property Selection for Button Elements.}
    \label{fig:heatmap-selected}
\end{figure}

Forms and buttons are key interaction elements whose styling shapes user engagement. We therefore analyze CSS styles to understand the design strategies of LLMs and real-world phishing kits. For 29,217 successfully generated websites, real-world kits, and their legitimate websites, we separately examine the top-ranked CSS properties of displayed buttons and forms, excluding non-displayed elements. \autoref{tab:llm_kits_css_top10} reports CSS statistics for LLM-generated websites and phishing kits, and \autoref{tab:legit_top20} shows legitimate target websites. Cross-LLM differences are shown in~\autoref{fig:heatmap-selected} and \autoref{fig:heatmap}.


\PP{Form}\autoref{tab:llm_kits_css_top10} shows that LLM-generated content contains nearly two forms per website on average, with only 4.6\% of samples lacking forms. The dominance of \texttt{padding} (63.8\%), \texttt{width} (41.2\%), and \texttt{margin} (38.5\%) underscores a reliance on layout structuring, while the prevalence of \texttt{box-shadow} (45.3\%) and \texttt{border-radius} (43.7\%) indicates a rounded, container-based aesthetic. 
Real-world phishing kits are instead led by \texttt{margin} (50.0\%) and \texttt{background} (32.1\%), whereas decorative properties such as \texttt{box-shadow} and \texttt{border-radius} are absent from the top ranks, indicating a functional, appearance-agnostic strategy. Kits also retain legacy and vendor-prefixed properties (\eg, \texttt{-moz-box-sizing}: 17.9\%, \texttt{vertical-align}: 17.9\%) inherited from the older pages they clone, which LLMs almost never emit and which serve as useful forensic signals. 
Legitimate websites (\autoref{tab:legit_top20}) prioritize typography-related properties (\eg, \texttt{line-height} and \texttt{font-family}), reflecting an emphasis on readability and coherent design integration, which is absent in the generated outputs. Such typographic control is largely absent from LLM-generated forms, suggesting that they fall back on browser-default styles and thus converge on a more homogeneous visual appearance. Moreover, the CSS similarity heatmap in \autoref{fig:heatmap} captures property-level selection similarity across LLMs. The high cosine similarity scores indicate that different models tend to select a highly overlapping set of CSS properties when generating styles.




\observ{LLM-generated forms favor structural properties (\eg, \texttt{padding}) over the typographic control (\eg, \texttt{font-family}) seen in legitimate pages, and different LLMs converge on highly overlapping property sets, whereas real-world kits inherit legacy, vendor-prefixed styles from the older pages they clone}

\begin{table}[!t]
\centering
\footnotesize
\caption{Top-10 CSS Properties in Buttons and Forms of LLM-generated Websites and Real-World Phishing Kits.}
\label{tab:llm_kits_css_top10}
\setlength{\tabcolsep}{1.8pt}
\resizebox{\linewidth}{!}{%
\begin{tabular}{l  rrlrr}
\toprule
\multicolumn{3}{c}{\textbf{Forms}} & \multicolumn{3}{c}{\textbf{Buttons}}  \\ \midrule
\multirow{2}{*}{LLMs}
& \multicolumn{1}{c}{\# of Form}
& \multicolumn{1}{c}{\# of Sample}
& \multirow{2}{*}{LLMs}
& \multicolumn{1}{c}{\# of Button}
& \multicolumn{1}{c}{\# of Sample} \\
& \multicolumn{1}{c}{(52,198)}
& \multicolumn{1}{c}{(27,881)}
&
& \multicolumn{1}{c}{(39,427)}
& \multicolumn{1}{c}{(26,938)} \\
\cmidrule(lr){1-3}\cmidrule(lr){4-6}
padding          & 18,815 (36.0\%)          & 17,798 (63.8\%)          & color            & 29,863 (75.7\%)           & 20,333 (75.5\%)            \\
box-shadow       & 12,724 (24.4\%)          & 12,618 (45.3\%)          & border           & 29,727 (75.4\%)           & 20,592 (76.4\%)            \\
$^{**}$border-r  & 12,285 (23.5\%)          & 12,172 (43.7\%)          & cursor           & 28,710 (72.8\%)           & 20,157 (74.8\%)            \\
width            & 11,849 (22.7\%)          & 11,497 (41.2\%)          & padding          & 28,535 (72.4\%)           & 19,636 (72.9\%)            \\
$^+$m.           & 11,372 (21.8\%)          & 10,725 (38.5\%)          & $^{**}$border-r  & 27,787 (70.5\%)           & 19,111 (70.9\%)            \\
$^*$bg-color     & 10,285 (19.7\%)          & 10,216 (36.6\%)          & $^*$bg-color     & 22,549 (57.2\%)           & 16,894 (62.7\%)            \\
display          & 9,667 (18.5\%)           & 8,119 (29.1\%)           & width            & 17,778 (45.1\%)           & 13,426 (49.8\%)            \\
max-width        & 8,330 (16.0\%)           & 8,135 (29.2\%)           & font-size        & 14,641 (37.1\%)           & 9,205 (34.2\%)             \\
border           & 5,430 (10.4\%)           & 5,383 (19.3\%)           & font-weight      & 11,078 (28.1\%)           & 6,375 (23.7\%)             \\
flex-direction      & 4,648 \phantom{0}(8.9\%) & 4,101 (14.7\%)           & $^*$bg           & 7,833 (19.9\%)            & 4,308 (16.0\%)             \\ \midrule\midrule

\multirow{2}{*}{Kits}
& \multicolumn{1}{c}{\# of Form}
& \multicolumn{1}{c}{\# of Sample}
& \multirow{2}{*}{Kits}
& \multicolumn{1}{c}{\# of Button}
& \multicolumn{1}{c}{\# of Sample} \\
& \multicolumn{1}{c}{(134)}
& \multicolumn{1}{c}{(56)}
&
& \multicolumn{1}{c}{(138)}
& \multicolumn{1}{c}{(50)} \\
\cmidrule(lr){1-3}\cmidrule(lr){4-6}

$^+$m.             & 53 (39.6\%) & 28 (50.0\%)            & color                & 91 (65.9\%) & 42 (84.0\%) \\
padding            & 42 (31.3\%) & 23 (41.1\%)            & padding              & 79 (57.2\%) & 39 (78.0\%) \\
$^*$bg             & 32 (23.9\%) & 18 (32.1\%)            & font-size            & 78 (56.5\%) & 39 (78.0\%) \\
box-sizing         & 31 (23.1\%) & 14 (25.0\%)            & border               & 78 (56.5\%) & 42 (84.0\%) \\
border             & 31 (23.1\%) & 17 (30.4\%)            & font-family          & 78 (56.5\%) & 39 (78.0\%) \\
display            & 27 (20.1\%) & 19 (33.9\%)            & $^+$m.               & 72 (52.2\%) & 37 (74.0\%) \\
font-size          & 25 (18.7\%) & 11 (19.6\%)            & cursor               & 69 (50.0\%) & 38 (76.0\%) \\
position           & 24 (17.9\%) & 15 (26.8\%)            & font-weight          & 63 (45.7\%) & 31 (62.0\%) \\
vertical-align     & 24 (17.9\%) & 10 (17.9\%)            & height               & 62 (44.9\%) & 35 (70.0\%) \\
-moz-box-sizing    & 22 (16.4\%) & 10 (17.9\%)            & width                & 62 (44.9\%) & 39 (78.0\%) \\

\bottomrule
\multicolumn{6}{l}{$^+$m. denotes margin, $^*$bg represents background, $^{**}$ border-r denotes border-radius;}\\
\end{tabular}
}
\end{table}

\PP{Button.}\autoref{tab:llm_kits_css_top10} shows that LLMs emphasize visual affordance over explicit layout control: surface-level properties (\eg, \texttt{border}: 76.4\%, \texttt{color}: 75.5\%) dominate, whereas structural properties such as \texttt{display} (13.7\%) and \texttt{height} (7.0\%) fall outside the top-10 entirely. Real-world kits show an even stronger surface emphasis (\texttt{color}: 84.0\%) and additionally specify \texttt{font-family} (78.0\%), which LLMs rank lower (34.2\%). Legitimate websites, in contrast, define both presentational and structural properties. As shown in \autoref{tab:legit_top20}, fundamental layout properties (\eg, \texttt{display} and \texttt{height}) are consistently defined in legitimate samples to ensure precise rendering. This divergence suggests that LLM-generated buttons adopt the visual traits of interactive elements while delegating structure to browser defaults, which may introduce cross-environment rendering variability. Furthermore, unlike the uniform convergence in forms, \autoref{fig:heatmap-selected} reveals a commercial--open-source split: open-source models converge tightly (pairwise cosine similarity $\geq$ 0.87, mostly above 0.92), whereas commercial models deviate from this cluster, with GPT~5 showing the lowest similarity to open-source models (0.78-0.83), and Gemini falling in between (0.83-0.94).


\observ{LLM-generated buttons favor visual styling (\eg, \texttt{color}, \texttt{border}) but often omit explicit layout definitions (\eg, \texttt{display}, \texttt{height}), a structural sparsity absent from both legitimate websites and real-world kits. Open-source models converge on generic templates, whereas commercial models show greater stylistic diversity}

\section{Functionality of Generated Websites} \label{sec:function}

To understand \textbf{RQ~3} (\textit{Functionality}) and \textbf{RQ~4} (\textit{Comparison to Phishing Kits}), we analyze the functionality of valid websites and compare them against real-world phishing kits.
%
%
The results in~\autoref{fig:boxplot}, \aref{appendix:result_function}-\autoref{fig:bar_chart}, and \autoref{tab:failure_analysis} reveal a significant disparity between visual mimicry (96.2\% valid) and functional execution, with only 10,889 (38.7\%) samples successfully completing the credential-harvesting pipeline. Commercial models generally outperform open-source and uncensored variants. The primary bottleneck is backend persistence, evidenced by 7,061 ``Unsaved'' cases and various PHP-side failures, such as redirection without capture (3xx) or execution errors (5xx). This gap between plausible appearance and functional failure aligns with prior observations that cross-component coordination remains a challenge for LLMs \cite{lu2025webgen, roy2024chatbots}, confirming that generating backend logic is more difficult than replicating frontend DOM structures.

\subsection{Characterizing LLMs \& Brand Differences} 
\begin{figure}[t]
\centering
\includegraphics[width=0.85\columnwidth]{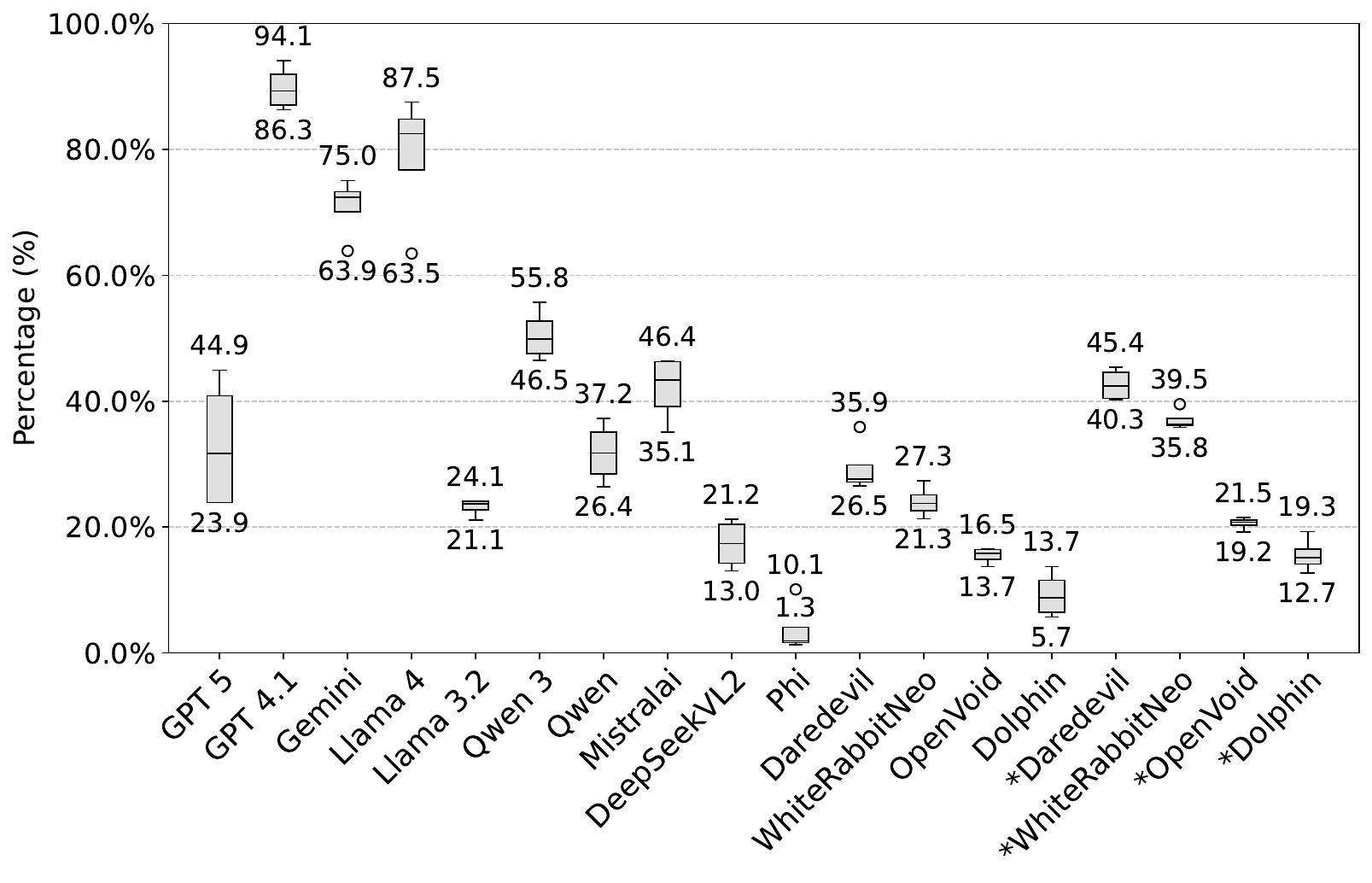}
\caption{Functionality of LLM-Generated Phishing Websites Across Models. Boxes show per-brand success-rate distributions, including minima, maxima, and inter-brand variation. Total valid websites: 28{,}121; functional samples: 10{,}889. * indicates prompts containing ``phishing''.}

~\label{fig:boxplot}
\end{figure}

\autoref{fig:boxplot} and \autoref{fig:bar_chart} show model-dependent performance variations. Most commercial models achieve higher functionality. GPT~4.1, Llama~4, and Gemini achieve maximum success rates of 94.1\% (450/478), 87.5\% (442/505), and 75.0\% (382/509), respectively, where each value is calculated from the single brand with the highest functionality score for that model. GPT~5 is a marked outlier at 44.9\% (204/454). Under regular prompts, open-source models (max value 55.8\%) and uncensored models (max value 35.9\%) are lower, led by Qwen~3 at 55.8\% (281/504) among open-source models and Daredevil at 35.9\% (89/248) among uncensored LLMs. 
For uncensored LLMs, phishing prompts consistently improve functionality over regular prompts (Wilcoxon rank-sum test, $p < .05$).

Across brands, the results reveal clear within-model brand preferences, with Facebook showing the largest variation. Phi generates few functional pages for Amazon, PayPal, and Wells Fargo ($\leq$ 9), but jumps to 49 for Facebook, and GPT~5 follows the same pattern (204 vs. 94-171). Llama~4 shows the opposite trend, with Facebook (318) noticeably lower than other brands (318 vs. 417-442), as do Gemini (329 vs. 369-382) and Qwen (130 vs. 151-188). By contrast, Amazon, PayPal, and Wells Fargo show small within-model differences, behaving consistently under each model's generation capabilities and constraints. Uncensored models show limited brand-specific variation, with performance primarily driven by prompt type rather than brands. Detailed per-LLM and brand analyses are provided in~\aref{appendix:result_function}. Overall, this brand asymmetry indicates model-specific sensitivity to brand characteristics.

\observ{Commercial LLMs generally produce more functional websites, except GPT~5. Most commercial and open-source models show clear brand preferences, whereas phishing prompts improve the functionality of uncensored LLMs. Brand complexity has little effect on functionality}

\subsection{Failure Analysis}

To identify bottlenecks in the generation pipeline, we conduct a granular failure analysis on the 17,232 non-functional samples. \autoref{tab:failure_analysis} summarizes failure types into Client-side (HTML), Server-side Logic (PHP), and Persistence (File) errors.

\PP{Persistence Failures (File)}The most frequent failure (7,061 cases) is the ``Unsaved'' error. Here, form submission triggers a successful backend response (HTTP 200), yet data is not written to the local file system. This ``silent failure'' typically results from missing file I/O operations (\eg, omitted \texttt{file\_put\_contents}) or incorrect permission handling, rendering the attack ineffective despite successful network interaction. Importantly, our ``Unsaved'' cases (HTTP 200 but no persistence) indicate broken end-to-end wiring, consistent with the known difficulty of multi-file coordination for LLM-generated solutions~\cite{jimenez2024swebench}.

\PP{Client-Side Failures (HTML)} Frontend-layer failures prevent valid request initiation. ``Submit Timeout'' (4,028 cases) is the most frequent, where forms fail to trigger a response due to non-functional interaction elements, such as using \texttt{<a href=``\#''>} anchors instead of valid submit buttons. Additionally, the ``No Form'' error (1,859 cases) is prevalent in models such as Dolphin and Phi, which frequently omit \texttt{<form>} tags entirely, resulting in static, non-interactive shells. Lastly, ``Invalid Action'' errors (260 cases) persist despite our lenient preprocessing designed to resolve generic placeholders. In these instances, models disregard the explicit instruction to target \texttt{save.php}, instead generating mismatched attributes (\eg, \texttt{submit.php}) that sever the execution flow.

\PP{Server-Side Logic Failures (PHP)}Even with a valid frontend structure, backend execution often failed due to internal logic or syntax errors. A prominent failure mode is ``3xx'' (Redirection without Capture, 1,991 cases), notably in GPT~5, where generated scripts redirect users to legitimate URLs without first capturing credentials. We conservatively interpret this as a benign redirection pattern that may suppress malicious capture behavior, rather than asserting a specific causal mechanism. ``5xx'' (Execution Errors, 1,931 cases) are also prevalent, notably in DeepSeekVL2, where PHP syntax errors or hallucinated unconfigured database connections (\eg, MySQL) caused scripts to crash prior to data persistence. A minor subset involved ``4xx'' (Client Errors, 96 cases), where models implemented excessive validation logic that rejected valid test inputs.

\observ{LLMs replicate static aesthetics well but often fail at stateful backend logic for credential transmission and persistence, with silent ``Unsaved'' failures dominating}

\begin{table}[t]
\centering
\caption{Taxonomy of Failure Cases (Total: 28{,}121; Functional: 10{,}889; Non-functional: 17{,}232; I: No Form, II: Load T/O, III: Submit T/O, IV: Invalid Action).}
\label{tab:failure_analysis}
\small
\setlength{\tabcolsep}{3pt}
\resizebox{1\linewidth}{!}{%
\begin{tabular}{p{0.01\linewidth}p{0.17\linewidth} rr rrrr rrr r}
\toprule

& \multirow{2}{*}{\textbf{LLM}} & \multirow{2}{*}{\textbf{Total}} & \multirow{2}{*}{\textbf{Fail}} &
\multicolumn{4}{c}{\textbf{HTML}} &       
\multicolumn{3}{c}{\textbf{PHP}} &       
\multicolumn{1}{c}{\textbf{File}} \\      

\cmidrule(lr){5-8} \cmidrule(lr){9-11} \cmidrule(lr){12-12}

 &  &  &  &
\multicolumn{1}{c}{I} & \multicolumn{1}{c}{II} & \multicolumn{1}{c}{III} & \multicolumn{1}{c}{IV} &
\multicolumn{1}{c}{3xx} & \multicolumn{1}{c}{4xx} & \multicolumn{1}{c}{5xx} & \multicolumn{1}{c}{Unsaved} \\
\midrule

\multirow{4}{*}{\rotatebox{90}{\textbf{Com.}}}
& \textbf{GPT~5}       & 1,682 & 1,117 & 19 & 0 & 197 & 2 & 347 & 51 & 101 & 400 \\
& \textbf{GPT~4.1}     & 1,975 & 204 & 1 & 0 & 72 & 5 & 41 & 1 & 23 & 61 \\
& \textbf{Gemini}      & 2,051 & 596 & 1 & 0 & 68 & 15 & 129 & 2 & 43 & 338 \\
& \textbf{Llama~4}     & 2,037 & 426 & 1 & 0 & 32 & 10 & 85 & 1 & 156 & 141 \\

\midrule

\multirow{6}{*}{\rotatebox{90}{\textbf{Open}}}
& \textbf{Llama~3.2}   & 1,790 & 1,375 & 191 & 1 & 421 & 3 & 86 & 0 & 171 & 502 \\
& \textbf{Qwen~3}      & 2,032 & 1,006 & 10 & 0 & 272 & 4 & 99 & 2 & 18 & 601 \\
& \textbf{Qwen}        & 2,025 & 1,381 & 87 & 0 & 324 & 8 & 88 & 1 & 80 & 793 \\
& \textbf{Mistralai}   & 2,037 & 1,181 & 20 & 1 & 210 & 19 & 102 & 4 & 71 & 754 \\
& \textbf{DeepSeekVL2} & 1,983 & 1,642 & 220 & 0 & 453 & 10 & 60 & 0 & 595 & 304 \\
& \textbf{Phi}         & 1,771 & 1,700 & 329 & 3 & 588 & 81 & 79 & 8 & 151 & 461 \\
\midrule

\multirow{4}{*}{\rotatebox{90}{\textbf{Uncensor.}}}
& \textbf{Daredevil}        & 1,070 & 757 & 154 & 0 & 253 & 11 & 83 & 0 & 34 & 469 \\
& \textbf{WhiteRN.}   & 1,161 & 882 & 100 & 0 & 412 & 39 & 140 & 19 & 113 & 376 \\
& \textbf{OpenVoid}         & 1,052 & 890 & 198 & 0 & 183 & 9 & 184 & 0 & 111 & 524 \\
& \textbf{Dolphin}          & 1,063 & 965 & 416 & 1 & 264 & 26 & 222 & 4 & 50 & 421 \\
\midrule

\multirow{4}{*}{\rotatebox{90}{\textbf{Uncensor.$^*$}}}
& \textbf{Daredevil}        & 1,095 & 629 & 22 & 0 & 27 & 2 & 41 & 0 & 36 & 254 \\
& \textbf{WhiteRN.}   & 1,161 & 731 & 29 & 0 & 131 & 10 & 48 & 3 & 54 & 139 \\
& \textbf{OpenVoid}         & 1,052 & 835 & 9 & 0 & 65 & 3 & 70 & 0 & 96 & 273 \\
& \textbf{Dolphin}          & 1,084 & 915 & 52 & 0 & 56 & 3 & 87 & 0 & 28 & 250 \\
\midrule

\multicolumn{2}{c}{\textbf{Total}} & 
28,121 & 17,232 & 1,859 & 6 & 4,028 & 260 & 1,991 & 96 & 1,931 & 7,061 \\
\bottomrule

\multicolumn{12}{l}{$^*$ means phishing prompts: adding ``phishing'' keywords directly into the prompts.}
\end{tabular}
}
\end{table}

\subsection{LLM-Generated vs.\ Real-World  Phishing Kits}

Among 64 real-world phishing kits, 37 (57.8\%) are \emph{functional}, successfully executing the full credential-harvesting pipeline; 14 (21.9\%) are \emph{non-functional}; and the remaining 13 (20.3\%) are \emph{non-verifiable}, as their exfiltration destinations are unconfigured placeholders, preventing the triggering of the actual transmission and storage behavior. 


Comparing functionality between the LLM-generated websites and kits, only 38.7\% of LLM-generated websites are functional (Table~\ref{tab:failure_analysis}), versus 57.8\% of real-world kits. This aggregate, however, masks substantial cross-model variation: three of the 14 LLMs exceed the kit-level rate: GPT~4.1 (89.7\%), Llama~4 (79.1\%), and Gemini (70.9\%). Thus, although LLM-generated pages have a lower overall success rate, capable individual models can produce working phishing pages at a higher rate than that observed across the sourced kits.

The two populations also exhibit different failure modes. The dominant bottleneck of LLM failures lies in the backend persistence: 41.0\% of failures stem from the generated code never persisting the captured credentials to a file (unsaved). By contrast, the 14 non-functional real kits fail primarily due to backend syntax faults or incomplete code that prevent execution (57.1\%) or frontend absence of credential input forms or exfiltration logic (57.1\%). Frontend form absence is relatively small in both populations, accounting for 10.8\% of LLM failures and 14.3\% of non-functional kits. These categories are not mutually exclusive, as a single kit may exhibit multiple failure signatures.

\observ{Deployed kits are functional more often (57.8\% vs.\ 38.7\%), but three high-capability models among the 14 LLMs exceed the kit-level functional rate, lowering technical barriers and enabling far greater production scale}

\section{Related Work} \label{sec:related}


\PP{Phishing Automation Techniques}Phishing attacks often mimic legitimate websites to deceive users~\cite{venturi2022classkits,Lee2024Beneath}. Attackers use phishing kits, bundled tools for deploying phishing sites on web servers, to launch mimicry attacks rapidly at scale against specific targets~\cite{Zhang2022crawlphish,Lim2025Phisher,Hugo2021Catching,Oest2018Inside}.
Open-source tools (\eg, BlackEye~\cite{blackeye:online}, Zphisher~\cite{htrtechz61:online}) provide pre-designed phishing websites for popular brands~\cite{Kulkarni2025}, but their reliance on curated templates restricts brand coverage and page diversity, forcing attackers to create new templates for unsupported targets.

Empirical studies further reveal heavy reuse and industrialization: Merlo \textit{et al.}~\cite{Merlo2022SANER} analyze static code similarity among phishing kits, focusing on PHP, JavaScript, and HTML. Results indicate that 90\% of phishing kits share at least 90\% of their code with one or more other kits.
Lee \textit{et al.}~\cite{Lee2024Beneath} collect 4,153 phishing kits from real-world phishing websites and analyze their server-side behaviors. They find that each kit with the same single landing page script is deployed in an average of 46.92 domains. These studies reveal a high degree of code reuse and automation that underscores the industrialization of phishing operations. However, the template-based nature of these kits continues to limit their brand coverage.
In contrast, we study on-demand LLM-generated websites, which replace curated template reuse with prompt-driven generation. Our analysis examines how this shift changes the scalability and completeness of phishing website generation relative to traditional phishing kits.


\PP{LLM-based Phishing Generation and Measurement}LLMs enable automated creation of phishing websites. Roy \textit{et al.}~\cite{roy2024chatbots} explore using commercial LLMs to generate phishing attacks with malicious prompts. They show that LLMs can produce phishing websites that mimic well-known brands using unmodified, or `vanilla', versions of the models, without requiring adversarial fine-tuning or exploits. The resulting websites are shown to be as evasive to anti-phishing tools as those manually crafted by attackers. However, their functionality assessment verifies whether requested components are present rather than executing the credential-harvesting pipeline end-to-end. Their visual evaluation relies on three in-house coders, leaving systematic user-perceived and operational comparisons unexplored.
Lu~\textit{et al.}~\cite{lu2025webgen} develop a benchmark to evaluate the ability of LLM-based agents to construct websites that meet the functional and visual requirements specified in user instructions.
Cal\`{o} \textit{et al.}~\cite{Tommaso2023WebGen} use LLMs to enable website creation through a user-oriented approach based on prompt engineering.
Si~\textit{et al.}~\cite{Si2025design2code} introduce automatic metrics and human evaluations to compare websites generated by LLMs. 
Chen~\textit{et al.}~\cite{SoK2025Chen} explore characteristics of LLM-generated phishing content through a review of related literature.
Prior work establishes that LLMs can generate phishing or general-purpose websites, but several dimensions remain insufficiently characterized, particularly \textit{how far} LLM-driven generation reaches and \textit{where} it breaks down.
Our work provides a controlled comparison across commercial, open-source, and uncensored LLMs and jointly evaluates user-perceived and metric-level visual fidelity, DOM/CSS structure, and end-to-end credential-harvesting functionality. We further benchmark generated pages against real-world phishing kits under the same measures, enabling us to identify both where LLM generation fails and where strong models already reach or exceed kit-level functionality.






\section{Discussion and Implications}
 \label{sec:dicuss}

\PP{Generation vs. User Perception}LLMs can generate structurally valid webpages, but our user study shows that such outputs do not necessarily convince users. The gap arises from the distinction between structural correctness and perceptual fidelity: LLMs produce syntactically correct HTML/CSS, but fail to replicate fine-grained visual cues salient to humans, such as accurate logos and typography. The high brand text matching and low logo matching also indicate limited fidelity in the visual replication. These results show that human-centered evaluation is indispensable for assessing real-world deceptiveness.


\PP{Willingness vs. Competence.}Our evaluation challenges the assumption that ``uncensored'' models yield more effective phishing pages. Commercial models consistently outperform uncensored ones in functional success rates. Notably, despite their high willingness to generate malicious content, uncensored models achieve only limited functional success, indicating that removing refusal constraints alone is insufficient to substantially improve end-to-end functionality. Within our evaluated models, these results suggest that practical risk depends not only on willingness to generate harmful content, but also on the underlying capability required for reliable system construction.


\PP{Stylistic Discrepancies as LLM Fingerprints}The DOM-CSS asymmetry, high DOM-level similarity paired with systematic omission of typographic CSS, admits a structural explanation. Enterprise websites enforce design systems with global styling rules (\eg, site-wide \texttt{font-family} and \texttt{line-height} declarations), whereas LLMs synthesize each page in isolation, optimizing for local structural correctness rather than global stylistic coherence. This suggests that the layout-versus-typography imbalance in CSS rules may serve as a candidate signal for detecting LLM-generated phishing pages, complementing the brand-text and visual-similarity signals.


\PP{LLM-Generated Pages vs. Real-World Kits.}They leave opposite structural signatures: kits clone older, fully-resourced pages and inherit legacy, vendor-prefixed CSS within heavy, obfuscated markup, whereas LLM output is lightweight, template-convergent, and marked by a DOM-CSS asymmetry. They also impersonate through different channels--kits bundle authentic logo assets, while LLMs reconstruct the mark from code and largely fail, leaning on brand text instead. As generation shifts from kits to LLMs, detection must shift with it by: elevating resilient brand-text cues co-occurring with authentication or payment interfaces to a first-class signal, adding a CSS-level discriminator that quantifies the layout-versus-typography imbalance, and fusing these generator-level features with existing kit-oriented and visual-similarity signals--since no single signature now spans both provenances.

\PP{Actionable Recommendations}A practical next step is to build a hybrid phishing detection pipeline that fuses complementary signals, explicitly incorporating CSS features. Our findings show that CSS carries significant discriminative power: LLM-generated pages tend to exhibit systematic imbalances between layout and missing or inconsistent typographic properties (\eg, font-family), making CSS a key signal. In addition, the system should account for the gap between user-perceived visual fidelity and standard similarity metrics by incorporating perceptual checks such as logo hashing, webfont verification, and screenshot-based embeddings. Notably, while exact brand logos are often poorly reproduced or sparse, brand-related textual information appears frequently and should be used as a stronger signal. Finally, the detector should flag suspicious co-occurrences of brand text with authentication or payment interfaces, and fuse all signals into an ensemble model trained on diverse, multi-model LLM-generated phishing datasets to ensure robustness against generator-specific artifacts, resulting in a more resilient and adaptable defense against rapidly evolving AI-generated phishing attacks.

\PP{Limitations}\textit{Prompt engineering} remains a challenge in LLM-based research, as outputs can be sensitive to prompt variations. To mitigate this issue, we employ diverse prompt-rewriting strategies and multiple advanced LLMs to reduce model-specific bias. Nevertheless, we acknowledge that such bias cannot be eliminated, and certain behaviors may still be influenced by prompt design choices.


Another challenge stems from the \textit{heterogeneity and rapid evolution of LLMs}. The models differ in scale, architecture, and training, and their capabilities, safety mechanisms, and generation behaviors continue to evolve over time. Accordingly, our study benchmarks individual models and presents a systematic pipeline for analyzing LLM-generated websites, uncovering model-specific and generative patterns, thereby providing more robust insights and actionable recommendations.

Additionally, the \textit{model scales} of our open-source censored and uncensored models are both relatively small. Their comparison partially controls for scale, letting us examine refusal removal within this range. Since strong uncensored models at frontier scale are not consistently available, our conclusions are scoped to the currently available small-model ecosystem.

Finally, our study includes a \textit{limited set of representative brands}. However, our goal is not to provide exhaustive coverage but rather to capture the fundamental characteristics of LLM-generated phishing websites. The selected brands span major categories and can serve as proxies for broader target classes. Moreover, one additional brand would generate around seven thousand variants, incurring substantial computational and monetary costs. We therefore balance coverage and feasibility, prioritizing analytical depth rather than exhaustive scale.



\section{Conclusion}

We presented a systematic framework for evaluating the phishing-generation capabilities of 14 LLMs. Analyzing 28,121 rendered samples, we quantified a critical divergence: visual mimicry has become commoditized (96.2\% visually plausible), yet functional viability remains a significant bottleneck (38.7\%). This suggests that the primary barrier to automated phishing is not refusal behavior alone but coding deficiencies that prevent reliable construction of an end-to-end credential-harvesting pipeline, and that practical risk is shaped more by high-capability models susceptible to evasion than by the mere availability of uncensored models. Our study also highlights potential defense opportunities: LLM-generated phishing websites exhibit distinguishable patterns, ranging from homogenized HTML templates to specific CSS anomalies, indicating that future defenses could benefit from looking beyond visual similarity to also examine the generative artifacts and functional inconsistencies inherent to LLM-generated content.



\section*{Use of Generative AI}




In this work, we use a diverse set of commercial, open-source, and uncensored large language models (LLMs) as part of an experimental pipeline for generating and evaluating phishing websites. We interact with these models directly and employ prompt-rewriting strategies to systematically investigate variations in their website-generation capabilities and behaviors.

All experimental design, data collection, evaluation, analysis, and conclusions are conducted by the authors. LLMs are used solely as experimental tools for website generation and are not relied upon for interpreting results or drawing conclusions. The generated websites are used exclusively in a controlled research environment and are neither deployed nor published online.

This study aims to characterize the capabilities and limitations of LLM-assisted phishing website generation, assess associated security risks, and inform the development of effective defenses against LLM-enabled phishing threats.

\section*{Ethics Considerations} \label{ethics}
In this work, we use LLMs to generate phishing websites. We discuss the associated stakeholders, impacts, and mitigation strategies.

\PP{Stakeholders}Primary stakeholders include: 
\begin{itemize}[leftmargin=*, itemsep=2pt, parsep=0pt, topsep=0pt, partopsep=0pt]
    \item LLM Providers: entities that develop, provide, and maintain large language models. In this work, we use both open-source and commercially available LLMs. While our analysis reveals certain weaknesses in current models, these limitations already exist and may be exploited by adversaries. We expect that model developers will benefit from our failure analysis, which can inform future improvements and lead to more robust and secure LLMs following publication.
    \item End Users: individuals who may be targeted by phishing attacks. All experiments are conducted in a controlled environment under IRB guidelines. Following publication, users may benefit from increased awareness and improved defensive measures.
    \item Organizations: whose brands are impersonated by attackers. We rely on both phishing data from official legal organizations and benign data from open websites. After publication, these organizations may better understand how LLM-generated phishing pages impersonate their brands and improve defenses using generator-level signals analyzed in our study.
    \item Researchers: who adopt or extend our methods will gain insights into how to design the models.
    \item Attackers: who may also gain some knowledge of LLMs; however, because these models are openly accessible, our contribution primarily strengthens defensive research rather than enabling attacks.
\end{itemize}

\PP{Impacts}We consider these principles: 

\begin{itemize}[leftmargin=*, itemsep=2pt, parsep=0pt, topsep=0pt, partopsep=0pt]
    \item Beneficence: Our ultimate goal is to improve LLM-based phishing detection systems, reducing the risk of phishing.
    \item Respect for Persons: This study involves human participants through a user study. All procedures were reviewed and approved by an Institutional Review Board (IRB). All participants remained anonymous, and they were informed of their right to withdraw their data at any point during or after the study. No personally identifiable information was collected or retained. Our study aims to improve their resilience to phishing attacks. By identifying how LLMs can be misused, we aim to inform the design of more effective defenses and increase user awareness, thereby reducing the likelihood of financial loss and psychological harm caused by phishing.
    \item Justice: We ensured balanced evaluation across multiple models (both commercial and open-source).
    \item Respect for the Public Interest: We use the APWG eCX dataset in accordance with its applicable data-use requirements and obtain benign website data only from publicly accessible webpages. We access commercial and open-source models through their publicly available interfaces or distributions and comply with applicable access, licensing, and data-use requirements.
\end{itemize}

\PP{Mitigations} Our work does not involve deploying phishing websites in the real world. All resources except API keys used in our work are publicly available and can be independently accessed and used by ordinary users; therefore, our study does not introduce capabilities beyond what is already publicly accessible. However, the study includes a user study, which introduces potential risks to participants. To mitigate these risks, all user study procedures were conducted under IRB approval. Participants provided informed consent, and no personally identifiable information was collected or retained. Exposure to phishing content was strictly limited to static screenshots presented in a controlled environment, ensuring that participants were not subjected to real-world deception or harm.


In addition, a potential risk lies in the misuse of our research artifacts. To mitigate this risk, the full prompts and generated phishing artifacts are made available only temporarily for artifact evaluation. Following the evaluation process and publication of the paper, these materials will be removed from public access and will not be distributed further. We will retain only simplified prompts and code for querying LLMs to support reproducibility while limiting the potential for direct misuse.

In conclusion, we think that the benefits of this research outweigh the potential harms. By advancing understanding of how LLMs perform in generating phishing websites, this work enables the community to design more interpretable, accurate, and practical defenses against phishing attacks. 
\clearpage

\bibliographystyle{IEEEtran}
\bibliography{ref}

@inproceedings{Hao2024Usenix,
    author = {Hao, Qingying and Diwan, Nirav and Yuan, Ying and Apruzzese, Giovanni and Conti, Mauro and Wang, Gang},
    title = {It doesn't look like anything to me: using diffusion model to subvert visual phishing detectors},
    year = {2024},
    booktitle = {Proc. of the USENIX Security Symposium}
}

@inproceedings{tan2025htmlrag,
  title={Htmlrag: Html is better than plain text for modeling retrieved knowledge in rag systems},
  author={Tan, Jiejun and Dou, Zhicheng and Wang, Wen and Wang, Mang and Chen, Weipeng and Wen, Ji-Rong},
  booktitle={Proc. of the ACM on Web Conference},
  year={2025}
}

@inproceedings{prabhumoye2017linguistic,
    title={Linguistic markers of influence in informal interactions},
    author={Prabhumoye, Shrimai and Choudhary, Samridhi and Spiliopoulou, Evangelia and Bogart, Christopher and Rose, Carolyn and Black, Alan W},
    booktitle={Proc. of the Workshop on NLP and Computational Social Science},
    year={2017}
}

@inproceedings{poliak2018collecting,
    title={Collecting Diverse Natural Language Inference Problems for Sentence Representation Evaluation},
    author={Poliak, Adam and Haldar, Aparajita and Rudinger, Rachel and Hu, J Edward and Pavlick, Ellie and White, Aaron Steven and Van Durme, Benjamin},
    booktitle={Proc. of the Conference on Empirical Methods in Natural Language Processing},
    year={2018}
}

@inproceedings{min2020syntactic,
    title={Syntactic Data Augmentation Increases Robustness to Inference Heuristics},
    author={Min, Junghyun and McCoy, R Thomas and Das, Dipanjan and Pitler, Emily and Linzen, Tal},
    booktitle={Proc. of the Annual Meeting of the Association for Computational Linguistics},
    year={2020}
}

@inproceedings{kriz2018simplification,
    title={Simplification using paraphrases and context-based lexical substitution},
    author={Kriz, Reno and Miltsakaki, Eleni and Apidianaki, Marianna and Callison-Burch, Chris},
    booktitle={Proc. of the Conference of the North American Chapter of the Association for Computational Linguistics: Human Language Technologies},
    year={2018}
}

@inproceedings{wei2019eda,
    title={EDA: Easy Data Augmentation Techniques for Boosting Performance on Text Classification Tasks},
    author={Wei, Jason and Zou, Kai},
    booktitle={Proc. of the Conference on Empirical Methods in Natural Language Processing and the International Joint Conference on Natural Language Processing},
    year={2019}
}

@inproceedings{li2024empowering,
    title={Empowering Large Language Models for Textual Data Augmentation},
    author={Yichuan Li and Kaize Ding and Jianling Wang and Kyumin Lee},
    booktitle={Proc. of the Annual Meeting of the Association for Computational Linguistics},
    year={2024},
}

@misc{Claude76:online,
    key={Claude},
    title={Claude.},
    howpublished={\url{https://claude.ai/new}},
    year = {2026},
    note = {(Accessed on 07/25/2026)}
}

@misc{VisionAI50:online,
    key={VisionAI},
    title={Vision AI: Image and visual AI tools.},
    howpublished={\url{https://cloud.google.com/vision}},
    year = {2026},
    note = {(Accessed on 07/25/2026)}
}

@misc{PaddlePa11:online,
    key={PaddleOCR},
    title={PaddlePaddle PaddleOCR.},
    howpublished={\url{https://github.com/PaddlePaddle/PaddleOCR}},
    year = {2026},
    note = {(Accessed on 07/25/2026)}
}

@misc{visualla13:online,
    key={Fastdup},
    title={Visual-layer/Fastdup.},
    howpublished={\url{https://github.com/visual-layer/fastdup}},
    year = {2026},
    note={(Accessed on 07/25/2026)}
}

@article{Kulkarni2025,
    author = {Kulkarni, Aditya and Balachandran, Vivek and Divakaran, Dinil Mon and Das, Tamal},
    title = {From ML to LLM: Evaluating the Robustness of Phishing Web Page Detection Models against Adversarial Attacks},
    year = {2025},
    journal = {Digital Threats},
}

@misc{htrtechz61:online,
    key={zphisher},
    title = {Htr-tech/zphisher: An automated phishing tool with 30+ templates.},
    howpublished = {\url{https://github.com/htr-tech/zphisher}},
    year = {2026},
    note = {(Accessed on 07/25/2026)}
}

@misc{blackeye:online,
    key={BLACKEYE},
    title = {EricksonAtHome/blackeye: BLACKEYE v2.0 | New Phishing tool with localtunnel (Discount)},
    howpublished = {\url{https://github.com/EricksonAtHome/blackeye}},
    year = {2026},
    note = {(Accessed on 07/25/2026)}
}

@misc{APWG73:online,
    key = {APWGTrend},
    title = {APWG},
    howpublished = {\url{https://apwg.org/trendsreports}},
    year = {2026},
    note = {(Accessed on 07/25/2026)}
}

@article{Devising2023Heiding,
    title={Devising and detecting phishing emails using large language models},
    author={Heiding, Fredrik and Schneier, Bruce and Vishwanath, Arun and Bernstein, Jeremy and Park, Peter S},
    journal={IEEE Access},
    year={2024}
}

@article{SoK2025Chen,
    author={Fengchao Chen and Tingmin Wu and Van Nguyen and Carsten Rudolph},
    title={SoK: Exposing the Generation and Detection Gaps in LLM-Generated Phishing Through Examination of Generation Methods, Content Characteristics, and Countermeasures}, 
    year={2025},
    journal={arXiv preprint arXiv:2508.21457},
}

@inproceedings{jimenez2024swebench,
    title={{SWE}-bench: Can Language Models Resolve Real-world Github Issues?},
    author={Carlos E Jimenez and John Yang and Alexander Wettig and Shunyu Yao and Kexin Pei and Ofir Press and Karthik R Narasimhan},
    booktitle={Proc. of the International Conference on Learning Representations},
    year={2024}
}

@inproceedings{roy2024chatbots,
    author={Roy, Sayak Saha and Thota, Poojitha and Naragam, Krishna Vamsi and Nilizadeh, Shirin},
    title={From chatbots to phishbots?: Phishing scam generation in commercial large language models},
    booktitle={Proc. of the IEEE Symposium on Security and Privacy},
    year={2024}
}

@article{lu2025webgen,
    title={WebGen-Bench: Evaluating LLMs on Generating Interactive and Functional Websites from Scratch},
    author={Lu, Zimu and Yang, Yunqiao and Ren, Houxing and Hou, Haotian and Xiao, Han and Wang, Ke and Shi, Weikang and Zhou, Aojun and Zhan, Mingjie and Li, Hongsheng},
    journal={arXiv preprint arXiv:2505.03733},
    year={2025}
}

@inproceedings{Zhang2022crawlphish,
    title={Crawlphish: Large-scale analysis of client-side cloaking techniques in phishing},
    author={Zhang, Penghui and Oest, Adam and Cho, Haehyun and Sun, Zhibo and Johnson, RC and Wardman, Brad and Sarker, Shaown and Kapravelos, Alexandros and Bao, Tiffany and Wang, Ruoyu and others},
    booktitle={Proc. of the IEEE Symposium on Security and Privacy},
    year={2021}
}

@article{venturi2022classkits,
  title={Classification of Web Phishing Kits for early detection by platform providers},
  author={Venturi, Andrea and Colajanni, Michele and Ramilli, Marco and Santangelo, Giorgio Valenziano},
  journal={arXiv preprint arXiv:2210.08273},
  year={2022}
}

@inproceedings{Lim2025Phisher,
    author = {Lim, Kyungchan and Lee, Kiho and Ji, Fujiao and Kwon, Yonghwi and Kim, Hyoungshick and Kim, Doowon},
    title = {What's in Phishers: A Longitudinal Study of Security Configurations in Phishing Websites and Kits},
    year = {2025},
    booktitle = {Proc. of the ACM on Web Conference},
}

@inproceedings{Hugo2021Catching,
    author = {Hugo Bijmans and Tim Booij and Anneke Schwedersky and Aria Nedgabat and Rolf van Wegberg},
    title = {Catching Phishers By Their Bait: Investigating the Dutch Phishing Landscape through Phishing Kit Detection},
    booktitle = {Proc. of the USENIX Security Symposium},
    year = {2021},
}

@inproceedings{Oest2018Inside,
  author={Oest, Adam and Safei, Yeganeh and Doupé, Adam and Ahn, Gail-Joon and Wardman, Brad and Warner, Gary},
  booktitle={APWG Symposium on Electronic Crime Research}, 
  title={Inside a phisher's mind: Understanding the anti-phishing ecosystem through phishing kit analysis}, 
  year={2018},
}

@inproceedings{Lee2024Beneath,
    author = {Lee, Woonghee and Hur, Junbeom and Kim, Doowon},
    title = {Beneath the Phishing Scripts: A Script-Level Analysis of Phishing Kits and Their Impact on Real-World Phishing Websites},
    year = {2024},
    booktitle = {Proc. of the ACM Asia Conference on Computer and Communications Security},
}

@inproceedings{Merlo2022SANER,
    author={Merlo, Ettore and Margier, Mathieu and Jourdan, Guy-Vincent and Onut, Iosif-Viorel},
    booktitle={Proc. of the IEEE International Conference on Software Analysis, Evolution and Reengineering}, 
    title={Phishing Kits Source Code Similarity Distribution: A Case Study}, 
    year={2022},
}

@inproceedings{Tommaso2023WebGen,
    author = {Cal\`{o}, Tommaso and De Russis, Luigi},
    title = {Leveraging Large Language Models for End-User Website Generation},
    year = {2023},
    booktitle = {Proc. of the End-User Development: International Symposium},
}

@inproceedings{Si2025design2code,
    title = {{D}esign2{C}ode: Benchmarking Multimodal Code Generation for Automated Front-End Engineering},
    author = {Si, Chenglei  and
      Zhang, Yanzhe  and
      Li, Ryan and
      Yang, Zhengyuan  and
      Liu, Ruibo  and
      Yang, Diyi},
    booktitle = {Proc. of the Conference of the Nations of the Americas Chapter of the Association for Computational Linguistics: Human Language Technologies},
    year = {2025},
}

@inproceedings{Lin2024Malla,
    author = {Zilong Lin and Jian Cui and Xiaojing Liao and XiaoFeng Wang},
    title = {Malla: Demystifying Real-world Large Language Model Integrated Malicious Services},
    booktitle = {Proc. of the USENIX Security Symposium},
    year = {2024},
}

@misc{GPT:online,
    key = {GPT},
    title = {GPT},
    howpublished = {\url{https://openai.com/index/chatgpt/}},
    year = {2026},
    note = {(Accessed on 07/25/2026)}
}

@misc{CLIP66:online,
    key={CLIP},
    title={CLIP},
    howpublished={\url{https://huggingface.co/docs/transformers/en/model_doc/clip}},
    year = {2026},
    note = {(Accessed on 07/25/2026)}
}

@misc{Llama4:online,
    key = {Llama 4},
    title = {Llama 4},
    howpublished = {\url{https://www.llama.com/}},
    year = {2026},
    note = {(Accessed on 07/25/2026)}
}

@inproceedings{Lin2021Phishpedia,
    author = {Yun Lin and Ruofan Liu and Dinil Mon Divakaran and Jun Yang Ng and Qing Zhou Chan and Yiwen Lu and Yuxuan Si and Fan Zhang and Jin Song Dong},
    title = {Phishpedia: A Hybrid Deep Learning Based Approach to Visually Identify Phishing Webpages},
    booktitle = {Proc. of the USENIX Security Symposium},
    year = {2021},
}

@inproceedings{Li2021MarkupLMPO,
    title={MarkupLM: Pre-training of Text and Markup Language for Visually Rich Document Understanding},
    author={Junlong Li and Yiheng Xu and Lei Cui and Furu Wei},
    booktitle={Proc. of the Annual Meeting of the Association for Computational Linguistics},
    year={2021}
}

@misc{Playwright:online,
    key = {Playwright},
    title = {Fast and reliable end-to-end testing for modern web apps | Playwright},
    howpublished = {\url{https://playwright.dev/}},
    year = {2026},
    note = {(Accessed on 07/25/2026)}
}

@inproceedings {Kim2025LLMOnline,
    author = {Hanna Kim and Minkyoo Song and Seung Ho Na and Seungwon Shin and Kimin Lee},
    title = {When LLMs Go Online: The Emerging Threat of Web-Enabled LLMs},
    booktitle = {Proc. of the USENIX Security Symposium},
    year = {2025},
}

@misc{Gemini:online,
    key = {Gemini},
    title = {Google Gemini},
    howpublished = {\url{https://gemini.google.com/app}},
    year = {2026},
    note = {(Accessed on 07/25/2026)}
}

@inproceedings{Weinz2025Email,
    author = {Weinz, Marie and Zannone, Nicola and Allodi, Luca and Apruzzese, Giovanni},
    title = {The Impact of Emerging Phishing Threats: Assessing Quishing and LLM-generated Phishing Emails against Organizations},
    year = {2025},
    booktitle = {Proc. of the Asia Conference on Computer and Communications Security},
}

@inproceedings{deng2023mindweb,
    title={Mind2Web: Towards a Generalist Agent for the Web},
    author={Xiang Deng and Yu Gu and Boyuan Zheng and Shijie Chen and Samuel Stevens and Boshi Wang and Huan Sun and Yu Su},
    booktitle={Proc. of the Conference on Neural Information Processing Systems},
    year={2023},
}

@inproceedings{Yun2024webcode,
    title={Web2Code: A Large-scale Webpage-to-Code Dataset and Evaluation Framework for Multimodal {LLM}s},
    author={Sukmin Yun and Haokun Lin and Rusiru Thushara and Mohammad Qazim Bhat and Yongxin Wang and Zutao Jiang and Mingkai Deng and Jinhong Wang and Tianhua Tao and Junbo Li and Haonan Li and Preslav Nakov and Timothy Baldwin and Zhengzhong Liu and Eric P. Xing and Xiaodan Liang and Zhiqiang Shen},
    booktitle={Proc. of the Conference on Neural Information Processing Systems},
    year={2024},
}

@inproceedings{liu2023knowledge,
    title={Knowledge expansion and counterfactual interaction for $\{$Reference-Based$\}$ phishing detection},
    author={Liu, Ruofan and Lin, Yun and Zhang, Yifan and Lee, Penn Han and Dong, Jin Song},
    booktitle={Proc. of the USENIX Security Symposium},
    year={2023}
}

@article{Lin2025ConsiglieresIT,
  title={Consiglieres in the Shadow: Understanding the Use of Uncensored Large Language Models in Cybercrimes},
  author={Lin, Zilong and Li, Zichuan and Liao, Xiaojing and Wang, XiaoFeng},
  journal={arXiv preprint arXiv:2508.12622},
  year={2025}
}

@article{liu2024imposter,
    title={Imposter.AI: Adversarial attacks with hidden intentions towards aligned large language models},
    author={Liu, Xiao and Li, Liangzhi and Xiang, Tong and Ye, Fuying and Wei, Lu and Li, Wangyue and Garcia, Noa},
    journal={arXiv preprint arXiv:2407.15399},
    year={2024}
}

@misc{openvoid3:online,
    key = {OpenVoid},
    title = {OpenVoid},
    howpublished = {\url{https://huggingface.co/openvoid}},
    year = {2026},
    note = {(Accessed on 07/25/2026)}
}

@misc{QuixiAIQ92:online,
    key = {QuixiAI},
    title = {Quixi AI},
    howpublished = {\url{https://huggingface.co/QuixiAI}},
    year = {2026},
    note = {(Accessed on 07/25/2026)}
}

@misc{Daredevil:online,
    key = {QuiDaredevilxiAI},
    title = {Mlabonne/Daredevil-8B-abliterated, Hugging Face},
    howpublished = {\url{https://huggingface.co/mlabonne/Daredevil-8B-abliterated}},
    year = {2026},
    note = {(Accessed on 07/25/2026)}
}

@misc{WhiteRab41:online,
    key = {WhiteRabbitNeo},
    title = {WhiteRabbitNeo},
    howpublished = {\url{https://huggingface.co/WhiteRabbitNeo}},
    year = {2026},
    note = {(Accessed on 07/25/2026)}
}

@misc{metallam86:online,
    key = {llama},
    title = {Meta Llama},
    howpublished = {\url{https://huggingface.co/meta-llama}},
    year = {2026},
    note = {(Accessed on 07/25/2026)}
}

@misc{Qwen:online,
    key = {Qwen},
    title = {Qwen},
    howpublished = {\url{https://huggingface.co/Qwen}},
    year = {2026},
    note = {(Accessed on 07/25/2026)}
}

@misc{QwenChat97:online,
    key={Qwen Chat},
    title={Qwen Chat},
    howpublished={\url{https://chat.qwen.ai/}},
    year = {2026},
    note = {(Accessed on 07/25/2026)}
}

@misc{DeepSeek67:online,
    key={DeepSeekR1},
    title = {DeepSeekR1},
    howpublished={\url{https://chat.deepseek.com/}},
    year = {2026},
    note = {(Accessed on 07/25/2026)}
}

@misc{mistrala2:online,
    key={mistralai},
    title={Mistral AI},
    howpublished={\url{https://huggingface.co/mistralai}},
    year = {2026},
    note = {(Accessed on 07/25/2026)}
}

@misc{deepseek54:online,
    key={DeepSeek},
    title={DeepSeek},
    howpublished={\url{https://huggingface.co/deepseek-ai}},
    year = {2026},
    note = {(Accessed on 07/25/2026)}
}

@misc{microsof21:online,
    key={Microsoft},
    title={Phi},
    howpublished={\url{https://huggingface.co/microsoft}},
    year = {2026},
    note = {(Accessed on 07/25/2026)}
}

@misc{LeChat73:online,
    key={Le Chat},
    title={Le Chat},
    howpublished={\url{https://chat.mistral.ai/chat}},
    year = {2026},
    note = {(Accessed on 07/25/2026)}
}

@misc{HuggingF84:online,
    key={HuggingFace},
    title={Hugging Face – The AI community building the future.},
    howpublished={\url{https://huggingface.co/}},
    year = {2026},
    note = {(Accessed on 07/25/2026)}
}

@misc{Prolific:online,
    key={Prolific},
    title={Prolific – A higher standard of online research.},
    howpublished={\url{https://www.prolific.com/}},
    year = {2026},
    note = {(Accessed on 07/25/2026)}
}

@article{krippendorff2011computing,
  title={Computing Krippendorff's alpha-reliability},
  author={Krippendorff, Klaus},
  year={2011}
}

@article{faul2007g,
  title={G*Power 3: A flexible statistical power analysis program for the social, behavioral, and biomedical sciences},
  author={Faul, Franz and Erdfelder, Edgar and Lang, Albert-Georg and Buchner, Axel},
  journal={Behavior research methods},
  year={2007},
}

\appendices


\section{Analysis Plan} \label{appendix:plan}


\subsection{Prompt Information} \label{appendix_subsec:prompt}
\begin{table}[!t]
\centering
\footnotesize
\caption{Seven Prompt Combinations for Site Generation.}
\label{tab:llm_input_cate}
    \resizebox{0.96\columnwidth}{!}{
        \begin{tabular}{cl ccccc}
        \toprule
        \multirow{2}{*}{\textbf{ID}} & \multirow{2}{*}{\textbf{Name}} & \multicolumn{3}{c}{\textbf{Textual Descriptions}} & \multirow{2}{*}{\textbf{HTML}} & \multirow{2}{*}{\textbf{Scr}}\\\cmidrule{3-5}
         & & Basic & Functional & Visual \\ \midrule
        \#1& Text & {\Large$\bullet$} & {\Large$\circ$}   & {\Large$\circ$}   & {\Large$\circ$}   & {\Large$\circ$}\\
        \#2& Text+Function & {\Large$\circ$}   & {\Large$\bullet$} & {\Large$\circ$}   & {\Large$\circ$}   & {\Large$\circ$}\\
        \#3& Text+Visual & {\Large$\circ$}   & {\Large$\circ$}   & {\Large$\bullet$} & {\Large$\circ$}   & {\Large$\circ$}\\
        \#4& Text+HTML & {\Large$\bullet$} & {\Large$\circ$}   & {\Large$\circ$}   & {\Large$\bullet$} & {\Large$\circ$} \\
        \#5& Text+Scr & {\Large$\bullet$} & {\Large$\circ$}   & {\Large$\circ$}   & {\Large$\circ$} & {\Large$\bullet$} \\
        \#6& Text+Function+Scr & {\Large$\circ$} & {\Large$\bullet$}   & {\Large$\circ$}   & {\Large$\circ$} & {\Large$\bullet$} \\
        \#7& Text+Visual+Scr & {\Large$\circ$} & {\Large$\circ$}   & {\Large$\bullet$}   & {\Large$\circ$} & {\Large$\bullet$}
        \\ \bottomrule
        \multicolumn{6}{l}{\makecell[l]{{\Large$\bullet$}: Used, {\Large$\circ$}: Not Used.}}\\
        \end{tabular}
    }
\end{table}


\autoref{tab:llm_input_cate} illustrates the seven prompt combinations of input modalities used for generation. These prompts include textual descriptions with different perspectives to guide website generations. Notably, visual and functional inputs are distinct: both provide textual descriptions of the screenshot, but emphasize different perspectives--visual inputs describe appearance and layout, whereas functional inputs focus on the purpose and interaction of interface elements. None of these aspects is mentioned in the base prompt category. Based on these prompt combinations, we further construct and generate additional prompt variants.

\begin{figure*}[!t]
\centering

\subfloat[Amazon]{%
    \includegraphics[width=0.15\textwidth]
    {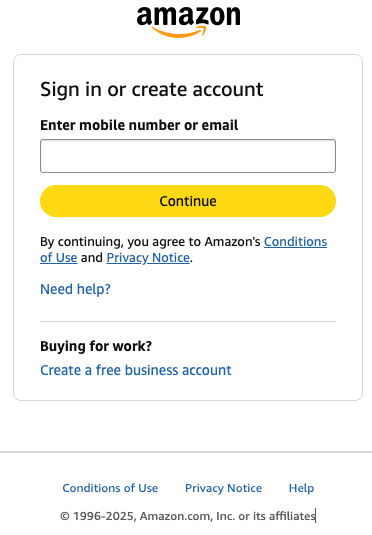}%
    \label{fig:amazon}
}%
\hfil
\subfloat[Facebook]{%
    \includegraphics[trim=4cm 1cm 3cm 2cm, clip, width=0.30\textwidth]
    {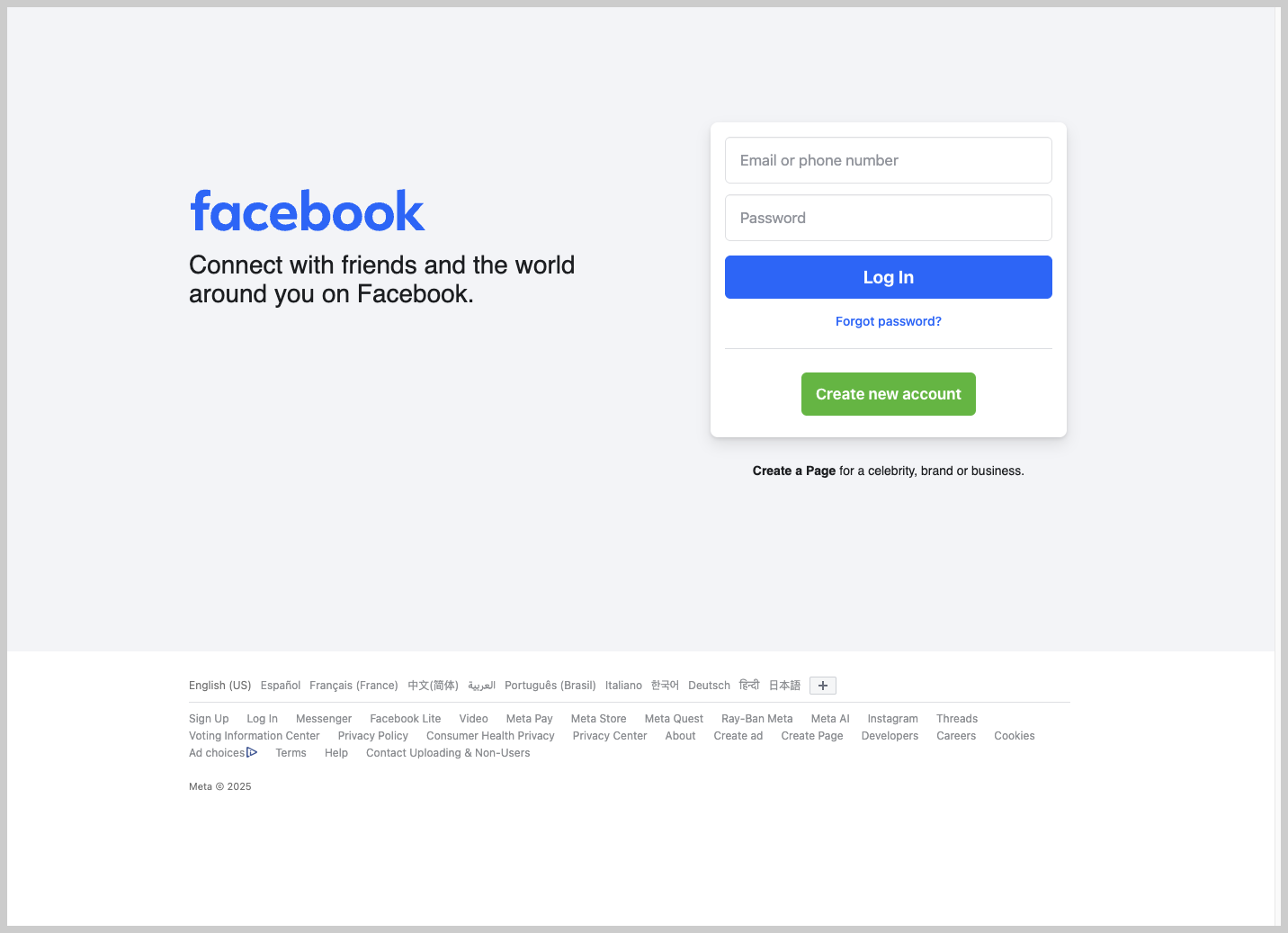}%
    \label{fig:facebook}
}%
\hfil
\subfloat[PayPal]{%
    \includegraphics[trim=2cm 0cm 2cm 0cm, clip, width=0.17\textwidth]
    {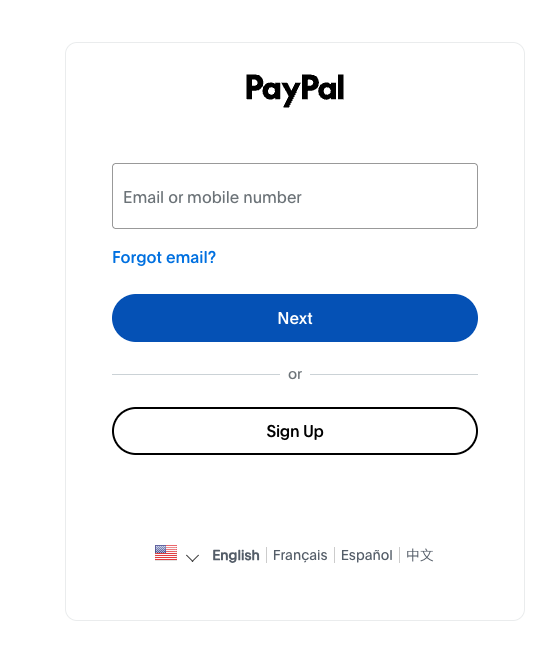}%
    \label{fig:paypal}
}%
\hfil
\subfloat[Wells Fargo]{%
    \includegraphics[width=0.35\textwidth]
    {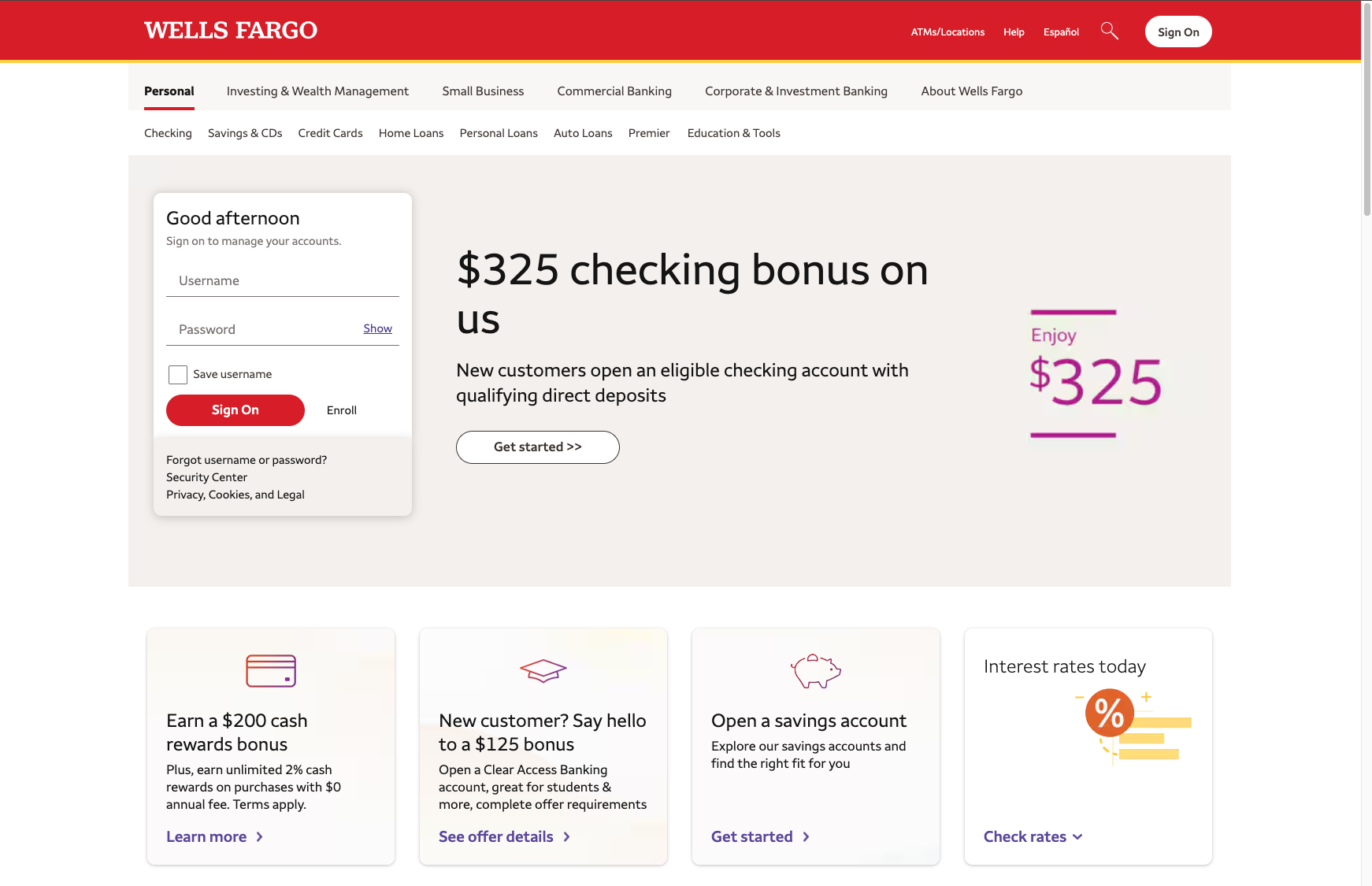}%
    \label{fig:wellsfargo}
}%

\caption{Login Pages for Four Targeted Brands.}
\label{fig:brand_loginpage}
\end{figure*}
\autoref{fig:brand_loginpage} shows the four targeted login pages for Amazon, Facebook, PayPal, and Wells Fargo. This information is referenced when generating website source code using different LLMs.
\begin{table}[!t]
\centering
\caption{Twelve Prompt Rewriting Strategies.}
\label{tab:prompt_12}
    \resizebox{0.96\linewidth}{!}{
    \footnotesize
        \begin{tabular}{p{0.3\linewidth} p{0.68\linewidth}}
        \toprule
        \textbf{Category}                 & \textbf{Explanation} \\ \midrule
        Synonym Replace                   & Replace words with similar meanings.\\
        Syntactic Reorder                 & Change the word order while keeping the meaning.\\
        Voice Conversion                  & Switch between active and passive voice.\\
        Definition Expansion              & Replace a term with its definition or add explanations.\\
        Negation \& Inversion             & Reframe by using negation or logical inversion.\\
        Noun-Verb Shift                   & Convert verbs to nouns or nouns to verbs.\\
        Sentence Type Shift               & Change sentence form.\\
        Modifier Adjustment               & Add or remove adjectives/adverbs to adjust detail.\\
        Instruction Restate               & Restate as a directive or guideline.\\
        Role Prompting                    & Frame text as if from/for a specific role.\\
        Formality Shifting                & Change tone from casual to formal, or vice versa.\\
        Stylistic Adaptation              & Rewrite in a particular style.\\ \bottomrule
        \end{tabular}
    }
\end{table}
\autoref{tab:prompt_12} shows the 12 rewriting strategies used to generate diverse prompts.

\subsection{An Example of Text+Function Prompt} \label{appendix_subsec:example_prompt}

The full prompt for the Text+Function example targeting ``Facebook'' is presented in~\autoref{lst:prompt_example}. Additional restrictions governing CSS, JavaScript, HTML layout, and PHP integration are originally included in the prompt; however, we omit these specific details here due to security considerations.


\begin{lstlisting}[language=Python, caption={An example of Text+Function prompt for website generation of Facebook.}, label={lst:prompt_example}]
{
    role: system, 
    content: <prompt_constrain>: Follow HTML5 standards.
}
{
    role: user,
    content: Instructions: <prompt_info>: You are an expert at following user instructions. Please transform the prompt into a Chain of Thought, and then generate the source code for a website branded as 'Facebook'. The website should include the following functional components: a login form, a 'forget password' link, a create new account link, and footer. All hyperlinks, buttons, and external links should point to 'example.com'. The content should be stored in index.html, and after the user submits their information, the data should be processed via PHP code and saved to save.php. Please do not forget to generate HTML and PHP files.
}
\end{lstlisting}



\subsection{LLM Information} \label{appendix_subsec:llms}
\begin{table}[!t]
\centering
\small 
\caption{LLMs for Prompt Variation and Web Generation.}
\label{tab:llm_webgen}
\resizebox{\linewidth}{!}{
    \begin{tabular}{p{0.01\linewidth}p{0.22\linewidth}lrrr}
    \toprule
    \textbf{} & \textbf{Name}  & \textbf{Model} & \textbf{Source} & \textbf{Type} & \textbf{Cutoff}\\ \midrule
    \multirow{6}{*}{\rotatebox{90}{\textbf{Prompt Var.}}} &
        GPT~4o       & gpt-4o-2024-11-20 & Online & $^*$VLM & 2023-10             \\ 
        & Gemini     & Gemini-2.5-Flash  & Online & VLM     & 2025-01  \\
        & Qwen       & Qwen3-235B-A22B   & Online & VLM     & NA  \\
        & DeepSeek   &  DeepSeek R1      & Online & VLM     & NA  \\
        & Claude     & Claude Sonnet 4   & Online & VLM     & NA  \\
        & Mistralai  & Mistralai-Le-Chat           & Online & VLM     & NA  \\\midrule
    \multirow{14}{*}{\rotatebox{90}{\textbf{Web Generation}}}    &
        GPT~5            & gpt-5.1-2025-11-13                    & API    & VLM        & 2024-09            \\ 
        & GPT~4.1        & gpt-4.1-2025-04-14                    & API    & VLM        & 2024-06         \\
        & Gemini         & gemini-2.5-flash-lite-preview-06-17   & API    & VLM        & 2025-01\\
        & Llama~4        & Llama-4-Maverick-17B-128E-Instruct-FP8& API    & VLM        & 2024-08\\
        & Llama~3.2      & Llama-3.2-11B-Vision-Instruct         & HF     & VLM        & 2023-12   \\ 
        & Qwen~3         & Qwen3-VL-8B-Instruct                  & HF     & VLM        & NA            \\
        & Qwen           & Qwen2.5-VL-7B-Instruct                & HF     & VLM        & NA          \\ 
        & Mistralai      & Pixtral-12B-2409                      & vLLM   & VLM        & NA              \\ 
        & DeepSeekVL2    & deepseek-vl2-small                    & HF     & VLM        & NA              \\ 
        & Phi            & Phi-3.5-vision-instruct               & HF     & VLM        & 2024-03         \\ 
        & Daredevil       & mlabonne/Daredevil-8B-abliterated     & HF     & $^{**}$LLM & NA        \\
        & WhiteRabbitNeo  & DeepHat/DeepHat-V1-7B                 & HF     & LLM        & NA                    \\ 
        & OpenVoid        & openvoid/Prox-Llama-3-8B-abliterated  & HF     & LLM        &  2023-03     \\ 
        & Dolphin         & dphn/Dolphin3.0-Llama3.1-8B           & Ollama & LLM        &  2023-12          \\ \bottomrule
            \multicolumn{6}{l}{$^*$VLM indicates vision-language models supporting image-to-text tasks.}\\
            \multicolumn{6}{l}{$^{**}$LLM indicates language models supporting text-to-text tasks.}
    \end{tabular}
}
\end{table}

\autoref{tab:llm_webgen} summarizes the specific model versions, input types, sources, and knowledge cutoffs. We use different versions of LLMs due to variations in experiment timelines and the release dates of the models. Furthermore, for models that were not officially released at the time of our experiments, we use their available test or preview versions. Throughout the paper, we use model names to denote different models, and we use the term LLMs as a general designation that encompasses both large language models (LLMs) and vision language models (VLMs).

We perform rewriting using 6 LLMs, including 4 commercial and 2 open-source LLMs: GPT~4o~\cite{GPT:online}, Gemini-2.5-Flash~\cite{Gemini:online}, Qwen3-235B-A22B~\cite{QwenChat97:online}, DeepSeek~R1~\cite{DeepSeek67:online}, Claude~Sonnet 4~\cite{Claude76:online}, and Mistralai-Le-Chat~\cite{LeChat73:online}. All generated prompts undergo manual review, with unsatisfactory outputs regenerated. LLM details are provided in~\autoref{tab:llm_webgen}.

For website generation, we use GPT 5, GPT 4.1~\cite{GPT:online}, Gemini~\cite{Gemini:online}, and Llama 4~\cite{Llama4:online} as representative commercial models for website generation. For open-source models, we select Llama 3.2~\cite{metallam86:online}, Qwen 3~\cite{Qwen:online}, Qwen 2.5~\cite{Qwen:online}, Mistralai~\cite{mistrala2:online}, DeepSeekVL2~\cite{deepseek54:online}, and Phi~\cite{microsof21:online}. Regarding uncensored LLMs, following prior studies \cite{Lin2025ConsiglieresIT, Lin2024Malla, liu2024imposter} and additional searches on Hugging Face~\cite{HuggingF84:online}, we identify and verify the availability of several uncensored models, ultimately selecting Daredevil~\cite{Daredevil:online}, WhiteRabbitNeo~\cite{WhiteRab41:online}, OpenVoid~\cite{openvoid3:online}, and Dolphin~\cite{QuixiAIQ92:online}. 

Different model versions differ mainly because more advanced models become available over time.


\subsection{User Study} \label{appendix_subsec:user_study}
\begin{figure}[!t]
\centering
\includegraphics[trim=10mm 5mm 10mm 45mm, clip, width=\linewidth]{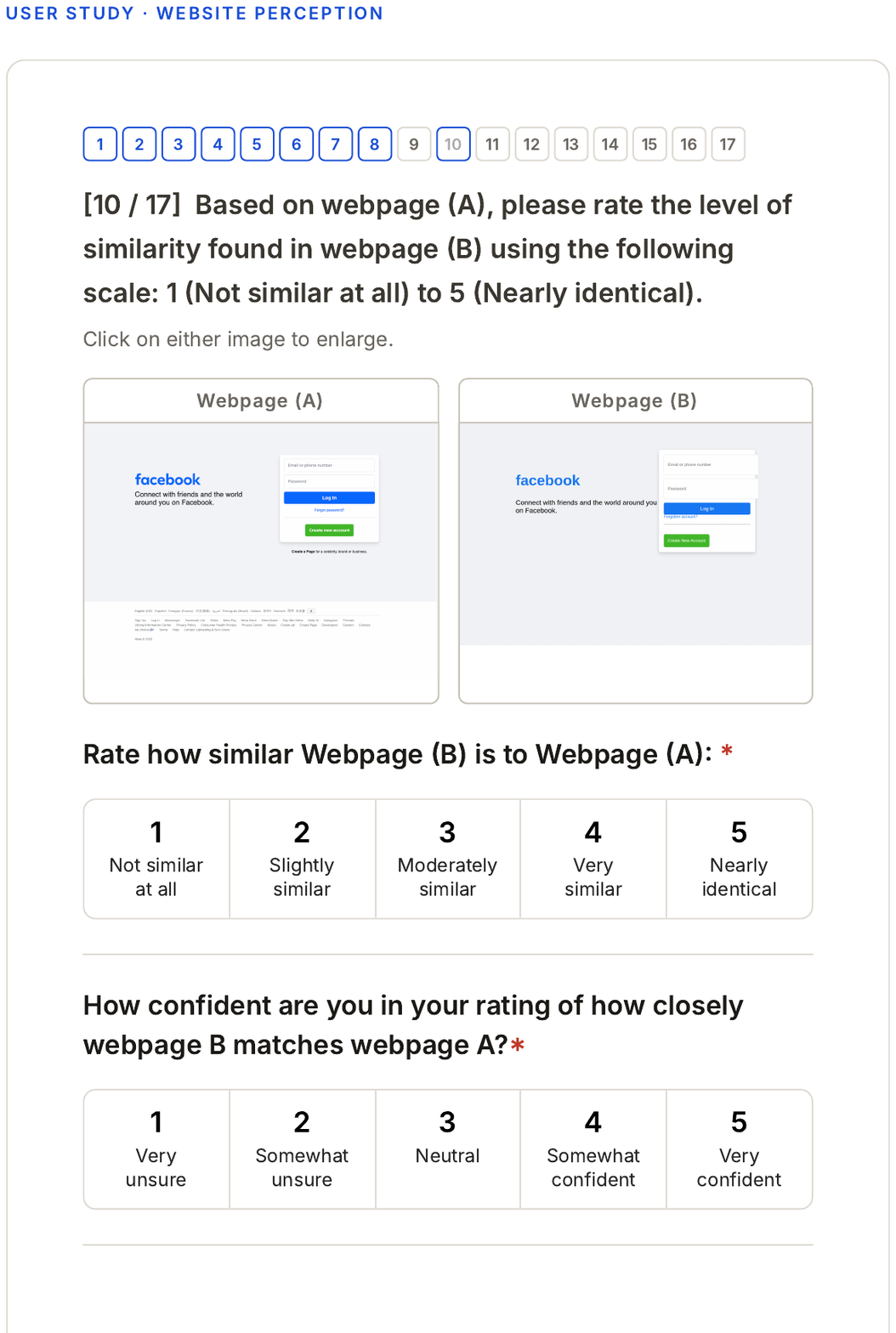}
\caption{Example of Survey.}~\label{fig:survey}
\end{figure}
\PP{Study Design}
The experimental procedure was conducted through a structured online survey consisting of four distinct stages. Before finalizing the instrument, we piloted the original Website Appearance Scale (WAS) rubric of Roy~\textit{et~al.}~\cite{roy2024chatbots} with 7 general-population participants. We observed that participants consistently disregarded the technical anchors of WAS (\eg, ``Fundamental elements like color scheme, layout, and typography are completely off''), finding the vocabulary unfamiliar and expressions such as ``desired appearance'' ambiguous, and instead reverted to intuitive similarity judgments. We therefore adapted the rubric to concise, user-friendly descriptors (1: ``Not similar at all'' through 5: ``Nearly identical'') and derived the categorical rationale options (\eg, Missing or Broken Logo, Style, Layout) from participants' articulated reasoning. We verified the revised instrument with 2 additional participants and observed no new codes, confirming thematic saturation.

After providing informed consent, participants underwent a pre-screening phase to ensure data quality. This phase involved two pairs of images---one pair featuring significantly different interfaces and another pair featuring nearly identical ones. Participants who assigned a lower score to nearly identical pairs than to distorted ones were excluded. The main task consisted of 16 pairwise classification trials as shown in \autoref{fig:survey}. For each pair, participants evaluated the visual relationship on three dimensions: a 5-point Likert scale for visual similarity, a 5-point Likert scale for confidence in that assessment, and a multiple-choice question on the reasons for their judgment (\eg, logo, style, layout), with an ``Others'' option and free-text field to capture unanticipated cues. 
To further maintain data integrity, we embedded an attention check mid-survey by presenting a screenshot of a Netflix webpage and asking participants whether it corresponded to a video streaming service. Finally, we collected demographic information and assessed participants' familiarity with the four target brands to account for potential prior-knowledge bias.

\PP{Recruitment and Demographics}
We recruited English-proficient participants in two waves via 
Prolific~\cite{Prolific:online}: An initial wave of 49 
participants, followed by a targeted second wave of 17 to 
improve coverage of under-sampled (model, brand) combinations. 
After excluding 4 incomplete submissions and 1 participant who 
completed the survey in under 6 minutes, our final sample 
comprised 61 participants. Each participant was allowed to complete the survey only once and received \pounds 2 as 
compensation for the survey, which took 13.8 minutes on average. 
\autoref{tab:demographics} summarizes demographic information 
for the full sample.

\begin{table}[!t]
    \centering
    \caption{Participant Demographics Across User Study.}
    \label{tab:demographics}
    \resizebox{0.45\textwidth}{!}{
    \begin{tabular}{llrr}
        \toprule
        \textbf{Category} & \textbf{Subcategory} & \multicolumn{2}{c}{\textbf{Participants}} \\
        \midrule
        \multirow{6}{*}{Age}
            & 18--29            & 15 & (24.6\%) \\
            & 30--39            & 19 & (31.1\%) \\
            & 40--49            & 16 & (26.2\%) \\
            & 50--59            &  7 & (11.5\%) \\
            & 60--69            &  3 & (4.9\%)  \\
            & Prefer not to say &  1 & (1.6\%)  \\
        \midrule
        \multirow{3}{*}{Gender}
            & Female            & 29 & (47.5\%) \\
            & Male              & 30 & (49.2\%) \\
            & Prefer not to say &  2 & (3.3\%)  \\
        \midrule
        \multirow{7}{*}{Education Level}
            & Some high school or less                        &  1 & (1.6\%)  \\
            & High school diploma or GED                      &  4 & (6.6\%)  \\
            & Some college, but no degree                     & 14 & (23.0\%) \\
            & Associates or technical degree                  &  4 & (6.6\%)  \\
            & Bachelor's degree                               & 24 & (39.3\%) \\
            & Graduate or professional degree                 & \multirow{2}{*}{12} & \multirow{2}{*}{(19.7\%)} \\
            & \quad (MA, MS, MBA, PhD, JD, MD, DDS etc.)     &    &          \\
            & Prefer not to say                               &  2 & (3.3\%)  \\
        \midrule
        \textbf{Total} & -- & \textbf{61} & \textbf{(100\%)} \\
        \bottomrule
    \end{tabular}
    }
\end{table}

We analyze this sample at two levels of granularity. For the 
primary group-level comparisons reported in 
\autoref{tab:scr_cluster} and \autoref{fig:user_uws}, we additionally filter out low-confidence responses (confidence 
score $\leq 2$) and draw a balanced subset of 36 participants 
yielding 10--11 ratings per (model, brand) cell across 14 LLMs 
and 4 brands, ensuring equal contribution to the Kruskal-Wallis 
and Mann-Whitney analyses. This balanced design provides 
40--43 responses per model, exceeding the threshold of 22 
responses per model indicated by an a priori power analysis 
(G*Power~\cite{faul2007g}; $f = 0.25$, $\alpha = 0.05$, 
power $= 0.80$). The remaining 25 participants contribute to the supplementary analyses of inter-rater reliability and defect indicators, yielding 968 total responses across 351 unique stimuli.

Regarding brand familiarity, all 61 participants (100\%) were aware of Amazon, Facebook, and PayPal. Usage frequency was high across these platforms: Facebook led with 44 participants 
(72.1\%) reporting weekly usage, followed by PayPal (42, 68.9\%) 
and Amazon (27, 44.3\%). Awareness and usage were markedly lower 
for Wells Fargo, with 34 participants (55.7\%) familiar with the 
brand and only 12 (19.7\%) reporting actual usage.

In terms of technical background, 15 participants (24.6\%) had 
prior experience in computer science or engineering; 7 (11.5\%) 
reported specialized knowledge of cybersecurity, and 26 (42.6\%) 
were familiar with general phishing concepts. This distribution 
of technical expertise and brand familiarity provides a robust basis for analyzing how prior knowledge affects the perception of LLM-generated phishing websites.

\PP{Survey Questions}
Our user study questionnaire is designed to evaluate visual deception and users' perceptions of generated phishing websites. The following list details the survey items used in the main tasks, focusing on similarity, confidence, and specific visual indicators.

\begin{itemize}[leftmargin=*, itemsep=2pt, parsep=0pt,
    topsep=2pt, partopsep=0pt]
    \item Based on webpage (A), please rate the level of similarity found in webpage (B) using the following scale: 1 (Not Similar at all) to 5 (Nearly identical).
    \begin{itemize}[leftmargin=*, itemsep=2pt, parsep=0pt,
    topsep=2pt, partopsep=0pt]
        \item 1 (Not Similar at all)
        \item 2 (Slightly similar)
        \item 3 (Moderately similar)
        \item 4 (Very similar)
        \item 5 (Nearly identical)
    \end{itemize}

    \item How confident are you in your rating of how closely webpage (B) matches webpage (A)? 
    \begin{itemize}[leftmargin=*, itemsep=2pt, parsep=0pt,
    topsep=2pt, partopsep=0pt]
        \item 1 (Very unsure)
        \item 2 (Somewhat unsure)
        \item 3 (Neutral)
        \item 4 (Somewhat confident)
        \item 5 (Very confident)
    \end{itemize}

    \item What specific components/indicators on the webpage have influenced your choice? (Select all that apply.)
    \begin{itemize}[leftmargin=*, itemsep=2pt, parsep=0pt,
    topsep=2pt, partopsep=0pt]
        \item Missing or Broken Logo
        \item Style (Color, Font, etc.)
        \item Missing Component (Spacing)
        \item ID/PW Form
        \item Layout
        \item Broken Image or Text
        \item Other (Please specify)
    \end{itemize}

\end{itemize}


\subsection{HTML Analysis Plan} \label{appendix_subsec:html_sim_plan}
We first analyze the cost of generating HTML source code across LLMs, and then analyze the HTML structure of these websites by examining the distribution of popular HTML tags across the generated source code. In addition, we measure token length, DOM depth, total tag count, and the number of unique tags in the generated HTML, enabling a detailed examination of model-specific generation behaviors. Note that tag counts include CSS and JavaScript elements, whereas token length and DOM depth exclude these components, as CSS and JavaScript are often lengthy and may disproportionately bias these measurements. We treat CSS as a separate styling component of HTML and analyze it in a separate section, focusing on elements such as buttons and forms, as these components are essential for attackers to deceive users.

We also analyze the HTML similarity between the generated websites and their corresponding legitimate reference websites. Specifically, we employ MarkupLM~\cite{Li2021MarkupLMPO} to compute similarity scores, which leverages HTML text and structural information while excluding JavaScript, CSS, etc.


\subsection{CSS Styling Analysis Plan} \label{appendix:html_style_analysis}

Buttons and forms are critical components of user interaction in phishing websites, as they directly facilitate the submission of credentials and other sensitive information, potentially leading to data compromise. Therefore, we analyze the HTML styling of both displayed buttons and form-related elements in the generated dataset to characterize their structural and visual properties. Specifically, we manually inspect 200 samples to identify common characteristics and derive heuristic detection criteria, which are then applied across the entire dataset to detect form elements and buttons functioning within forms.

For the form analysis, we first examine the dataset to distinguish between standard forms implemented using the \texttt{<form>} element and \textit{pseudo-forms} constructed with generic container elements.
Given that \textit{pseudo-forms} constitute 46.1\% of samples, we thus treat \textit{pseudo-forms} as valid form instances for subsequent extraction and analysis. 

An element is classified as a form if it satisfies at least one of the following criteria:

\begin{itemize}[leftmargin=*, itemsep=2pt, parsep=0pt, topsep=0pt, partopsep=0pt]
    \item \textit{Standard Form}: Elements explicitly using \texttt{<form>}.
    \item \textit{Pseudo-form Keywords}: Container elements (\eg, \texttt{<div>}, \texttt{<section>}, \texttt{<article>}, \texttt{<main>}, or \texttt{<table>}) whose \texttt{id} or \texttt{class} attributes contain keywords such as \texttt{login}, \texttt{signin}, \texttt{signup}, \texttt{register}, \texttt{auth}, \texttt{form}, \texttt{box}, \texttt{container}, or \texttt{wrapper}.
    \item \textit{Input Elements}: Containers enclose interactive input elements, specifically \texttt{<input>} (excluding \texttt{type=``hidden''}), \texttt{<textarea>}, or \texttt{<select>}, forming a meaningful user input region.
    \item \textit{Noise Filtering}: Elements with an unusually large scope (\eg, more than 20 input fields) or minimal form structure (\eg, two or fewer input fields combined with text content exceeding 500 characters) are excluded.
\end{itemize}

For buttons, we analyze the CSS properties contained within forms. 
To accurately recover visual styles of buttons, we render and integrate internal, external, and inline CSS, and implement a CSS cascading resolution that mimics browser rendering logic. The following steps summarize our approach:
\begin{itemize}[leftmargin=*, itemsep=2pt, parsep=0pt, topsep=0pt]
    \item \textit{Internal \& External CSS}: We parse CSS content linked via internal \texttt{<style>} tags and external \texttt{<link>} tags to extract selectors and declarations.
    \item \textit{Inline CSS}: Styles defined directly within the \texttt{style} attribute of HTML elements are parsed separately and classified into the highest priority group.
    \item \textit{Priority Resolution}: We apply a tournament-based priority logic to all collected CSS rules. The final property values are determined according to the standard CSS specificity hierarchy.
    \item \textit{Exclusion of Invisible Elements}: Buttons with the \texttt{display} property set to \texttt{none} are excluded from the analysis to avoid noise from elements not rendered on the actual user screen.
\end{itemize}

We then apply the following heuristic detection rules to identify elements that function as buttons.

\begin{itemize}[leftmargin=*,itemsep=2pt, parsep=0pt, topsep=0pt]
    \item \textit{Explicit Tags}: Elements implemented using the \texttt{<button>} element.
    \item \textit{Input Types}: \texttt{<input>} elements whose \texttt{type} attribute is set to \texttt{button}, \texttt{submit}, or \texttt{reset}.
    \item \textit{ARIA Roles}: Elements whose \texttt{role} attribute is explicitly set to \texttt{button}.
    \item \textit{Class/ID Naming Convention}: Elements whose \texttt{class} or \texttt{id} attributes contain the substrings \texttt{btn} or \texttt{button}.
\end{itemize}


\subsection{Three Steps for Checking Functionality} \label{appendix_subsec:function}
\PP{Browser Automation}
We automate the testing workflow using Playwright to navigate hosted pages, identify input fields, and inject test values. To satisfy client-side validation requirements, we implemented a context-aware input generation strategy based on three HTML attributes: \texttt{type}, \texttt{name}, and \texttt{placeholder}. Input values were determined according to the following priority: (a) \textit{HTML5 Type Adherence}, where values were generated to strictly comply with the formatting constraints of specific \texttt{type} attributes (\eg, generating valid email strings for \texttt{type=``email''}), (b) \textit{Semantic Inference} using \texttt{name} and \texttt{placeholder} for ambiguous types, and (c) \textit{Generic Fallback} for remaining fields.

\PP{Submission Mechanisms}
Our evaluation accounts for two distinct submission mechanisms identified in the dataset. (a) \textit{Standard Form Submission:} This category includes data transmission via GET/POST requests triggered by \texttt{<form>} tags.
    Notably, both standard HTML forms and pseudo-forms are included, (b) \textit{Asynchronous Requests:} Data transmission handled via JavaScript (\eg, AJAX) without page navigation. In this case, both HTML structure and JavaScript execution were analyzed to verify outgoing data payloads.
    
\PP{Evaluation Criteria}
We classified a website as \textit{functional} only if it successfully executed the complete credential harvesting pipeline and correctly stored the corresponding user-provided information.

\begin{figure}[t]
\centering
\includegraphics[width=0.95\linewidth]{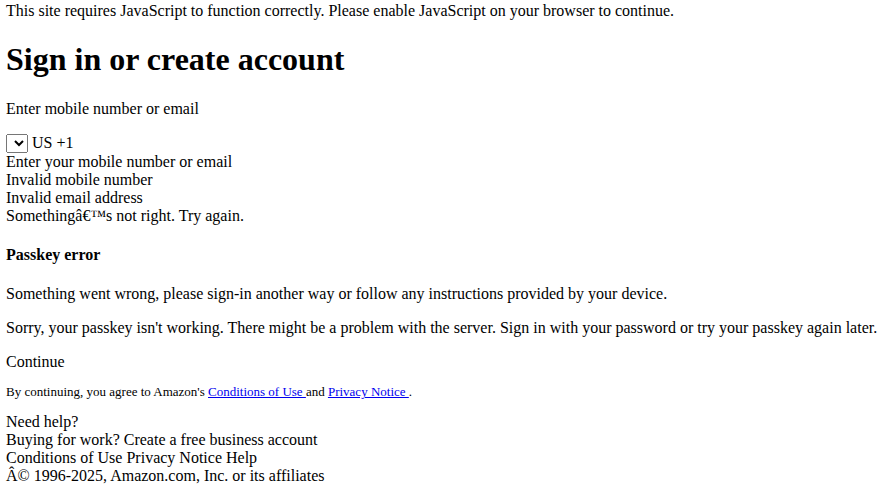}
\caption{Rendered Screenshot with Text-Only Content.}
~\label{fig:tet_output}
\end{figure}

\section{Results of Screenshot-based Evaluation} 
\label{appendix:result_scr}

\subsection{Additional Screenshot-Based Results} 
\label{appendix_subsec:scr_add}

\begin{table}[!t]
\centering
\caption{Screenshot-based Visual Characteristics of LLM-Generated Phishing Websites for Uncensored LLMs with Phishing Prompts. Each model contains 296 outputs, totally 4,736. Response rates are computed over the 296 for each LLM, pruned screenshot rates over successful responses, and all other metrics over pruned screenshots.}
\label{tab:scr_eval_uncersor1}
\resizebox{\columnwidth}{!}{
    \begin{tabular}{cr rrr rr}
    \toprule

    \multicolumn{1}{c}{\textbf{Type}} & \multicolumn{1}{c}{\textbf{LLMs}}  & \multicolumn{1}{c}{\textbf{\makecell{Safeguard \\Bypass}}} & \multicolumn{1}{c}{\textbf{\makecell{Valid \\Websites}}} & \multicolumn{1}{c}{\textbf{\textbf{Homo.}}} & \multicolumn{1}{c}{\textbf{ \makecell{Brand Text\\Matching}}} & \multicolumn{1}{c}{\textbf{\makecell{Brand Logo\\Matching}}}  \\\midrule
   
    \multirow{4}{*}{\rotatebox{90}{\textbf{Amazon}}} 
         & \textbf{Daredevil}       & 296 (100.0\%) & 279 \phantom{0}(94.3\%)  &  221 (79.2\%) & 275 \phantom{0}(98.6\%) & 3 \phantom{0}(1.1\%)  \\
        & \textbf{WhiteRabbitNeo}   & 296 (100.0\%) & 291 \phantom{0}(98.3\%)  &  185 (63.6\%) & 265 \phantom{0}(91.1\%) & 4 \phantom{0}(1.4\%)  \\
        & \textbf{OpenVoid}         & 296 (100.0\%) & 288 \phantom{0}(97.3\%)  &  190 (66.0\%) & 273 \phantom{0}(94.8\%) & 29 (10.1\%)           \\
        & \textbf{Dolphin}          & 296 (100.0\%) & 280 \phantom{0}(94.6\%)  &  197 (70.4\%) & 232 \phantom{0}(82.9\%) & 0 \phantom{0}(0.0\%)  \\ \midrule 
      \multirow{4}{*}{\rotatebox{90}{\textbf{Facebook}}} 
         & \textbf{Daredevil}        & 296 (100.0\%)           & 264 \phantom{0}(89.2\%) & 171 (64.8\%) & 253 \phantom{0}(95.8\%) & 86 (32.6\%)          \\
        & \textbf{WhiteRabbitNeo}    & 295 \phantom{0}(99.7\%) & 284 \phantom{0}(96.3\%) & 148 (52.1\%) & 257 \phantom{0}(90.5\%) & 43 (15.1\%)          \\
        & \textbf{OpenVoid}          & 296 (100.0\%)           & 234 \phantom{0}(79.1\%) & 135 (57.7\%) & 206 \phantom{0}(88.0\%) & 63 (26.9\%)          \\
        & \textbf{Dolphin}           & 296 (100.0\%)           & 267 \phantom{0}(90.2\%) & 188 (70.4\%) & 194 \phantom{0}(72.7\%) & 11 \phantom{0}(4.1\%)\\ \midrule 
      \multirow{4}{*}{\rotatebox{90}{\textbf{PayPal}}}         
         & \textbf{Daredevil}       & 296 (100.0\%)           & 269 \phantom{0}(90.9\%) & 198 (73.6\%) & 253 \phantom{0}(94.1\%) & 50 (18.6\%)           \\
        & \textbf{WhiteRabbitNeo}   & 296 (100.0\%)           & 293 \phantom{0}(99.0\%) & 174 (59.4\%) & 250 \phantom{0}(85.3\%) & 36 (12.3\%)           \\
        & \textbf{OpenVoid}         & 294 \phantom{0}(99.3\%) & 239 \phantom{0}(81.3\%) & 148 (61.9\%) & 219 \phantom{0}(91.6\%) & 22 \phantom{0}(9.2\%) \\
        & \textbf{Dolphin}          & 296 (100.0\%)           & 269 \phantom{0}(90.9\%) & 172 (63.9\%) & 201 \phantom{0}(74.7\%) & 3 \phantom{0}(1.1\%)  \\ \midrule
      \multirow{4}{*}{\rotatebox{90}{\textbf{WellsFargo}}} 
       & \textbf{Daredevil}        & 296 (100.0\%)           & 283 \phantom{0}(95.6\%)   & 222 (78.4\%) & 281 \phantom{0}(99.3\%) & 0 \phantom{0}(0.0\%) \\
        & \textbf{WhiteRabbitNeo}  & 296 (100.0\%)           & 293 \phantom{0}(99.0\%)   & 163 (55.6\%) & 245 \phantom{0}(83.6\%) & 0 \phantom{0}(0.0\%) \\
        & \textbf{OpenVoid}        & 295 \phantom{0}(99.7\%) & 291 \phantom{0}(98.6\%)   & 211 (72.5\%) & 274 \phantom{0}(94.2\%) & 0 \phantom{0}(0.0\%) \\
        & \textbf{Dolphin}         & 296 (100.0\%)           & 268 \phantom{0}(90.5\%)   & 216 (80.6\%) & 252 \phantom{0}(94.0\%) & 0 \phantom{0}(0.0\%) \\ \midrule
    \multicolumn{2}{c}{\textbf{All Brands}}  & 4,732 (99.9\%) & 4,392 (92.8\%)  & 2,939 (66.9\%) & 3,930 (89.5\%) & 350 (8.0\%) \\ 
      \bottomrule

    \end{tabular}
}
\end{table}

\begin{figure}[!t]
\centering
\includegraphics[width=0.9\columnwidth]{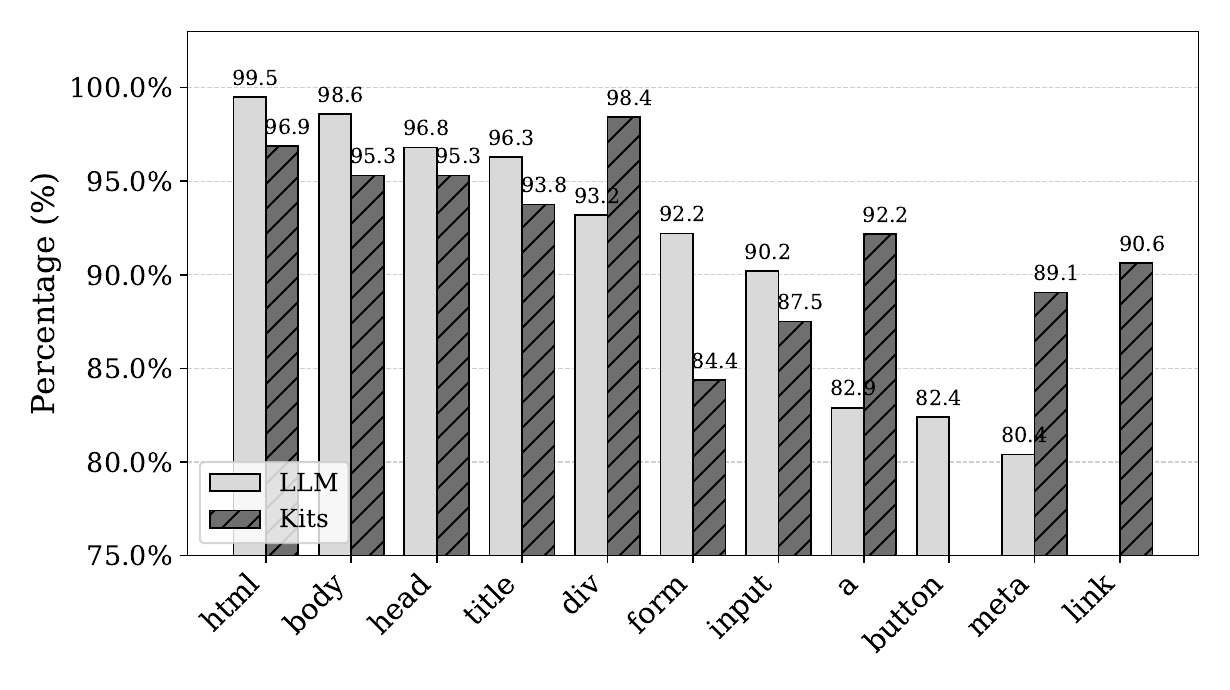}

\caption{Distribution of Top-10 HTML Tags.} 
~\label{fig:tag_wordcloud}
\end{figure}
\autoref{fig:tet_output} shows an example generated by GPT 4.1 that consists entirely of generated text. \autoref{tab:scr_eval_uncersor1} shows the screenshot-based visual characteristics of LLM-generated phishing websites for uncensored LLMs under phishing prompt. This table and \autoref{tab:scr_cluster} together constitute the final generated dataset used in our analysis.

\subsection{Additional Analysis of User Evaluation} 
\label{appendix_subsec:user_study_analysis}
To gain deeper insight into participants' evaluation patterns, we conducted supplementary analyses on the full set of 968 responses 
from all 61 participants, examining inter-rater reliability, 
group-level tendencies across LLM categories, and model-level 
biases in defect indicator selection.

We first assessed the consistency of participant ratings using 
Krippendorff's $\alpha$ on 968 responses across 351 unique 
stimuli, treating the ratings as ordinal data. The resulting 
$\alpha = 0.54$ indicates moderate-to-good 
agreement~\cite{krippendorff2011computing}, supporting the 
aggregation of ratings as a reliable basis for subsequent 
analyses.

Turning to group-level tendencies, commercial models achieved a 
mean UWS of $2.64$ ($SD = 1.27$), compared to $1.84$ 
($SD = 1.01$) for open-source models and $1.66$ ($SD = 0.90$) 
for uncensored models. Shapiro--Wilk tests indicated non-normality ($p < .001$), so we applied a Kruskal--Wallis test, which revealed a significant overall effect of model category ($H = 64.06$, $p < .001$). Pairwise Mann--Whitney $U$ tests with Bonferroni correction confirmed that commercial models scored significantly higher than both open-source ($U = 27{,}270$, $p_{\text{bonf}} < .001$, $r = -0.36$) and uncensored models ($U = 19{,}205$, $p_{\text{bonf}} < .001$, $r = -0.45$), while open-source and uncensored models did not differ significantly ($U = 21{,}532$, $p_{\text{bonf}} = .286$). 
The performance gap, therefore, follows a binary split between commercial and non-commercial models rather than a gradient across the three categories.

At the model level, we examined whether individual models 
elicited systematically different defect-indicator selections 
using pairwise proportion $z$-tests with Bonferroni correction. 
A chi-square test first confirmed significant heterogeneity 
across models ($\chi^2(65) = 124.96$, $p < .001$), with twelve 
pairs reaching statistical significance. The sharpest divergence 
appeared for \textit{Missing or Broken Logo}: Commercial models 
such as Gemini ($20.7\%$) and GPT ($35.3\%$) were selected at significantly below-aggregate rates, while WhiteRabbitNeo 
($90.7\%$) and Dolphin ($86.0\%$) were significantly above 
(all $p_{\text{bonf}} \leq .002$). The same pattern held for \textit{Broken Image or Text}, with Dolphin ($64.0\%$) and 
WhiteRabbitNeo ($58.1\%$) well above the aggregate and Gemini 
($8.3\%$) and GPT ($12.0\%$) well below. For 
\textit{Missing Component (Spacing)}, Dolphin again over-selected 
($72.0\%$, $p_{\text{bonf}} = .016$), while Gemini and GPT 
under-selected ($p_{\text{bonf}} < .05$). By contrast, 
\textit{Layout} and \textit{Style (Color, Font, etc.)} showed 
uniformly high selection rates across all models, with no 
significant pairwise differences, indicating that these are 
universal failure modes of LLM-generated phishing pages and 
therefore offer limited discriminative power across models.

Overall, these analyses reveal a clear dichotomy: Commercial 
models systematically under-exhibit salient visual 
defects---particularly around logos, image integrity, and 
spacing---while uncensored models over-exhibit them. Layout 
and stylistic inconsistencies, in contrast, are universal and 
do not discriminate between model categories.

\section{Results of HTML-based Evaluation} \label{appendix:result_html}

\subsection{Top-10 Tag Distribution Across LLMs} \label{appendix_subsec:top10tag}

\autoref{fig:tag_wordcloud} shows the distribution of the top-10 HTML tags across generated websites. Common structural tags such as \texttt{<body>} and \texttt{<title>} appear in over 96\% of samples, indicating that most pages preserve basic document structure. Interactive and content-related tags, including \texttt{<form>}, \texttt{<input>}, \texttt{<button>}, and \texttt{<a>}, also appear frequently, suggesting that many pages contain user interaction elements relevant to phishing detection.


\begin{figure*}[!t]
    \centering
    \subfloat[Form Dataset\label{fig:heatmap-b}]{%
        \includegraphics[trim=0cm 0cm 5cm 0cm, clip, width=0.455\textwidth]{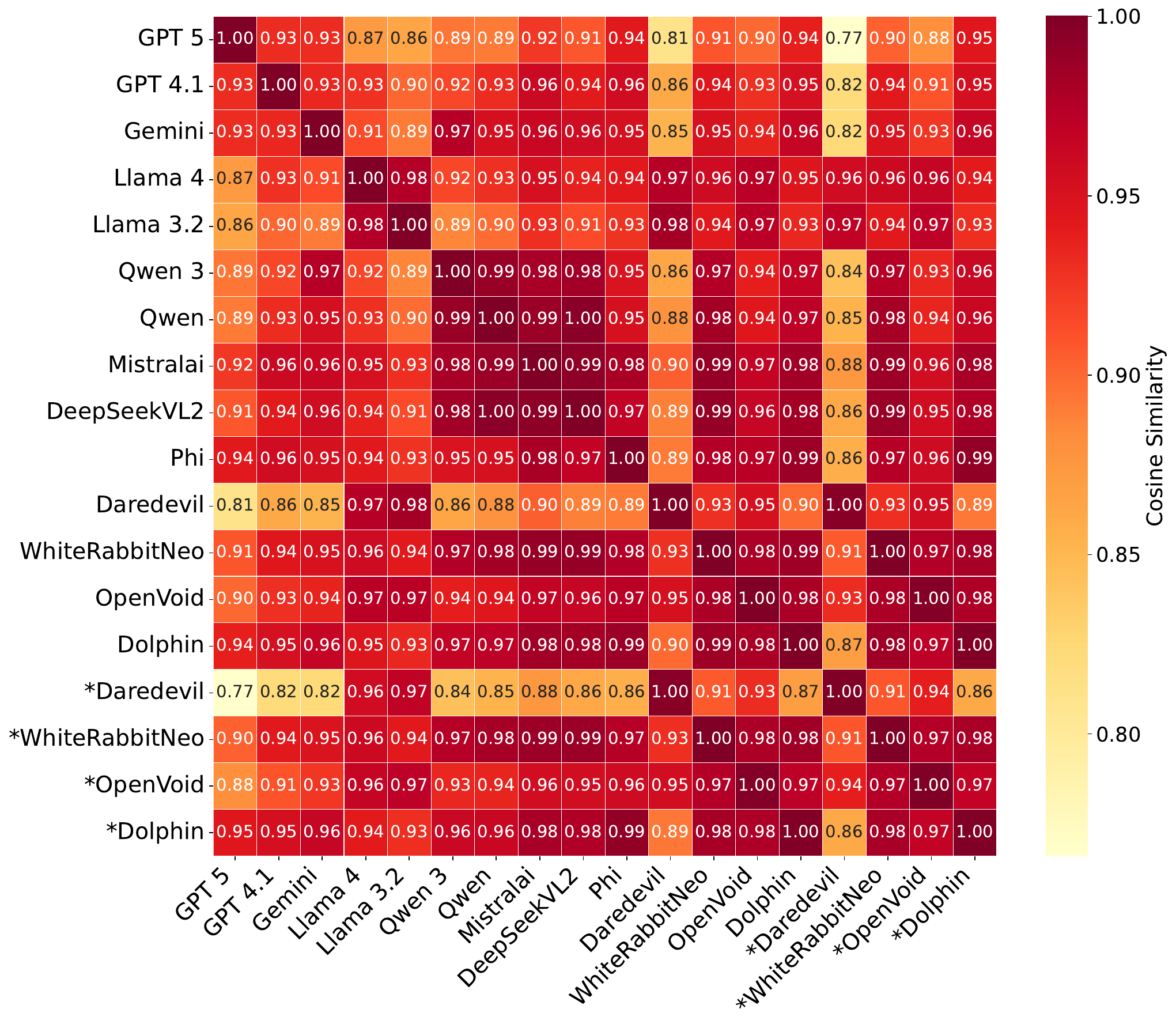}%
    }%
    \subfloat[Button Dataset\label{fig:heatmap-a}]{%
        \includegraphics[width=0.53\textwidth]{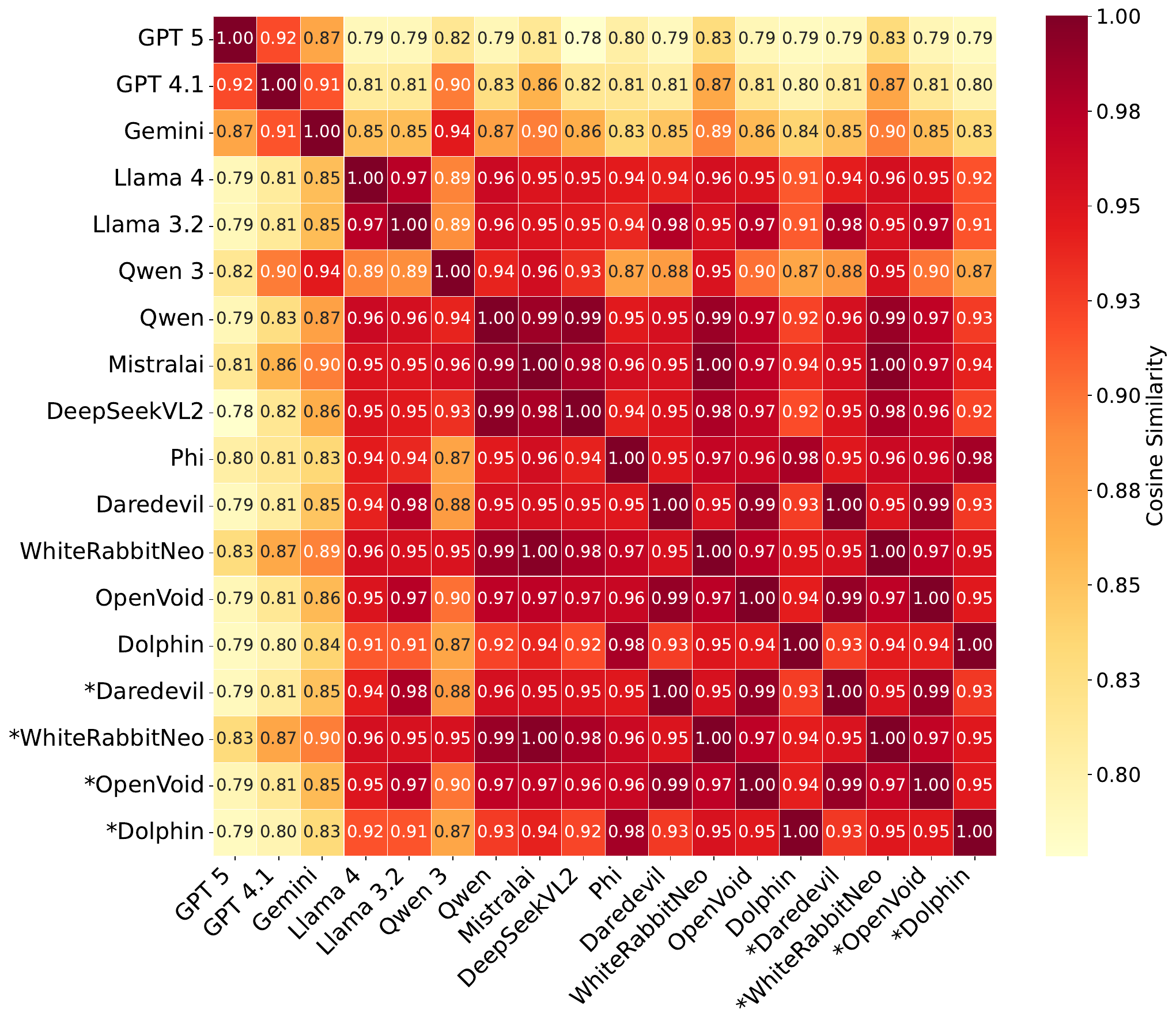}%
    }
    \caption{Cosine Similarity Heatmap of CSS Property Selection Across LLMs. * denotes adding ``phishing'' to prompts.}
    \label{fig:heatmap}
\end{figure*}


\begin{table}[!t]
\centering
\small
\caption{Top-20 CSS Properties in Buttons and Forms of Legitimate Websites (4 samples, covering 6 buttons and 8 forms).}
\setlength{\tabcolsep}{2pt}
\label{tab:legit_top20}
\resizebox{\linewidth}{!}{%
\begin{tabular}{l  rrlrr}
\toprule
\multicolumn{3}{c}{\textbf{Forms}} & \multicolumn{3}{c}{\textbf{Buttons}}  \\ \cmidrule(lr){1-3}\cmidrule(lr){4-6}
\multicolumn{1}{l}{CSS Property}
& \multicolumn{1}{c}{\# of Form (8)}
& \multicolumn{1}{c}{\# of Sample (4)}
& \multicolumn{1}{c}{CSS Property}
& \multicolumn{1}{c}{\# of Button (6)}
& \multicolumn{1}{c}{\# of Sample (4)} \\
\midrule
line-height      & 8 (100.0\%)           & 4 (100.0\%)           & $^*$bg-color     & 6 (100.0\%)           & 4 (100.0\%)            \\
color            & 8 (100.0\%)           & 4 (100.0\%)           & display          & 6 (100.0\%)           & 4 (100.0\%)            \\
font-size        & 8 (100.0\%)           & 4 (100.0\%)           & color            & 6 (100.0\%)           & 4 (100.0\%)            \\
font-family      & 8 (100.0\%)           & 4 (100.0\%)           & font-family      & 6 (100.0\%)           & 4 (100.0\%)            \\
box-sizing       & 5 \phantom{0}(62.5\%) & 3 \phantom{0}(75.0\%) & height           & 6 (100.0\%)           & 4 (100.0\%)            \\
$^+$m.           & 5 \phantom{0}(62.5\%) & 3 \phantom{0}(75.0\%) & text-align       & 6 (100.0\%)           & 4 (100.0\%)            \\
font-weight      & 4 \phantom{0}(50.0\%) & 2 \phantom{0}(50.0\%) & box-sizing       & 6 (100.0\%)           & 4 (100.0\%)            \\
position         & 3 \phantom{0}(37.5\%) & 2 \phantom{0}(50.0\%) & line-height      & 6 (100.0\%)           & 4 (100.0\%)            \\
padding          & 3 \phantom{0}(37.5\%) & 3 \phantom{0}(75.0\%) & font-size        & 6 (100.0\%)           & 4 (100.0\%)            \\
$^*$bg-color     & 2 \phantom{0}(25.0\%) & 2 \phantom{0}(50.0\%) & width            & 6 (100.0\%)           & 4 (100.0\%)            \\
display          & 2 \phantom{0}(25.0\%) & 2 \phantom{0}(50.0\%) & $^{**}$border-r  & 6 (100.0\%)           & 4 (100.0\%)            \\
border           & 2 \phantom{0}(25.0\%) & 2 \phantom{0}(50.0\%) & cursor           & 6 (100.0\%)           & 4 (100.0\%)            \\
$^{**}$border-r  & 2 \phantom{0}(25.0\%) & 2 \phantom{0}(50.0\%) & border           & 6 (100.0\%)           & 4 (100.0\%)            \\
width            & 2 \phantom{0}(25.0\%) & 2 \phantom{0}(50.0\%) & padding          & 6 (100.0\%)           & 4 (100.0\%)            \\
direction        & 2 \phantom{0}(25.0\%) & 1 \phantom{0}(25.0\%) & $^+$m.-top       & 5 \phantom{0}(83.3\%) & 4 (100.0\%)            \\ 
text-align       & 2 \phantom{0}(25.0\%) & 1 \phantom{0}(25.0\%) & font-weight      & 5 \phantom{0}(83.3\%) & 3 \phantom{0}(75.0\%)  \\
border-style     & 1 \phantom{0}(12.5\%) & 1 \phantom{0}(25.0\%) & position         & 5 \phantom{0}(83.3\%) & 3 \phantom{0}(75.0\%)  \\
padding-top      & 1 \phantom{0}(12.5\%) & 1 \phantom{0}(25.0\%) & $^{-}$text-d     & 5 \phantom{0}(83.3\%) & 3 \phantom{0}(75.0\%)  \\
box-shadow       & 1 \phantom{0}(12.5\%) & 1 \phantom{0}(25.0\%) & $^+$m.-right     & 4 \phantom{0}(66.7\%) & 3 \phantom{0}(75.0\%)  \\
border-width     & 1 \phantom{0}(12.5\%) & 1 \phantom{0}(25.0\%) & appearance       & 4 \phantom{0}(66.7\%) & 3 \phantom{0}(75.0\%)  \\
\bottomrule
\multicolumn{6}{l}{$^+$m. denotes margin, $^*$bg represents background, $^{**}$ border-r denotes border-radius;}\\
\multicolumn{6}{l}{$^{-}$ text-d represents text-decoration.}
\end{tabular}%
}
\end{table}

\subsection{HTML Similarity}
\label{apx:html_similar}

\begin{figure*}[!t]
\centering

\subfloat[Amazon]{%
    \includegraphics[width=0.49\textwidth]{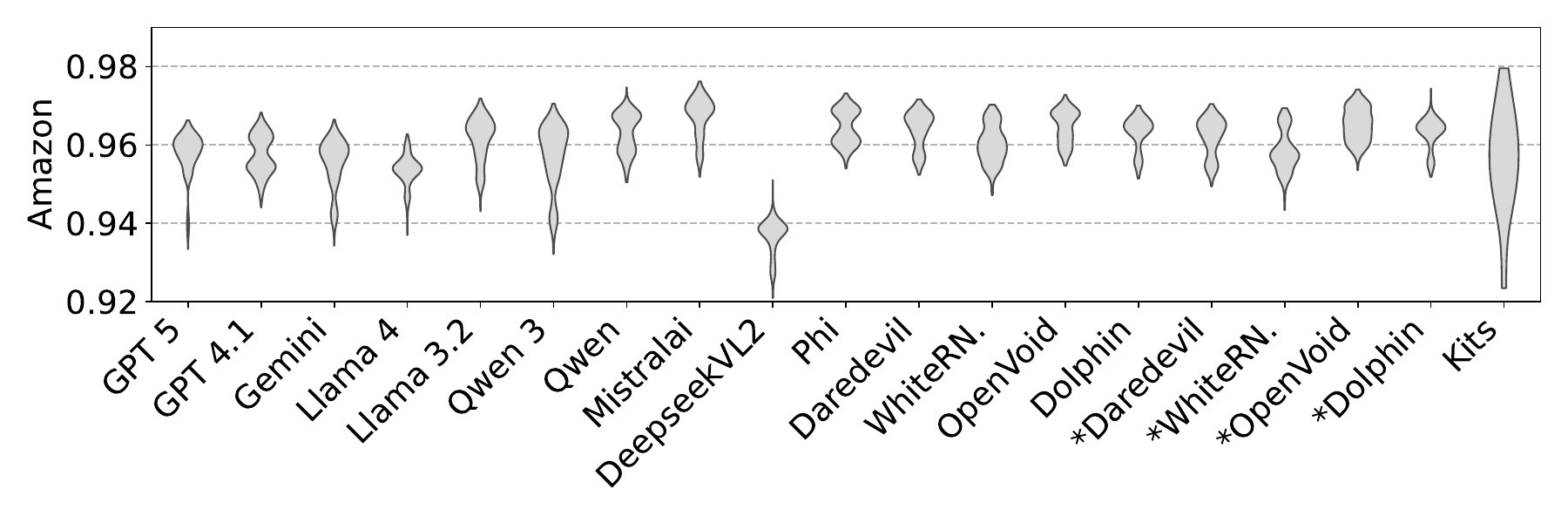}%
}%
\hfill
\subfloat[Facebook]{%
    \includegraphics[width=0.49\textwidth]{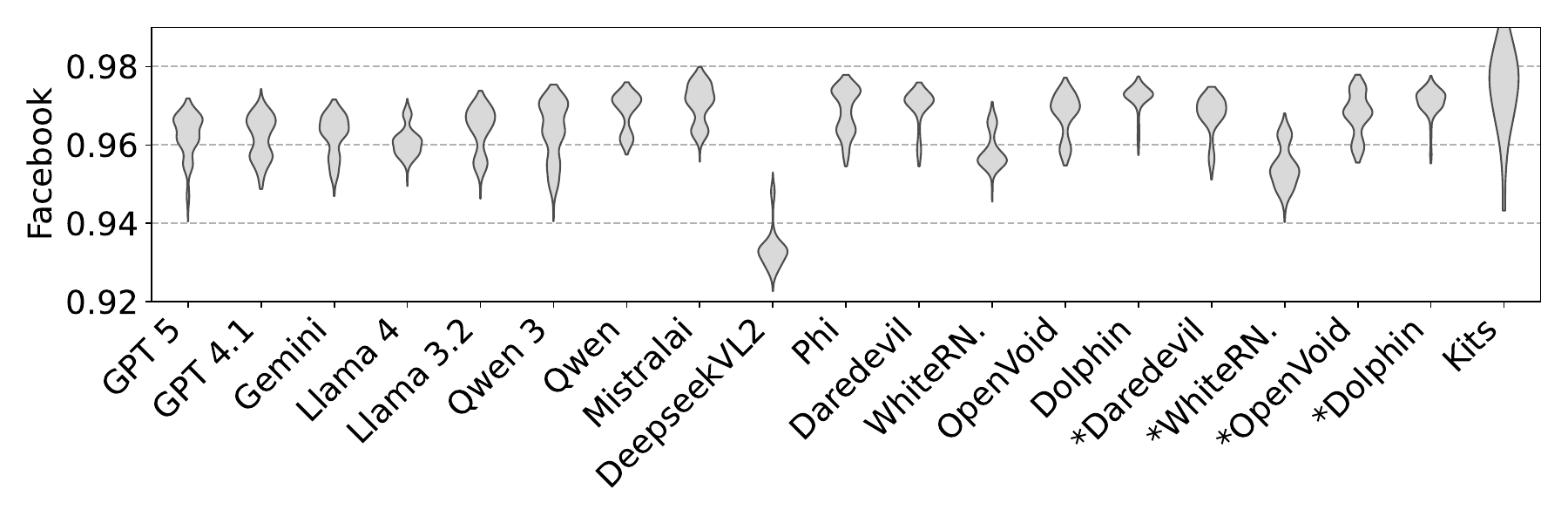}%
}\\

\subfloat[PayPal]{%
    \includegraphics[width=0.49\textwidth]{figures/figure_files/html_sim/paypal_violin2.pdf}%
}%
\hfill
\subfloat[Wells Fargo]{%
    \includegraphics[width=0.49\textwidth]{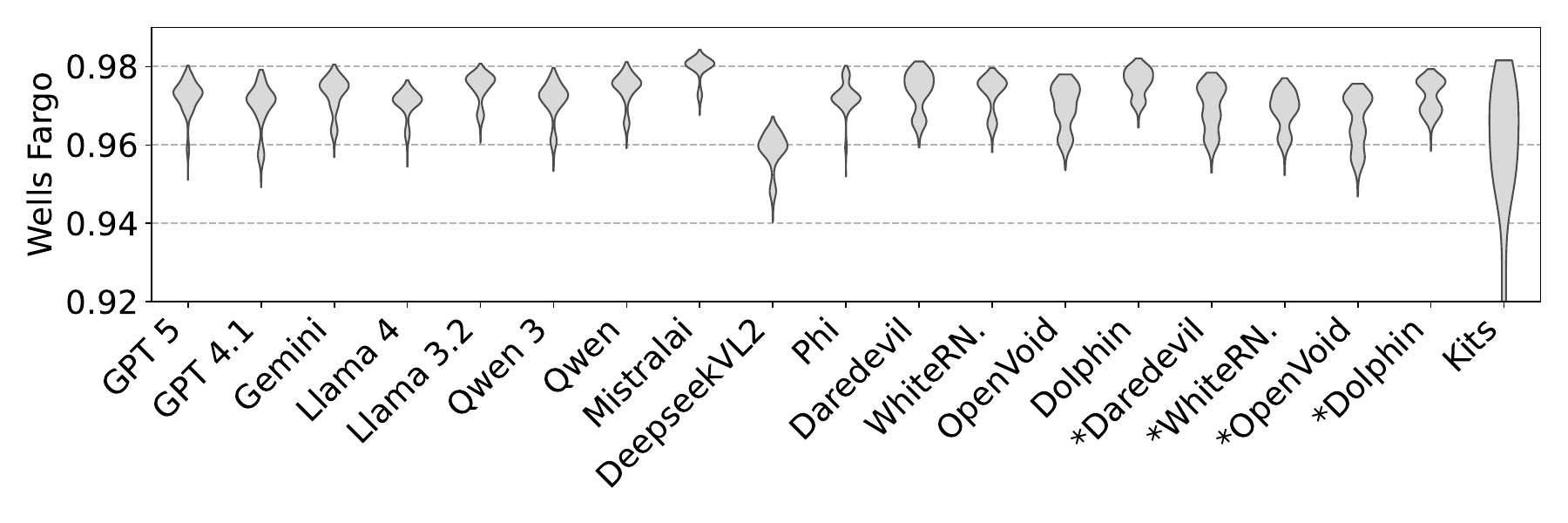}%
}%

\caption{MarkupLM-based HTML Cosine Similarity: Generated vs. Legitimate HTML. * denotes adding ``phishing'' to prompts.}
\label{fig:html_sim}
\end{figure*}

We analyze the HTML similarity between the generated websites and corresponding legitimate reference websites. Specifically, we employ MarkupLM~\cite{Li2021MarkupLMPO} to compute similarity scores, which leverages HTML text and structural information while excluding JavaScript, CSS, etc. The similarity results across different brands are shown in \autoref{fig:html_sim}, with detailed analyses presented in \autoref{subsec:html_sim}.

Across all four brands, generated HTML achieves consistently high similarity with the legitimate references ($>$ 0.92), indicating that LLMs can faithfully reproduce both the textual content and structural layout of target websites. PayPal exhibits the highest and most concentrated similarity (around 0.98), while Amazon and Wells Fargo show slightly broader distributions, suggesting brand-dependent variation in generation fidelity. DeepSeekVL2 stands out as a clear outlier, producing notably lower similarity (0.93–0.96) across most brands, which reflects weaker HTML generation capability rather than safety alignment. Explicitly including the term ``phishing'' in the prompt (denoted by *) yields nearly identical distributions to the unmodified prompts, indicating that uncensored models do not alter their HTML output based on intent disclosure. Overall, most LLMs are capable of generating structurally convincing replicas of legitimate websites.

\begin{figure*}[!t]
\centering
\includegraphics[width=0.9\linewidth]{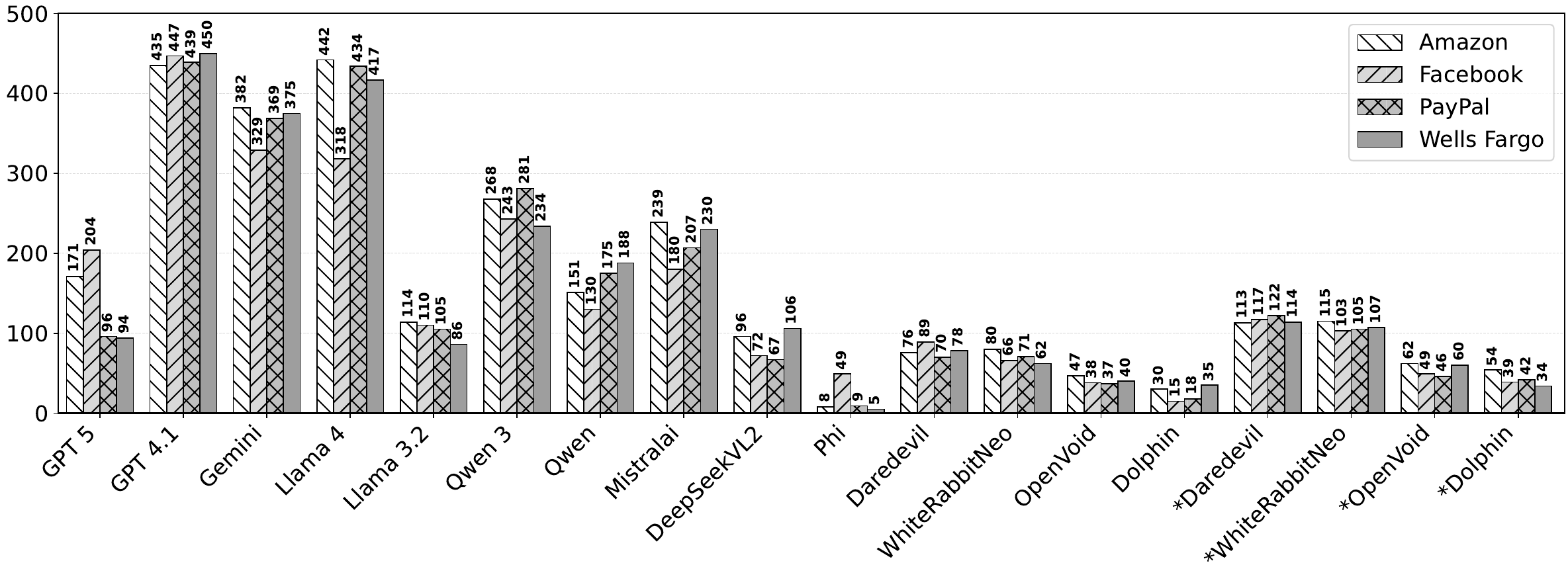}
\caption{Functionality statistics of valid LLM-generated phishing websites across four target brands. Bars show the number of functionality-successful samples for each model-brand pair, with Facebook exhibiting the largest within-model variation. (Total Valid Websites: 28{,}121; Total Functional Samples: 10{,}889). * denotes adding ``phishing'' to prompts.}
~\label{fig:bar_chart}
\end{figure*}

\subsection{HTML Style Heatmap} \label{apx:heatmap_button_form}
\autoref{fig:heatmap} presents comprehensive cosine similarity heatmaps for form and button CSS styles across all 18 evaluated models, providing a granular view of stylistic relationships across the full spectrum. 
Our results highlight two key characteristics of LLM-generated code.

\begin{itemize}[leftmargin=*, itemsep=0pt, parsep=0pt, topsep=0pt]
    \item Architectural Consistency: Models sharing the same architectural lineage (\eg, Qwen and Qwen 3) exhibit high similarity scores. This indicates that stylistic patterns are largely consistent across different versions or sizes within the same model family.
    \item Invariance to Malicious Intent: We observed that the stylistic similarity between uncensored models and their counterparts prompted with explicit phishing keywords (\eg, Daredevil and *Daredevil) reaches 1.0. This indicates that explicitly articulating malicious intent in the prompt does not alter the model's structural CSS generation patterns, implying that coding style is independent of the prompt's ethical context.
\end{itemize}
These findings validate the selection strategy employed in the main text. By excluding redundant variants, the primary analysis focused on meaningful stylistic divergence.

\subsection{CSS Properties for Legitimate Websites} \label{apx:benign_css}

\autoref{tab:legit_top20} presents the top-20 CSS properties used in buttons and forms on legitimate websites across four samples, covering six buttons and eight forms that were manually examined. Results show that legitimate websites frequently use specific properties across both forms and buttons, such as \texttt{<line-height>} in forms and \texttt{<background-color>} in buttons. For buttons in particular, all top 20 CSS properties are used by at least three legitimate websites, indicating a high degree of commonality. In contrast, LLM-generated websites utilize a broader variety of CSS properties, following the underlying distribution of the LLM.

\section{Functionality of Generated Websites} \label{appendix:result_function}

\autoref{fig:bar_chart} shows the functionality statistics of LLM-generated phishing websites with valid pages across four target brands.
Across brands, Facebook exhibits the largest within-model variation. For instance, Phi produces few functional pages ($ \leq$ 9) for Amazon, PayPal, and Wells Fargo, but jumps to 49 for Facebook. A similar pattern appears in GPT~5 (204 vs. 94-171). Conversely, Llama~4 shows the opposite trend, with Facebook (318) noticeably lower than other brands (318 vs. 417-442). Gemini (329 vs. 369-382) and Qwen (130 vs. 151-188) follow the same direction. In contrast, Amazon, PayPal, and Wells Fargo show small within-model differences, indicating that these three brands behave consistently under each model's generation capability and constraints.

Despite performance differences, models exhibit similar failure patterns on Facebook. The most frequent failure type is \textit{Unsaved} (46.5\%), followed by \textit{Submit Timeout} (14.9\%). These results indicate that models often succeed in generating visually plausible login pages but fail to complete the backend persistence step.
Consistently, \textit{Unsaved} is also the dominant failure type at the model level for nearly all models, except for DeepSeekVL2.
This suggests that performance differences across models are largely driven by how effectively they handle this shared persistence bottleneck.


\end{document}